\documentclass[review]{elsarticle}
\usepackage{graphics} 
\usepackage{epsfig} 
\usepackage{mathptmx} 
\usepackage{times} 
\usepackage{amsmath} 
\usepackage{amssymb}  
\usepackage{bm}
\usepackage{subfigure}
\usepackage{booktabs} 
\usepackage[left]{lineno}
\usepackage{array}
\usepackage{setspace}
\usepackage{makecell}
\usepackage{nomencl}
\usepackage{lscape}
\usepackage{float}
\usepackage{geometry}

\journal{ArXiv}

\begin{document}

\begin{frontmatter}

\title{Two points are enough}
\tnotetext[mytitlenote]{$^{\dagger}$ These authors contributed equally to this work: Hao Liu, Yanbin Zhao.}

\author[mymainaddress]{Hao Liu*$^{\dagger}$}
\cortext[mycorrespondingauthor]{Corresponding author}
\ead{haoliu7850052@zju.edu.cn}
\author[mymainaddress]{Yanbin Zhao*$^{\dagger}$}
\ead{zyb9825@zju.edu.cn}
\author[mysecondaryaddress]{Huarong Zheng}
\author[mythirdaddress]{Xiulin Fan}
\author[myfourthaddress]{Zhihua Deng}
\author[myfifthaddress]{Mengchi Chen}
\author[mysixthaddress]{Xingkai Wang}
\author[myfifthaddress]{Zhiyang Liu}
\author[mythirdaddress]{Jianguo Lu}
\author[myseventhaddress]{Jian Chen*}
\ead{chenj8@sustech.edu.cn}



\address[mymainaddress]{State Key Laboratory of Fluid Power Components and Mechatronic Systems, School of Mechanical Engineering, Zhejiang University, Hangzhou, China}
\address[mysecondaryaddress]{Ocean College, Zhejiang University, Hangzhou, China}
\address[mythirdaddress]{State Key Laboratory of Silicon and Advanced Semiconductor Materials, School of Materials Science and Engineering, Zhejiang University, Hangzhou, China}
\address[myfourthaddress]{Energy Research Institute at NTU (ERI@N), Nanyang Technological University, Singapore}
\address[myfifthaddress]{School of Control Science and Engineering, Zhejiang University, Hangzhou, China}
\address[mysixthaddress]{School of Materials Science and Engineering, Tianjin University, Tianjin, China}
\address[myseventhaddress]{School of System Design and Intelligent Manufacturing, Southern University of Science and Technology, Shenzhen, China}

\begin{abstract}
	
Prognosis and diagnosis play an important role in accelerating the development of lithium-ion batteries, as well as reliable and long-life operation. In this work, we answer an important question: What is the minimum amount of data required to extract features for accurate battery prognosis and diagnosis? Based on the first principle, we successfully extracted the best two-point feature (BTPF) for accurate battery prognosis and diagnosis using the fewest data points (only two) and the simplest feature selection method (Pearson correlation coefficient). The BTPF extraction method is tested on 820 cells from 6 open-source datasets (covering five different chemistry types, seven manufacturers, and three data types). It achieves comparable accuracy to state-of-the-art features in both prognosis and diagnosis tasks. This work challenges the cognition of existing studies on the difficulty of battery prognosis and diagnosis tasks, subverts the fixed pattern of establishing prognosis and diagnosis methods for complex dynamic systems through deliberate feature engineering, highlights the promise of data-driven methods for field battery prognosis and diagnosis applications, and provides a new benchmark for future studies.
 
\end{abstract}

\begin{keyword}
	
Lithium-ion battery, feature engineering, prognosis, diagnosis, first principle, remaining useful life, state of health, machine learning. 

\end{keyword}

\end{frontmatter}


\section{Introduction}

Lithium-ion (Li-ion) batteries have been commercialized on a large scale in portable electronic products, transportation cars, and stationary energy storage systems. These fields urgently need batteries with lower cost \cite{Link2024Rapidly, cano2018batteries}, higher energy density \cite{schmuch2018performance, viswanathan2022challenges}, shorter fast-charging time \cite{Wang2022Fast}, longer cycle life \cite{Janek2023Challenges, Gervillie2024Non}, and better safety \cite{Zhao2020Designing}. Prognosis \cite{hu2020battery} and diagnosis \cite{ng2020predicting} are vital for battery optimization. Prognosis can be utilized to accelerate the optimization of battery design \cite{Paulson2022Feature}, production \cite{Weng2021Predicting}, and management \cite{attia2020closed}, thereby saving a lot of time and economic costs. Diagnosis can be utilized to monitor the state of health (SoH), optimize management strategies, and perform reliable maintenance, thereby reducing the occurrence of safety incidents and extending the battery life \cite{Wang2022A, Zhang2023Realistic, Zhu2023A}. Recently, thanks to the open-source of many battery test datasets \cite{Reis2021Lithium-ion}, researchers have proposed many data-driven methods that combine machine learning (ML) and feature engineering to achieve high-precision battery prognosis and diagnosis \cite{Geslin2023Selecting, Huang2023Feature}, and have shown great potential in battery material development \cite{Paulson2022Feature, Ning2024Bridging}, production process optimization \cite{Weng2021Predicting}, aging mechanism research \cite{Vlijmen2024Interpretable, Wildfeuer2023Experimental}, charging protocol optimization \cite{attia2020closed}, health management \cite{roman2021machine, Zhang2023Realistic}, and recycling \cite{Tao2023Collaborative}.

Different features for battery prognosis and diagnosis are extracted from different data types, including charge/discharge data, impedance data, and relaxation data. The features extracted from charge/discharge data include various instantaneous features, statistical features, and model parameter features extracted from capacity vs. voltage ($Q/V$) data \cite{severson2019data, Jiang2021Bayesian, Guo2024Semi-supervised, Wang2023Large-scale, Lu2023Deep, Wang2024Physics}, incremental capacity ($dQ/dV$) data \cite{Dubarry2021Analysis, Dubarry2020Big, Dubarry2017State, Li2024Predicting, Tao2023Collaborative}, and differential voltage ($dV/dQ$) data \cite{Dubarry2021Analysis}. 

Regarding the features extracted from $Q/V$ data, Severson et al. \cite{severson2019data} proposed the feature for early battery life prediction based on the discharge $Q/V$ data of the 10th and 100th cycles, $var(\bigtriangleup Q_{100-10}/V(V))$, and applied it to accelerate the optimization of fast-charging protocols \cite{attia2020closed}. Furthermore, Attia et al. \cite{Attia2021Statistical} proposed the $IQR(\bigtriangleup Q_{100-10}/V(V))$ feature (the $IQR$ of $\bigtriangleup Q_{100-10}/V(V)$ represents the difference between the 75th and 25th percentiles of the capacity values in the $\bigtriangleup Q_{100-10}/V(V)$ vector) for early battery life prediction based on the same dataset in \cite{severson2019data}, which outperforms the $var(\bigtriangleup Q_{100-10}/V(V))$ feature. Note that the $IQR(\bigtriangleup Q_{100-10}/V(V))$ feature inspired the bold thinking of this paper to a great extent.
Paulson et al. \cite{Paulson2022Feature} extracted a total of 396 features from charge and discharge data and studied the impact of feature selection and chemistry type on early battery life prediction. To reduce the demand for charge/discharge data, Jiang et al. \cite{Jiang2021Bayesian} and Guo et al. \cite{Guo2024Semi-supervised} conducted improved early battery life prediction research based on the open-source dataset published in \cite{severson2019data}, further reducing the demand for test cycles and labeled data. Jiang et al. \cite{Jiang2021Bayesian} also proposed to extract the feature mean(square($\bigtriangleup V_{3-2}(Q)$) from the charge Q/V data for early-cycle classification of battery lifetime. Weng et al. \cite{Weng2021Predicting} found that the resistance feature measured at 5\% SOC after battery formation indicates lithium consumption during the formation process and can be used for cycle life prediction. In terms of reducing the demand for charge/discharge data for diagnostic features, Roman et al. \cite{roman2021machine}, Wang et al. \cite{Wang2024Physics}, and Lu et al. \cite{Lu2023Deep} used different ML models to achieve accurate estimation of SoH using partial charge data.

Since changes in battery voltage may be related to changes in the electrochemical reactions occurring at the two electrodes, voltage change features that characterize the thermodynamic and kinetic states can be utilized for battery prognosis and diagnosis \cite{Zheng2018Incremental, Weng2013On-board, Li2020State, Wei2022Remaining, Xia2023State, Wang2016On-board}. The features extracted from $dQ/dV$ data and $dV/dQ$ data are voltage change features \cite{Barai2019A}. Dubarry et al. \cite{Dubarry2021Analysis, Dubarry2020Big, Dubarry2017State} extracted interpretable instantaneous features, statistical features, and model parameter features from $dQ/dV$ curves and $dV/dQ$ curves for the prognosis and diagnosis of batteries with three different chemistry types (i.e., LFP, NCA, and NMC811). Tao et al. \cite{Tao2023Collaborative} extracted 30 instantaneous and statistical features from charge/discharge $Q/V$ curves and charge/discharge $dQ/dV$ curves for the classification and recycling of batteries with five different chemistry types (i.e., LCO, NMC, LFP, NCA, and NCM-LCO). To reduce the demand for $dQ/dV$ and $dV/dQ$ data in feature engineering, Li et al. \cite{Li2024Predicting} extracted the feature $mean(\bigtriangleup dQ/dV_{w3-w0}^(3.6V-3.9V)(V))$ for early battery life prediction under different usage conditions from the local (3.60V-3.90V range) $\bigtriangleup dQ/dV$ curve obtained by subtracting the $dQ/dV$ data in the reference performance test (RPT) of week 0 from the $dQ/dV$ data in RPT of week 3.

Electrochemical impedance spectroscopy (EIS) is a non-destructive method that can reveal the electrode kinetic processes at different time scales inside the battery, including charge transfer reactions, interface evolution, and mass diffusion \cite{Hu2023Application}. By correlating the impedance evolution with the battery degradation mechanism, EIS provides important insights into the evolution of internal electrochemical processes during battery aging and has become a powerful diagnostic and prognostic tool \cite{Iurilli2021On}. Impedance data features include instantaneous and statistical features extracted from EIS and distribution of relaxation times (DRT) data, as well as model parameter features obtained by equivalent circuit model (ECM) fitting. Zhu et al. \cite{Zhu2022Data} identified three resistance parameters for characterizing battery SoH by fitting EIS data using ECM. Jones et al. \cite{Jones2022Impedance} proposed using future cycling protocols and all raw EIS impedances after full discharge (114 real and imaginary impedances at 57 frequencies between 0.02 Hz and 20 kHz) as features for SoH estimation of batteries under uneven usage. Jiang et al. \cite{Jiang2022A} compared the performance of raw EIS impedance features, ECM parameter features, and impedance features at specific frequencies in battery SoH estimation. The comparative test results show that the impedance features at specific frequencies have comprehensive and excellent performance in battery SOH estimation, including accuracy, estimation coverage, and training time. Zhou et al. \cite{Zhou2023State} proposed to use the center coordinates and radius features of the EIS semicircular curve in the high-medium frequency ranges to estimate the battery SoH, and compared them with the ECM parameter features and impedance features at specific frequencies. Test results show the highest SoH estimation accuracy of the center coordinate and radius features. To reduce the demand for EIS data in feature engineering, Zhang et al. \cite{Zhang2020Identifying} determined the importance weights of EIS impedance features at different frequencies through the automatic relevance determination (ARD) of the Gaussian process regression (GPR) model. They found that the two imaginary impedance features in the low-frequency region (17.80 Hz and 2.16 Hz) are sufficient to estimate the battery SoH accurately. Note that this study also inspired the bold thinking of this paper to a great extent.

DRT can directly distinguish the time constants of the main electrochemical processes inside the battery, thereby simplifying EIS analysis \cite{Ning2024Bridging, Maradesa2024Advancing}. Since the time scale information of the dynamic process inside the battery is closely related to the battery SoH, DRT is becoming a promising technology for battery prognosis and diagnosis \cite{Zhao2022Investigation, Zhang2022Degradation, He2023Comparative, Chen2021Detection, Soni2022Lithium, Jung2024A}. Zhu et al. \cite{Zhu2023Adaptive} extracted 12 statistical features from the complete DRT curve and selected five strongly correlated features for battery SoH estimation through the Pearson correlation coefficient. Su et al. \cite{Su2024Modeling} extracted 17 features from the DRT curve, including ECM parameter features and statistical features, for battery SoH estimation. Then, eight strongly correlated features are selected through the Spearman correlation coefficient and converted into one indirect feature using the weighted principal component analysis.

In addition to the charge/discharge process, the battery relaxation process after the current is interrupted also contains important battery aging information \cite{Fan2023Battery}. After the battery is disconnected from the charge or discharge current, lithium ions continue to diffuse into the active materials, causing the battery voltage to gradually equilibrate with its open circuit voltage (OCV) \cite{Fernando2024Benchmark}. 
The change in relaxation voltage over relaxation time is closely related to loss of Li inventory ($LLI$) and loss of active material $LAM$ in battery aging \cite{Fan2023Battery, Chen2022Battery, Chen2021A}, is not limited by charge/discharge protocols \cite{Zhu2022Data, Jiang2023An}, and can be used for prognosis and diagnosis \cite{Xiang2024Two, Qian2019State, Reichert2013Influence}. Chen et al. \cite{Chen2022Battery, Chen2021A} extracted instantaneous and statistical features from the cycle-by-cycle cutoff relaxation voltage data after full charge and discharge to classify the dominant battery aging modes, including solid electrolyte interphase (SEI)-driven $LLI$, Li plating related $LLI$, and $LAM$ in positive electrode ($LAM_{PE}$). Zhu et al. \cite{Zhu2022Data, Jiang2023An} used three statistical features (i.e., variance, skewness, and maximum) extracted from the complete voltage relaxation data after full charge to estimate the SoH of batteries from three different manufacturers. To reduce the demand for DRT data in feature engineering, Fan et al. \cite{Fan2023Battery} used 10 seconds of relaxation voltage data during a 2-hour relaxation process to estimate the battery SoH.

Batteries are one of the hottest research topics in prognosis and diagnosis \cite{hu2020battery, ng2020predicting}, and many data-driven battery prognosis and diagnosis methods are published annually. To the best of the authors' knowledge, no data-driven method has answered this fundamental question: What is the minimum amount of data required to extract an accurate feature for battery prognosis and diagnosis? Solving this problem is very important for promoting the field deployment and online application of data-driven methods, especially for large battery systems consisting of hundreds or thousands of cells. Significantly reducing the data requirements for feature engineering means significantly reducing data collection, storage, and computational costs (including time and economic costs) of data-driven methods \cite{Sulzer2021The}. Taking the data-driven methods based on EIS data as an example, since the complete EIS measurement process includes impedance measurements at dozens of frequencies in a wide frequency range, the measurement process often takes tens of seconds to several minutes. If the input features need to be extracted from the complete EIS data, it will greatly hinder the field deployment and online application of data-driven methods. Suppose the input features can be extracted from the impedance data corresponding to two high frequencies. In that case, it will facilitate the field deployment and online application of data-driven methods, because some electric vehicles have integrated online impedance measurement functions on DC/DC, and can quickly measure the impedance values corresponding to the two high frequencies. Although hundreds of different battery prognosis and diagnosis features have been proposed for data-driven methods, most rely on a large amount of test data at a specific stage under specific conditions \cite{Huang2023Feature}. Although existing studies have made some progress in freeing data-driven methods from their dependence on specific test conditions and data at specific stages \cite{Wang2023Large-scale, Zhang2023Realistic, Steininger2023Automated}, the current progress in reducing the data requirements of data-driven methods is far from enough. This paper will conduct a first-principle study to answer the above fundamental question.

Based on the first principle, we start with the simplest single data point and feature selection method (Pearson correlation coefficient), gradually increasing the number of data points. When the number of data points increased to only two, we were pleasantly surprised to find that the best two-point feature (BTPF) obtained by simple subtraction of two data points achieved excellent performance in battery prognosis and diagnosis tasks after simple selection by the Pearson correlation coefficient. Furthermore, we proposed a general method to extract BTPFs for prognosis and diagnosis tasks from different types of data (including charge/discharge data, impedance data, and relaxation data), and conducted fair comparison tests with multiple state-of-the-art (SOAT) features on six open-source datasets (including 820 Li-ion cells from seven different manufacturers and five different chemistry types). The comparison results show that the accuracy of BTPFs in battery prognosis and diagnosis tasks is comparable to that of SOAT features. 

This work challenges the cognition of existing studies on the difficulty of battery prognosis and diagnosis tasks. Existing studies generally believe that accurate battery prognosis and diagnosis is challenging \cite{severson2019data, ng2020predicting} due to the complex coupling of many factors in battery design (significant variability in materials, chemistry types, and structures), production (significant variability in manufacturers, production equipment, and processes), and usage (significant variability in device, fields, and operating conditions) \cite{han2019review}. This paper proposes a general BTPF extraction method for battery prognosis and diagnosis tasks based on only two data points and the simplest Pearson correlation coefficient. We test the performance of BTPFs against existing SOAT features using different types of data from hundreds of battery cells. The results of this paper will prompt researchers to rethink and re-evaluate the difficulty of battery prognosis and diagnosis tasks.

This work overturns the fixed pattern of establishing data-driven prognosis and diagnosis methods for complex dynamical systems through deliberate feature engineering. Based on the first principle, this paper attempts to use the simplest single data point and feature selection method (Pearson correlation coefficient), gradually increasing the number of data points. It is ultimately determined that two data points are the minimum amount of data required to extract an accurate prognostic and diagnostic feature, answering the above fundamental question. The BTPF extraction method proposed in this paper provides a new direction for feature engineering of complex dynamical systems. Moreover, the BTPF extraction method provides a new benchmark for feature engineering of data-driven battery prognosis and diagnosis methods.

This work highlights the promise of data-driven field battery prognostic and diagnostic methods. Taking EIS data as an example, the BTPF dramatically reduces the dozens of frequencies required for a complete EIS measurement to two frequencies, significantly shortening the measurement time and making it possible to perform battery prognosis and diagnosis through online impedance measurement at two specific frequencies. It is worth mentioning that some electric vehicles have already integrated the online impedance measurement function on DCDC \cite{Hasegawa2016Development, Mizutani2017On}. In addition, for large battery systems consisting of hundreds or thousands of cells, such as those in passenger cars \cite{Winter2018Before} and energy storage stations \cite{Preger2020Degradation}, the use of two-point features can significantly reduce the data collection, storage, and computational costs required for prognosis and diagnosis of each cell, helping to promote the deployment of simpler and faster battery prognostic and diagnostic methods in different fields.

\section{Data}

In this paper, six open-source battery datasets are selected to test the proposed BTPF extraction method. These datasets contain 820 Li-ion cells with seven different manufacturers, five chemistry types, and three data types, as shown in \textbf{Supplementary Table 1}. For ease of reference, we number these open-source datasets as Dataset 1 to Dataset 6. Among them, Dataset 1 and Dataset 2 are used for the prognosis task. Dataset 1 has variable charging protocols and a constant discharging protocol, and Dataset 7 has variable charging and discharging protocols. The features used for the prognosis task in Dataset 1 are extracted from the discharge $Q/V$ data in different cycles, and the features used for the prognosis task in Dataset 7 are extracted from the discharge $dQ/dV$ and $dV/dQ$ data in different periodic RPTs. Dataset 3 and Dataset 4 are used for the diagnosis task. Dataset 3 has variable charging and discharging protocols, and Dataset 4 has constant charging and discharging protocols. The features used for the diagnosis task in Dataset 3 are extracted from the EIS data after full discharge in each cycle, and the features used for the diagnosis task in Dataset 4 are extracted from the EIS data after full charge in each cycle. Dataset 5 and Dataset 6 are used for the diagnosis task, both of which have variable charging and discharging protocols. The features used for the diagnosis task in Dataset 5 are extracted from the relaxation $V/t$ data after full charge in each cycle, and the features used for the diagnosis task in Dataset 6 are extracted from the relaxation V/t data after full discharge in each periodic RPT.

To compare the BTPF with SOAT features in recent publications, this paper also uses capacity to characterize the battery SoH. Meanwhile, the cycle number corresponding to the full discharge capacity decay to 80\% of the nominal capacity is used as the battery cycle life. For more detailed dataset and cell descriptions, please refer to the corresponding references in \textbf{Supplementary Table 1}, which will not be repeated here.

\section{Methods}

As shown in Figure 1, the BTPF extraction method are proposed for battery prognosis and diagnosis tasks, which mainly includes four steps: (1) Data collection. (2) Difference calculation. (3) Feature extraction. (4) Feature selection.

\begin{figure}
	\centering
	\includegraphics[width=7.8cm]{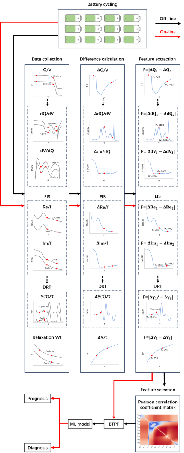}
	\caption{The BTPF extraction method for battery prognosis and diagnosis tasks.}
	\label{fig1}
\end{figure}


The detailed steps of the BTPF extraction method for the prognosis task are: (1) Data collection. Collect the target data in two specific battery cycles (such as the 10th and 100th cycles in \cite{severson2019data}). The data types include charge/discharge $Q/V$ data, charge/discharge $dQ/dV$ data, charge/discharge $dV/dQ$ data, EIS data, DRT data, and relaxation $V/t$ data after charge/discharge. (2) Difference calculation. Firstly, the target data collected in two specific cycles is standardized to facilitate subsequent calculations. The spline function is utilized to fit the target data and a unified linear interpolation is performed. For the charge/discharge $Q/V$ data, the capacity is fitted as a function of voltage and evaluated with same linearly spaced voltage points. For the charge/discharge $dQ/dV$ data, the incremental capacity is fitted as a function of voltage and evaluated with same linearly spaced voltage points. For the charge/discharge $dV/dQ$ data, the differential voltage is fitted as a function of capacity and evaluated with same linearly spaced capacity points. For the relaxation $V/t$ data after charge/discharge, the relaxation voltage is fitted as a function of relaxation time and evaluated with same linearly spaced time points. Then, subtract the fitted target data in the lower cycle from that of the higher cycle to obtain the difference curve, which can be one of the charge/discharge capacity difference ($\bigtriangleup Q/V$) curve, charge/discharge incremental capacity difference ($\bigtriangleup dQ/dV$) curve, charge/discharge differential voltage difference ($\bigtriangleup dV/dQ$) curve, and relaxation voltage difference ($\bigtriangleup V/t$) curve after charge/discharge. It should be mentioned that since the frequency and time data points of EIS and DRT curves in different cycles are originally the same, there is no need to perform spline function fitting and unified linear interpolation. We can directly subtract the original EIS and DRT data in the lower cycle from that in the higher cycle to obtain the real impedance difference ($\bigtriangleup Re/f$) curve, the imaginary impedance difference ($\bigtriangleup Im/f$) curve, and the DRT difference ($\bigtriangleup \gamma (\tau)/\tau$) curve. (3) Feature extraction. Extract candidate two-point features from the difference curve. Subtract the ordinate values corresponding to any two abscissa values on the differential curve and take the absolute value as a candidate two-point feature. Traverse all combinations of two abscissa values  on the differential curve, and a total of $(n^2-n)/2$ candidate two-point features can be obtained, where $n$ is the number of abscissa values on the differential curve. Further, all candidate two-point features are obtained by traversing all cells in the training set. (4) Feature selection. The Pearson correlation coefficient between each candidate two-point feature and the cycle life of all cells in the training set is calculated, and the feature corresponding to the Pearson correlation coefficient with the largest absolute value is selected as the BTPF. The BTPFs of all cells in the training set are used as input and the corresponding cycle life of all cells in the training set is used as output to train the ML regression model. When making an online prognosis, only two target data points in two specific cycles need to be collected to calculate the BTPF. Then, the BTPF is input into the trained ML regression model to predict the cycle life of the target cell.

The detailed steps of the two-point feature extraction method for the diagnosis task are: (1) Data collection. Collect targeted data in every cycle throughout the cell life. The data types include charge/discharge Q/V data, charge/discharge $dQ/dV$ data, charge/discharge $dV/dQ$ data, EIS data, DRT data, and relaxation $V/t$ data after charge/discharge. (2) Difference calculation. Firstly, the target data collected in each cycle is standardized to facilitate subsequent calculations. The spline function is utilized to fit the target data and a unified linear interpolation is performed. For the charge/discharge $Q/V$ data, the capacity is fitted as a function of voltage and evaluated with same linearly spaced voltage points. For the charge/discharge $dQ/dV$ data, the incremental capacity is fitted as a function of voltage and evaluated with same linearly spaced voltage points. For the charge/discharge $dV/dQ$ data, the differential voltage is fitted as a function of capacity and evaluated with same linearly spaced capacity points. For the relaxation $V/t$ data after charge/discharge, the relaxation voltage is fitted as a function of relaxation time and evaluated with same linearly spaced time points. Then, subtract the fitted target data in the first cycle from that of the higher cycle to obtain the difference curve, which can be one of the charge/discharge $\bigtriangleup Q/V$ curves, charge/discharge $\bigtriangleup dQ/dV$ curves, charge/discharge $\bigtriangleup dV/dQ$ curves, and relaxation $\bigtriangleup V/t$ curves after charge/discharge. It should be mentioned that since the frequency and time data points of EIS and DRT curves in different cycles are originally the same, there is no need to perform spline function fitting and unified linear interpolation. We can directly subtract the original EIS and DRT data in the first cycle from that in higher cycles to obtain the $\bigtriangleup Re/f$ curves, the $\bigtriangleup Im/f$ curves, and the $\bigtriangleup \gamma (\tau)/\tau$ curves.  (3) Feature extraction. Extract candidate two-point features from the difference curve. Subtract the ordinate values corresponding to any two abscissa values on the differential curve and take the absolute value as a candidate two-point feature. Traverse all combinations of two abscissa values on the differential curve, and a total of $(n^2-n)/2$ candidate two-point features can be obtained, where $n$ is the number of abscissa values on the differential curve. Further, all candidate two-point features are obtained by traversing all cells in the training set. (4) Feature selection. The Pearson correlation coefficient between each candidate two-point feature and the cycle capacities of all cells in the training set is calculated, and the feature corresponding to the Pearson correlation coefficient with the largest absolute value is selected as the BTPF. The BTPFs of all cells in the training set are used as input and the corresponding cycle capacities of all cells in the training set is used as output to train the ML regression model. When making an online diagnosis, only two target data points in the first cycle and the targe cycle need to be collected to calculate the BTPF. Then, the BTPF is input into the trained ML regression model to estimate the SoH of the target cell.

\section{Results}

The comparison results of the BTPF proposed in this paper and the recent SOAT features on different datasets are shown in Tables 1 to 6. The detailed results for each dataset are presented below. 

\begin{table}[h]
	\caption{Comparison results of the BTPF and SOAT features on 124 cells in Dataset 1. The BTPF (Q/V) respresents the feature $PTP_{DIS:2.735V-2.900V}\bigtriangleup Q_{100-10}/V(V)$ and the SOAT feature respresents the feature $var(\bigtriangleup Q_{100-10}/V(V))$ proposed in \cite{severson2019data}.}
	\label{table_1}
	\begin{center}
		\begin{tabular}{c|c|c|c|c|c|c}
			\hline
			 \multicolumn{7}{c}{Dataset 1}  \\
			\hline
			\multicolumn{1}{c}{} &  \multicolumn{2}{c}{Training set}  & \multicolumn{2}{c}{Primary test set} & \multicolumn{2}{c}{Secondary test set}  \\
			\hline
			Metrics & \makecell{BTPF \\ (Q/V)} & \makecell{SOAT \\ \cite{severson2019data}}  & \makecell{BTPF \\ (Q/V)}  & \makecell{SOAT \\ \cite{severson2019data}}  & \makecell{BTPF \\ (Q/V)} & \makecell{SOAT \\ \cite{severson2019data}} \\
			\hline
			MAE (cycles) & 42.6 & 44.9  & 91.1  & 73.9 & 148.4 & 138.4 \\
			\hline	
			MAPE (\%) & 5.3 & 5.8 & 12.4 & 9.3 & 13.9 & 12.6 \\
			\hline
			RMSE (cycles) & 73 & 75 & 129 & 117 & 197 & 191 \\
			\hline
			$R^{2} (-)$ & 0.94 & 0.95 & 0.89 & 0.91 & 0.58 & 0.61 \\
			\hline	
			\makecell{Data used \\in one cycle\\ (in two cycles)} & \makecell{2 \\ (4)} & \makecell{About 800\\ (1600)}  & \makecell{2 \\ (4)}  & \makecell{About 800\\ (1600)} & \makecell{2 \\ (4)} & \makecell{About 800\\ (1600)} \\
			\hline								
		\end{tabular}
	\end{center}
\end{table}

\begin{table*}[h]
	\caption{Comparison results of the BTPF and SOAT features on 225 cells in Dataset 2. The BTPF (dQ/dV) respresents the feature $PTP_{DIS:3.648V-4.080V}\bigtriangleup dQ_{w3-w0}/dV(V)$, the BTPF (dV/dQ) respresents the feature $PTP_{DIS:0.2016Ah-0.2324Ah}\bigtriangleup dV_{w3-w0}/dQ(Q)$, and the SOAT feature respresents the feature $mean(\bigtriangleup dQ/dV_{w3-w0}^{3.6V-3.9V}(V))$ proposed in \cite{Li2024Predicting}.}
	\label{table_2}
	\begin{center}
		\begin{tabular}{c|c|c|c|c|c|c}
			\hline
			\multicolumn{7}{c}{Dataset 2}  \\
			\hline
			\multicolumn{1}{c}{} &  \multicolumn{3}{c}{High-DoD training set}  & \multicolumn{3}{c}{High-DoD test set}   \\
			\hline
			Metrics & \makecell{BTPF \\ (dQ/dV)} & \makecell{BTPF \\ (dV/dQ)}  & \makecell{SOAT \\ \cite{Li2024Predicting}} & \makecell{BTPF \\ (dQ/dV)} & \makecell{BTPF \\ (dV/dQ)} & \makecell{SOAT \\ \cite{Li2024Predicting}}  \\
			\hline
			MAE (cycles) & 3.09 & 0.96 & 2.66 & 2.92 & 1.05 & 2.52  \\
			\hline	
			MAPE (\%) & 24.8 & 23.5 & 17.1 & 20.0 & 16.4 & 18.3  \\
			\hline
			RMSE (cycles) & 1.50 & 1.27 & 1.55 & 1.67 & 1.44 & 1.84  \\
			\hline
			$R^{2} (-)$ & 0.68 & 0.76 & 0.66 & 0.58 & 0.69 & 0.49  \\
			\hline	
			\makecell{Data used \\in one cycle\\ (in two cycles)} & 2 (4) & 2 (4)  & \makecell{250\\ (500)} & 2 (4) & 2 (4) & \makecell{250\\ (500)}  \\
			\hline
			\hline
			\multicolumn{1}{c}{} &  \multicolumn{3}{c}{Low-DoD test set}  & \multicolumn{3}{c}{}  \\
			\hline
			Metrics & \makecell{BTPF \\ (dQ/dV)} & \makecell{BTPF \\ (dV/dQ)}  & \makecell{SOAT \\ \cite{Li2024Predicting}} &  &  &   \\
			\hline
			MAE (cycles) & 8.19 & 6.76 & 7.71 &  &  &   \\
			\hline	
			MAPE (\%) & 30.6 & 33.2 & 28.6 &  &  &   \\
			\hline
			RMSE (cycles) & 5.11 & 8.03 & 5.27 &  &  &   \\
			\hline
			$R^{2} (-)$ & 0.18 & 0.25 & 0.03 &  &  &   \\
			\hline	
			\makecell{Data used \\in one cycle\\ (in two cycles)} & \makecell{2 \\ (4)} & \makecell{2 \\ (4)}  & \makecell{250\\ (500)} &  &  &  \\
			\hline								
		\end{tabular}
	\end{center}
\end{table*}

\begin{table}[h]
	\caption{Comparison results of the BTPF and SOAT features on 40 cells in Dataset 3. The BTPF (EIS) respresents the feature $DTP_{ADIS: 1922.08Hz-4.36Hz}\bigtriangleup Re_{n-1}/f(f)$, the BTPF (DRT) respresents the feature $DTP_{ADIS:1.23e^{-4}-5.41e^{-3}}\bigtriangleup \gamma (\tau)_{n-1}/\tau$, and the SOAT feature respresents the raw EIS impedances proposed in \cite{Jones2022Impedance}.}
	\label{table_3}
	\begin{center}
		\begin{tabular}{c|c|c|c|c|c|c}
			\hline
			\multicolumn{5}{c}{Dataset 3}  \\
			\hline
			\multicolumn{1}{c}{} &  \multicolumn{3}{c}{Training set}  & \multicolumn{3}{c}{Test set}  \\
			\hline
			Metrics & \makecell{BTPF \\ (EIS)} &  \makecell{BTPF \\ (DRT)} & \makecell{SOAT \\ \cite{Jones2022Impedance}}  & \makecell{BTPF \\ (EIS)} & \makecell{BTPF \\ (DRT)} & \makecell{SOAT \\ \cite{Jones2022Impedance}} \\
			\hline
			MAE (mAh) & 2.74 & 2.96 & 2.60 & 2.90 & 2.86 & 3.41 \\
			\hline	
			MAPE (\%) & 10.60 & 11.50 & 10.40 & 11.61 & 11.51 & 13.41 \\
			\hline
			RMSE (mAh) & 3.33 & 3.64 & 3.14 & 3.55 & 3.59 & 4.27 \\
			\hline
			$R^{2} (-)$ & 0.69 & 0.63 & 0.72 & 0.65 & 0.65 & 0.49 \\
			\hline	
			\makecell{Data used \\in one cycle\\ (in two cycles)} & \makecell{2 \\ (4)} & \makecell{2 \\ (4)}  & \makecell{57\\ (-)} & \makecell{2 \\ (4)} & \makecell{2 \\ (4)} & \makecell{57\\ (-)} \\
			\hline								
		\end{tabular}
	\end{center}
\end{table}

\begin{table}[h]
	\caption{Comparison results of the BTPF and SOAT features on 8 cells in Dataset 4. The BTPF (EIS) respresents the feature $DTP_{ACH:115.778Hz-11.145Hz}\bigtriangleup Im_{n-1}/f(f)$, the BTPF (DRT) respresents the feature $DTP_{ACH:0.003s-14.42s}\bigtriangleup \gamma (\tau)_{n-1}/\tau$, and the SOAT feature respresents the raw EIS impedances proposed in \cite{Zhang2020Identifying}.}
	\label{table_4}
	\begin{center}
		\begin{tabular}{c|c|c|c|c|c|c}
			\hline
			\multicolumn{5}{c}{Dataset 4}  \\
			\hline
			\multicolumn{1}{c}{} &  \multicolumn{3}{c}{Training set}  & \multicolumn{3}{c}{Test set}  \\
			\hline
			Metrics & \makecell{BTPF \\ (EIS)} &  \makecell{BTPF \\ (DRT)} & \makecell{SOAT \\ \cite{Zhang2020Identifying}}  & \makecell{BTPF \\ (EIS)} & \makecell{BTPF \\ (DRT)} & \makecell{SOAT \\ \cite{Zhang2020Identifying}} \\
			\hline
			MAE (mAh) & 0.58 & 0.73 & 0.87 & 2.95 & 2.33 & 2.45 \\
			\hline	
			MAPE (\%) & 2.08 & 2.59 & 3.08 & 12.82 & 10.06 & 11.32 \\
			\hline
			RMSE (mAh) & 0.86 & 0.92 & 0.81 & 3.77 & 3.38 & 3.57 \\
			\hline
			$R^{2} (-)$ & 0.90 & 0.89 & 0.89 & 0.29 & 0.53 & 0.35 \\
			\hline	
			\makecell{Data used \\in one cycle\\ (in two cycles)} & \makecell{2 \\ (4)} & \makecell{2 \\ (4)}  & \makecell{ 60 \\ (-)} & \makecell{2 \\ (4)} & \makecell{2 \\ (4)} & \makecell{ 60 \\ (-)} \\
			\hline								
		\end{tabular}
	\end{center}
\end{table}

\begin{table}[h]
	\caption{Comparison results of the BTPF and SOAT features on cells in Dataset 5. The BTPF (V/t) respresents the feature $DTP_{ACH:0s-936s}\bigtriangleup V_{n-1}/t(t)$ and the SOAT feature respresents the features [Var, Ske, Max] proposed in \cite{Zhu2022Data}.}
	\label{table_1}
	\begin{center}
		\begin{tabular}{c|c|c|c|c}
			\hline
			\multicolumn{5}{c}{Dataset 5}  \\
			\hline
			\multicolumn{1}{c}{} &  \multicolumn{2}{c}{Training set}  & \multicolumn{2}{c}{Test set}  \\
			\hline
			Metrics & \makecell{BTPF\\ (V/t)} & \makecell{SOAT\\ \cite{Zhu2022Data}}  & \makecell{BTPF\\ (V/t)} & \makecell{SOAT\\ \cite{Zhu2022Data}} \\
			\hline
			MAE (mAh) & 0.003 & 0.002 & 0.013  & 0.011 \\
			\hline	
			MAPE (\%) & 0.358 & 0.241 & 1.644 & 1.328 \\
			\hline
			RMSE (mAh) & 0.004 & 0.003 & 0.018 & 0.016 \\
			\hline
			$R^{2} (-)$ & 0.958 & 0.982 & 0.921 & 0.943 \\
			\hline	
			\makecell{Data used \\in one cycle\\ (in two cycles)} & \makecell{2 \\ (4)} & \makecell{ 15 \\ (-)}  & \makecell{2 \\ (4)}  & \makecell{ 15 \\ (-)} \\
			\hline								
		\end{tabular}
	\end{center}
\end{table}

\begin{table}[h]
	\caption{Comparison results of the BTPF and SOAT features on cells in Dataset 6. The BTPF (V/t) respresents the feature $DTP_{ADIS:6120s-6984s}\bigtriangleup V_{n-1}/t(t)$ and the SOAT feature respresents the features [Var, Ske, Max] proposed in \cite{Zhu2022Data}.}
	\label{table_1}
	\begin{center}
		\begin{tabular}{c|c|c|c|c}
			\hline
			\multicolumn{5}{c}{Dataset 6}  \\
			\hline
			\multicolumn{1}{c}{} &  \multicolumn{2}{c}{Training set}  & \multicolumn{2}{c}{Test set}  \\
			\hline
			Metrics & \makecell{BTPF\\ (V/t)} & \makecell{SOAT\\ \cite{Zhu2022Data}}  & \makecell{BTPF\\ (V/t)} & \makecell{SOAT\\ \cite{Zhu2022Data}} \\
			\hline
			MAE (mAh) & 0.14 &  0.15 & 0.16 & 0.16 \\
			\hline	
			MAPE (\%) & 0.19 & 0.28 & 0.14 & 0.16 \\
			\hline
			RMSE (mAh) & 0.22 & 0.23 & 0.22 & 0.23 \\
			\hline
			$R^{2} (-)$ & 0.86 & 0.85 & 0.87 & 0.85 \\
			\hline	
			\makecell{Data used \\in one cycle\\ (in two cycles)} & \makecell{2 \\ (4)} & \makecell{720 \\ (-)}  & \makecell{2 \\ (4)} & \makecell{720 \\ (-)} \\
			\hline								
		\end{tabular}
	\end{center}
\end{table}

\subsection{\textbf{Performance on the prognosis task}}

\subsubsection{\textbf{Dataset 1}}

For Dataset 1, the BTPF extraction method for battery prognosis task is shown in \textbf{Supplementary Figure 1}. To make a fair comparison with the SOAT feature $var(\bigtriangleup Q_{100-10}/V(V))$ proposed by Severson et al. \cite{severson2019data}, this paper uses the same training and test sets as \cite{severson2019data}. Specifically, Dataset 1 contains 169 LFP cells with nominal capacity of 1.1 Ah. Among them, the first 124 cells are published by Severson et al. \cite{severson2019data}, the remanining 45 cells are published by Attia et al. \cite{attia2020closed}. The 124 LFP cells are divided into three sub-datasets, namely the training set (41 cells), the primary test set (43 cells), and the secondary test set (40 cells). All remanining 45 cells are used as the third test set. The training set is utilized to train the ML regression model, and the test sets are utilized to verify and test the predictive performance of the trained ML regression model. For more details about the training and test sets, please refer to \cite{severson2019data, attia2020closed} and will not be repeated here.

\textbf{Data collection:} Similar to \cite{severson2019data}, the discharge $Q/V$ data in the 10th and 100th cycles are collected (the constant current-constant voltage (CC-CV) discharge mode is used, the discharge current in the CC stage is 4C, the discharge cut-off voltage in the CC stage is 2.0V, and the cut-off discharge current in the CV stage is C/50) for feature extraction. 

\textbf{Difference calculation:} The discharge $Q/V$ data collected in the 10th and 100th cycles are standardized to facilitate subsequent calculations. Specifically, the spline function is utilized to fit the discharge $Q/V$ data in the CC stage, and a unified linear interpolation is performed. The discharge capacity is fitted as a function of the discharge voltage, and the discharge voltage is linearly divided into 100 values between 3.5 V and 2.0 V at intervals of 15 mV, as shown in \textbf{Supplementary Figure 2(a)}. It should be noted here that in \cite{severson2019data}, the discharge voltage is linearly divided into 1000 values between 3.5 V and 2.0 V at intervals of 1.5 mV. To reduce the computational burden, this paper uses a larger voltage division interval of 15mV than 1.5mV in \cite{severson2019data}. Increasing the voltage interval does not affect the fairness of the comparison. The fitted discharge $Q/V$ data in the two cycles are subtracted (the discharge $Q/V$ data in the 100th cycle minus the discharge $Q/V$ data in the 10th cycle) to obtain the discharge $\bigtriangleup Q_{100-10}/V$ curve. As shown in \textbf{Supplementary Figure 2(b)}, the obtained discharge $\bigtriangleup Q_{100-10}/V$ curve is the shaded part between the two discharge $Q/V$ curves in the 10th and 100th cycles. The discharge $\bigtriangleup Q_{100-10}/V$ curves of first 124 cells in Dataset 1 \cite{severson2019data} are shown in \textbf{Supplementary Figure 2(c)}.

\textbf{Feature extraction:} The capacity difference ($\bigtriangleup Q_{100-10}$) values corresponding to any two voltage values on the discharge $\bigtriangleup Q_{100-10}/V$ curve are subtracted and the absolute value is taken as a candidate two-point feature, as shown in \textbf{Supplementary Figure 3}. Since the discharge voltage is linearly divided into 100 values between 3.5 V and 2.0 V in this paper, a total of $(100^2-100)/2=4950$ candidate two-point features can be obtained by traversing all combinations of two voltage values on the discharge $\bigtriangleup Q_{100-10}/V$ curve, as shown in Figure 2(a).

\textbf{Feature selection:} The Pearson correlation coefficient between each candidate two-point feature and the cycle life of 41 cells in the training set is calculated, and the candidate two-point feature corresponding to the correlation coefficient with the largest absolute value of 0.873 is selected as the BTPF $PTP_{DIS:2.735V-2.900V}\bigtriangleup Q_{100-10}/V(V)$, as shown in Figure 2(a). For $PTP_{DIS:2.735V-2.900V}\bigtriangleup Q_{100-10}/V(V)$, $PTP$ represents a two-point feature for the prognosis task, $DIS:2.735V-2.900V$ represents that the two-point feature extraction uses the two discharge $\bigtriangleup Q$ values corresponding to the two discharge voltages $2.735V$ and $2.900V$, $\bigtriangleup Q_{100-10}/V$ represents that the two-point feature extraction uses the $\bigtriangleup Q_{100-10}/V$ curve obtained by subtracting the $Q/V$ data in the 10th cycle from the $Q/V$ data in the 100th cycle, and $(V)$ represents that the $Q/V$ data is fitted as a function of voltage. The two data points on the $\bigtriangleup Q_{100-10}/V$ curve that utilized to calculate $PTP_{DIS:2.735V-2.900V}\bigtriangleup Q_{100-10}/V(V)$ are shown in \textbf{Supplementary Figure 2(c)}. 
The distribution relationship between the cycle life of all 124 cells in Dataset 1 and $PTP_{DIS:2.735V-2.900V}\bigtriangleup Q_{100-10}/V(V)$ is shown in Figure 2(b). The Pearson correlation coefficient is 0.851, which is lower than that of feature $var(\bigtriangleup Q_{100-10}/V(V))$ (0.93 in \cite{severson2019data}), as shown in \textbf{Supplementary Figure 4}.

\begin{figure}[htbp]
	\centering
	\subfigure[]{\label{fig2:subfig1}\includegraphics[width=1\textwidth]{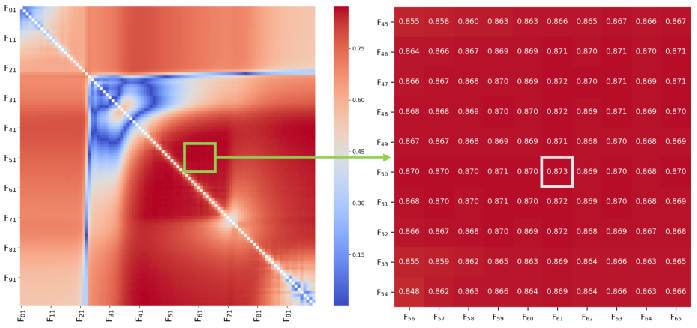}}
	\subfigure[]{\label{fig2:subfig2}\includegraphics[width=0.8\textwidth]{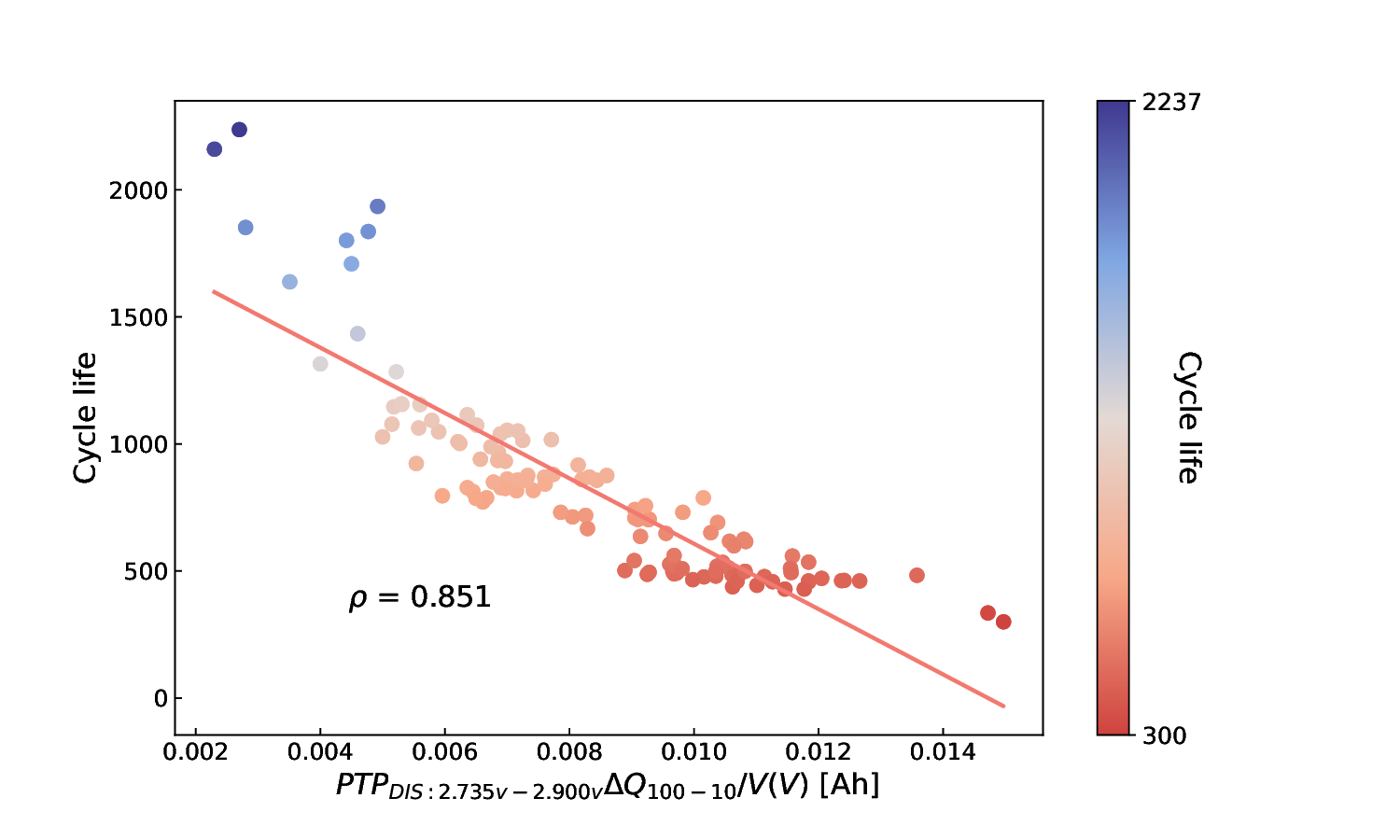}}
	\caption{(a) Selection of the BTPF using the Pearson correlation coefficient between each candidate two-point feature and the cycle life of 41 cells in the training set. F01 to F100 represent 100 values of discharge voltage from 2.0 V to 3.5 V, with an interval of 15 mV. The colors are determined based on the Pearson correlation coefficient values. (b) Cycle life of first 124 cells in Dataset 1 plotted as a function of $PTP_{DIS:2.735V-2.900V}\bigtriangleup Q_{100-10}/V(V)$, with a Pearson correlation coefficient of 0.851. The colors are determined based on the cycle life of cells.}
	\label{fig2}
\end{figure}


\textbf{Cycle life prediction:} Combining the BTPF $PTP_{DIS:2.735V-2.900V}\bigtriangleup Q_{100-10}/V(V)$ and the XGBoost regression model, the cell cycle life prediction results of first 124 cells in Dataset 1 are shown in Figure 3(a). For more detailed XGBoost regression model training and hyperparameter adjustment, please refer to \textbf{Supplementary Note 1}. Combining the feature $var(\bigtriangleup Q_{100-10}/V(V))$ \cite{severson2019data} and the XGBoost regression model of this paper (using exactly the same model training and hyperparameter adjustment method), the cell cycle life prediction results of first 124 cells in Dataset 1 are shown in Figure 3(b). The cell cycle life prediction results of different features on the training set, primary test set, and secondary test set are shown in Table 1.  More detailed results are shown in \textbf{Supplementary Tables 2 to 4}. in this paper, the matrics of mean absolute error (MAE), mean absolute percentage error (MAPE), root mean square error (RMSE), and $R^2$ are chosen to evaluate XGBoost regression model performance in the prognosis and diagnosis tasks. For more detailed matrics introduction, please refer to \textbf{Supplementary Note 2}.
It can be found that $PTP_{DIS:2.735V-2.900V}\bigtriangleup Q_{100-10}/V(V)$ achieves a comparable accuracy to that of $var(\bigtriangleup Q_{100-10}/V(V))$ \cite{severson2019data}, while $var(\bigtriangleup Q_{100-10}/V(V))$ \cite{severson2019data} uses about 800 discharge $Q/V$ data points in each cycle, which is about 400 times that of $PTP_{DIS:2.735V-2.900V}\bigtriangleup Q_{100-10}/V(V)$. Fewer data points mean less data collection, storage, and computational costs.

\begin{figure}[htbp]
	\centering
	\subfigure[]{\label{fig3:subfig1}\includegraphics[width=0.8\textwidth]{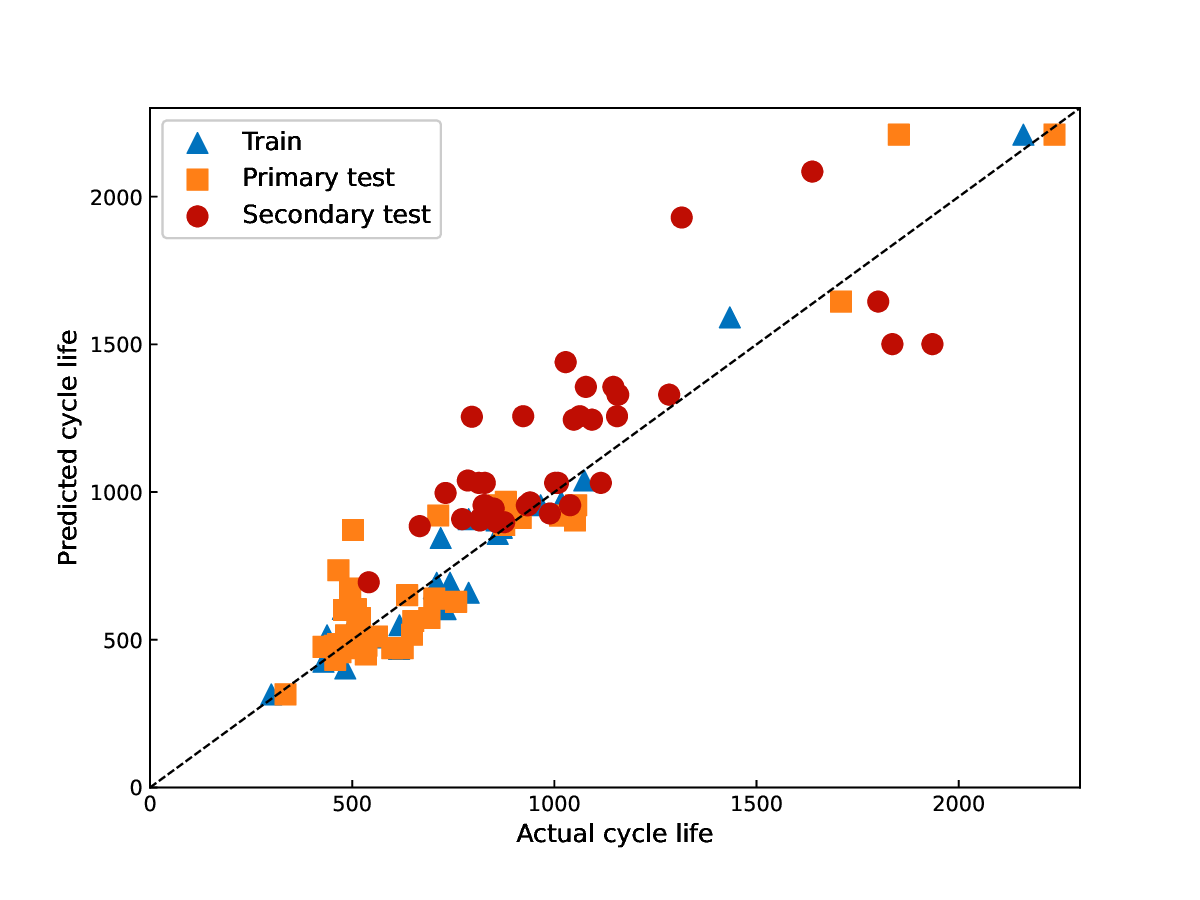}}
	\subfigure[]{\label{fig3:subfig2}\includegraphics[width=0.8\textwidth]{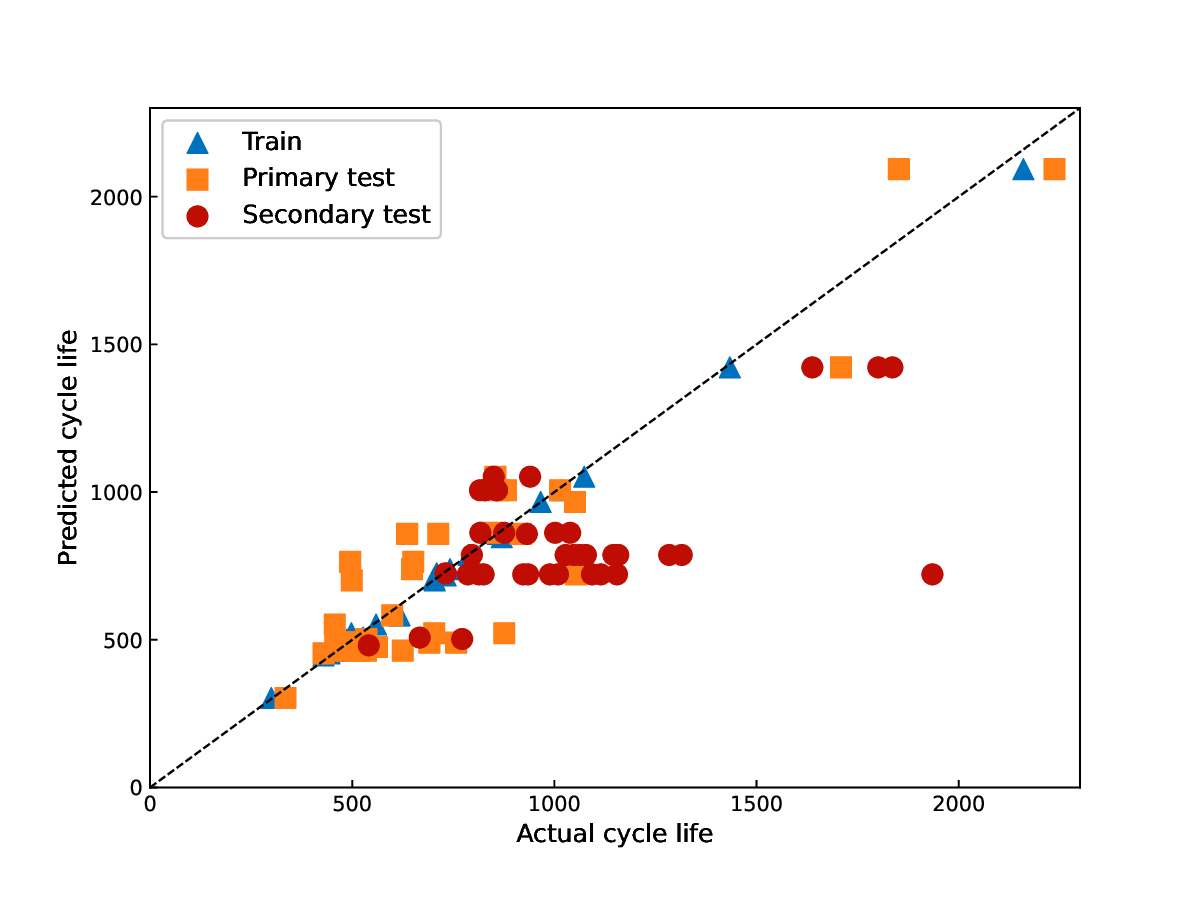}}
	\caption{Cycle life prediction results of of first 124 cells in Dataset 1. (a) Cycle life prediction results of $PTP_{DIS:2.735V-2.900V}\bigtriangleup Q_{100-10}/V(V)$. The input of XGBoost regression model is $PTP_{DIS:2.735V-2.900V}\bigtriangleup Q_{100-10}/V(V)$, and the output is the cell cycle life. (b) Cycle life prediction results of $var(\bigtriangleup Q_{100-10}/V(V))$ \cite{severson2019data}. The input of the XGBoost regression model is $var(\bigtriangleup Q_{100-10}/V(V))$, and the output is the cell cycle life.}
	\label{fig3}
\end{figure}


Furthermore, the third test set with 45 cells \cite{attia2020closed} in Dataset 1 is utilized to test the generalizability of $PTP_{DIS:2.735V-2.900V}\bigtriangleup Q_{100-10}/V(V)$, and the previously trained XGBoost regression model is directly used for prediction. The test results are shown in \textbf{Supplementary Figure 5} and \textbf{Supplementary Table 4}.

\subsubsection{\textbf{Dataset 2}}

(1) \quad BTPF extraction from $dQ/dV$ data

For Dataset 2, the BTPF extraction method for battery prognosis task is shown in \textbf{Supplementary Figure 6}. To make a fair comparison with the SOAT feature $mean(\bigtriangleup dQ/dV_{w3-w0}^{3.6V-3.9V}(V))$ proposed by Li et al. \cite{Li2024Predicting}, this paper uses the same training and test sets as \cite{Li2024Predicting}. Specifically, Dataset 2 containing 225 NMC cells is divided into three sub-datasets according to the depth of discharge (DOD), namely the high-DoD training set (116 cells), the high-DoD test set (60 cells), and the low-DoD test set (49 cells). The high-DoD training set is utilized to train the ML regression model, and the high-DoD test set and the low-DoD test set are utilized to verify and test the prediction performance of the trained ML regression model. For more details about the training and test sets, please refer to \cite{Li2024Predicting} and will not be repeated here.

\textbf{Data collection:} Similar to \cite{Li2024Predicting}, the discharge $Q/V$ data in the periodic RPTs of week 0 and week 3 are collected (CC discharge mode is used, the discharge current is C/5, and discharge cut-off voltage is 3.0V) and $dQ/dV$ curves are calculated for feature extraction. 

\textbf{Difference calculation:} Discharge $dQ/dV$ curves in two periodic RPTs are standardized to facilitate subsequent calculation calculations, as shown in \textbf{Supplementary Figure 7(a)}. The fitted discharge $dQ/dV$ curves in the two RPTs are subtracted to obtain the discharge $\bigtriangleup dQ_{w3-w0}/dV$ curve, as shown in \textbf{Supplementary Figure 7(b)}. The discharge $\bigtriangleup dQ_{w3-w0}/dV$ curves of all 225 cells in Dataset 2 are shown in \textbf{Supplementary Figure 7(c)}.

\textbf{Feature extraction:} The incremental capacity difference ($\bigtriangleup dQ_{w3-w0}$) values corresponding to any two voltage values on the discharge $\bigtriangleup dQ_{w3-w0}/dV$ curve are subtracted and the absolute value is taken as a candidate two-point feature, as shown in \textbf{Supplementary Figure 8}. A total of $(100^2-100)/2=4950$ candidate two-point features can be obtained by traversing all combinations of two voltage values on the discharge $\bigtriangleup dQ_{w3-w0}/dV$ curve.

\textbf{Feature selection:} The Pearson correlation coefficient between each candidate two-point feature and the week life of all 116 cells in the training set is calculated, and the candidate two-point feature corresponding to the correlation coefficient with the largest absolute value of 0.892 is selected as the BTPF $PTP_{DIS:3.648V-4.080V}\bigtriangleup dQ_{w3-w0}/dV(V)$, as shown in Figure 4(a). The two data points on the $\bigtriangleup dQ_{w3-w0}/dV$ curve that utilized to calculate $PTP_{DIS:3.648V-4.080V}\bigtriangleup dQ_{w3-w0}/dV(V)$ are shown in \textbf{Supplementary Figure 7(c)}. 
The distribution relationship between the week life of all 225 cells in Dataset 2 and $PTP_{DIS:3.648V-4.080V}\bigtriangleup dQ_{w3-w0}/dV(V)$ is shown in Figure 4(b). The Pearson correlation coefficient is 0.827, which is slightly lower than that of $log(mean(\bigtriangleup dQ/dV_{w3-w0}^{3.6V-3.9V}(V)))$ (0.848 in \cite{Li2024Predicting}), as shown in \textbf{Supplementary Figure 9}.

\begin{figure}[htbp]
	\centering
	\subfigure[]{\label{fig4:subfig1}\includegraphics[width=1.0\textwidth]{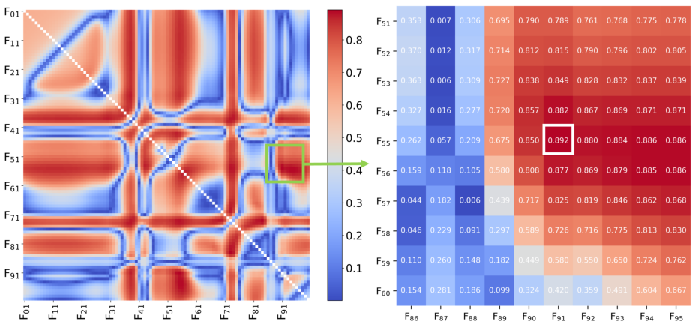}}
	\subfigure[]{\label{fig4:subfig2}\includegraphics[width=0.8\textwidth]{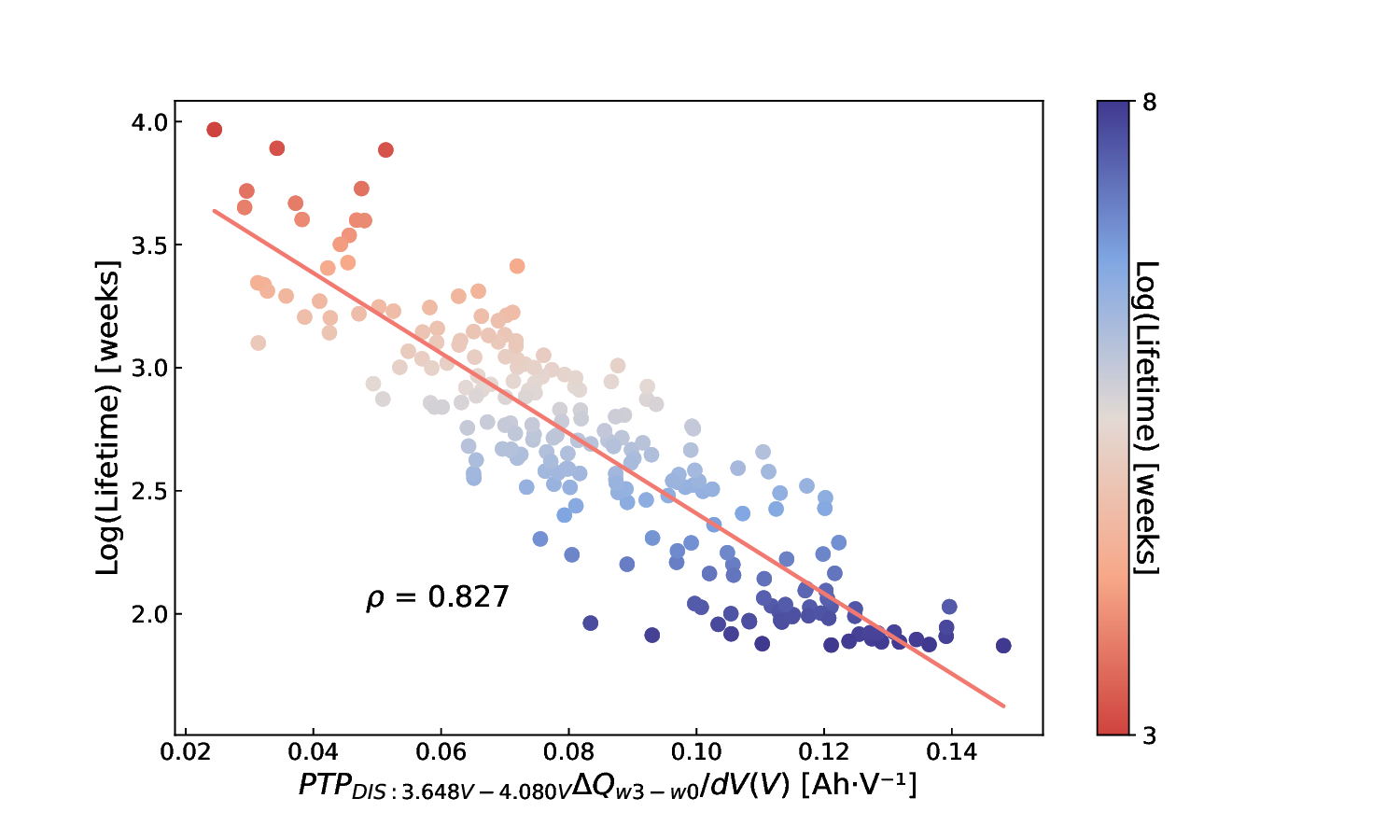}}
	\caption{(a) Selection of the BTPF using the Pearson correlation coefficient between each candidate two-point feature and the week life of 116 cells in the training set. F01 to F100 represent 100 values of discharge voltage from 3.0 V to 4.2 V at intervals of 12 mV. The colors are determined based on the Pearson correlation coefficient values. (b) Week life of 225 cells in Dataset 2 plotted as a function of $PTP_{DIS:3.648V-4.080V}\bigtriangleup dQ_{w3-w0}/dV(V)$, with a Pearson correlation coefficient of 0.827. The colors are determined based on the week life of cells.}
	\label{fig4}
\end{figure}


\textbf{Week life prediction:} Combining the BTPF $PTP_{DIS:3.648V-4.080V}\bigtriangleup dQ_{w3-w0}/dV(V)$ and the XGBoost regression model, the cell week life prediction results are shown in Figure 5(a). Combining the feature $mean(\bigtriangleup dQ/dV_{w3-w0}^{3.6V-3.9V}(V))$ \cite{Li2024Predicting} and the XGBoost regression model, the cell week life prediction results are shown in Figure 5(b).

\begin{figure}[htbp]
	\centering
	\subfigure[]{\label{fig5:subfig1}\includegraphics[width=0.8\textwidth]{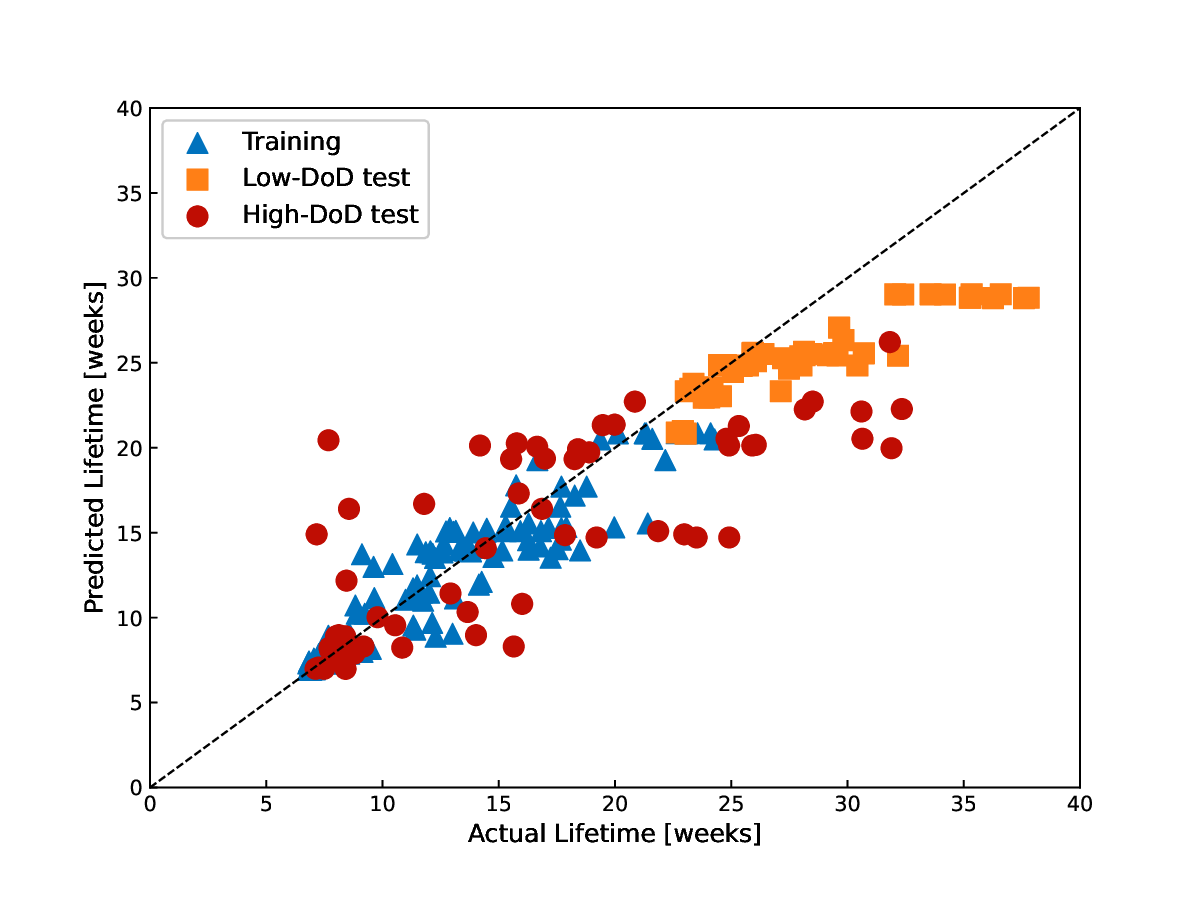}}
	\subfigure[]{\label{fig5:subfig2}\includegraphics[width=0.8\textwidth]{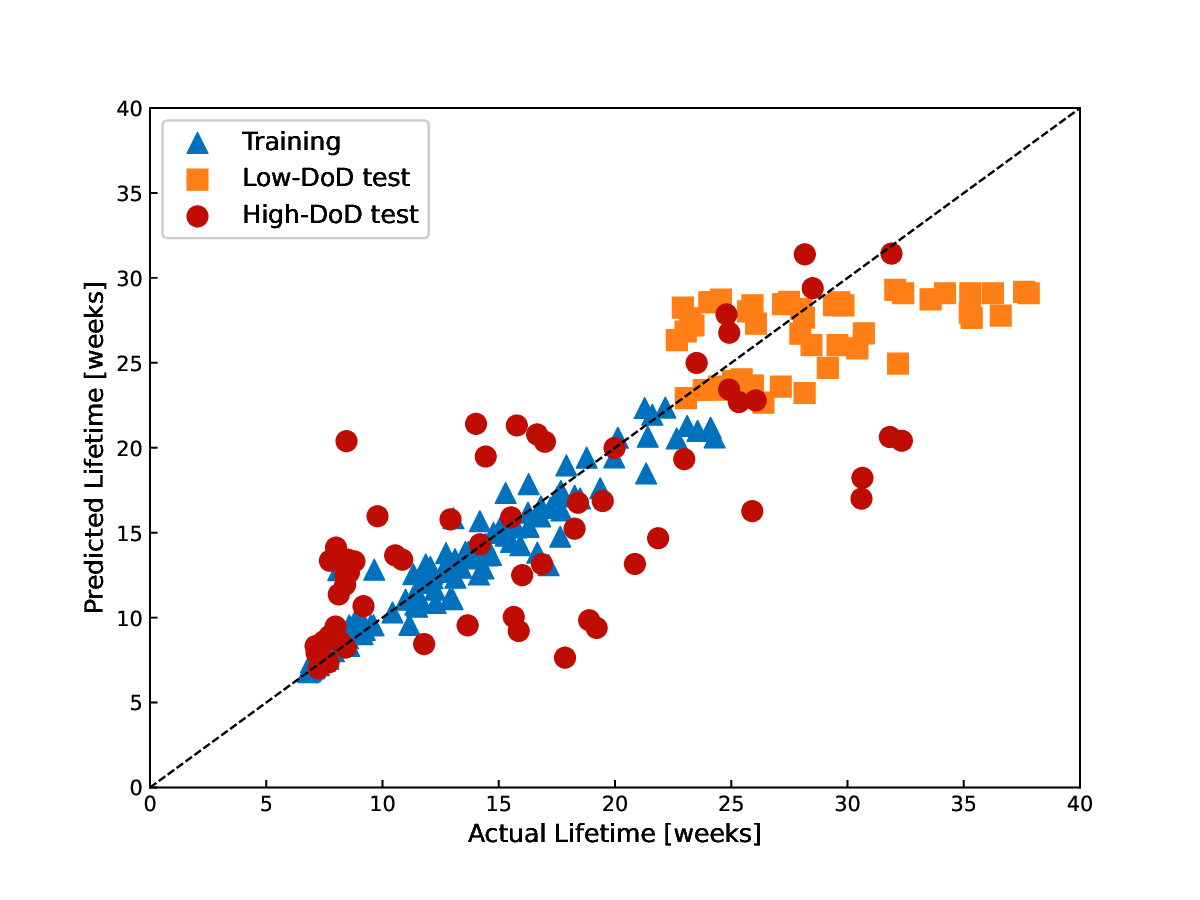}}
	\caption{Week life prediction results of 225 cells in Dataset 2. (a) The week life prediction results of $PTP_{DIS:3.648V-4.080V}\bigtriangleup dQ_{w3-w0}/dV(V)$. The input of XGBoost regression model is $PTP_{DIS:3.648V-4.080V}\bigtriangleup dQ_{w3-w0}/dV(V)$, and the output is the cell week life. (b) The week life prediction results of $log(mean(\bigtriangleup dQ/dV_{w3-w0}^{3.6V-3.9V}(V)))$ \cite{Li2024Predicting}. The input of the XGBoost regression model is $log(mean(\bigtriangleup dQ/dV_{w3-w0}^{3.6V-3.9V}(V)))$, and the output is the cell week life.}
	\label{fig5}
\end{figure}


The cell week life prediction results of different features on the high-DoD training set, high-DoD test set, and low-DoD test set are shown in Table 2. More detailed results are shown in \textbf{Supplementary Tables 6 to 8}. It can be found that $PTP_{DIS:3.648V-4.080V}\bigtriangleup dQ_{w3-w0}/dV(V)$ achieves a comparable accuracy to that of $log(mean(\bigtriangleup dQ/dV_{w3-w0}^{3.6V-3.9V}(V)))$ \cite{Li2024Predicting}, while $log(mean(\bigtriangleup dQ/dV_{w3-w0}^{3.6V-3.9V}(V)))$ uses about 250 discharge $Q/V$ data points, which is about 125 times that of $PTP_{DIS:3.648V-4.080V}\bigtriangleup dQ_{w3-w0}/dV(V)$. Fewer data points mean less data collection, storage, and computational costs.

Furthermore, $PTP_{DIS:3.648V-4.080V}\bigtriangleup dQ_{w3-w0}/dV(V)$ and $log(mean(\bigtriangleup dQ/dV_{w3-w0}^{3.6V-3.9V}(V)))$ are combined with the other two features proposed in \cite{Li2024Predicting}, log($|\bigtriangleup CV Time_{w0}|$) and $DoD$, to predict the week life of cells in Dataset 2. The prediction results are shown in \textbf{Supplementary Figure 10}. The prediction results of different features are shown in \textbf{Supplementary Tables 9 to 11}.

(2) \quad BTPF extraction from $dV/dQ$ data

Similar to the BTPF extracted from the discharge $dQ/dV$ data (\textbf{Supplementary Figure 6(a)}), we also can extract the BTPF from the discharge $dV/dQ$ data, as shown in \textbf{Supplementary Figure 6(b)}.

\textbf{Data collection:} Similar to \cite{Li2024Predicting}, the discharge $Q/V$ data in the periodic RPTs of week 0 and week 3 are collected and $dV/dQ$ curves are calculated for feature extraction. 

\textbf{Difference calculation:} Discharge $dV/dQ$ curves in two periodic RPTs are standardized to facilitate subsequent calculation calculations, as shown in \textbf{Supplementary Figure 11(a)}. The fitted discharge $dV/dQ$ curves in the two RPTs are subtracted to obtain the discharge $\bigtriangleup dV_{w3-w0}/dQ$ curve, as shown in \textbf{Supplementary Figure 11(b)}. The discharge $\bigtriangleup dV_{w3-w0}/dQ$ curves of all 225 cells in Dataset 2 are shown in \textbf{Supplementary Figure 11(c)}.

\textbf{Feature extraction:} The differential voltage difference ($\bigtriangleup dV_{w3-w0}$) values corresponding to any two capacity values on the discharge $\bigtriangleup dV_{w3-w0}/dQ$ curve are subtracted and the absolute value is taken as a candidate two-point feature, as shown in \textbf{Supplementary Figure 12}. A total of $(100^2-100)/2=4950$ candidate two-point features can be obtained by traversing all combinations of two capacity values on the discharge $\bigtriangleup dV_{w3-w0}/dQ$ curve.

\textbf{Feature selection:} The Pearson correlation coefficient between each candidate two-point feature and the week life of 116 cells in the high-DOD training set is calculated, and the candidate two-point feature corresponding to the correlation coefficient with the largest absolute value of 0.886 is selected as the BTPF $PTP_{DIS:0.2016Ah-0.2324Ah}\bigtriangleup dV_{w3-w0}/dQ(Q)$, as shown in Figure 6(a). The two data points on the $\bigtriangleup dV_{w3-w0}/dQ$ curve utilized to calculate $PTP_{DIS:0.2016Ah-0.2324Ah}\bigtriangleup dV_{w3-w0}/dQ(Q)$ are shown in \textbf{Supplementary Figure 11(c)}. 
The distribution relationship between the week life of all 225 cells in Dataset 2 and $PTP_{DIS:0.2016Ah-0.2324Ah}\bigtriangleup dV_{w3-w0}/dQ(Q)$ is shown in Figure 6(b). The Pearson correlation coefficient is 0.805.

\begin{figure}[htbp]
	\centering
	\subfigure[]{\label{fig6:subfig1}\includegraphics[width=1\textwidth]{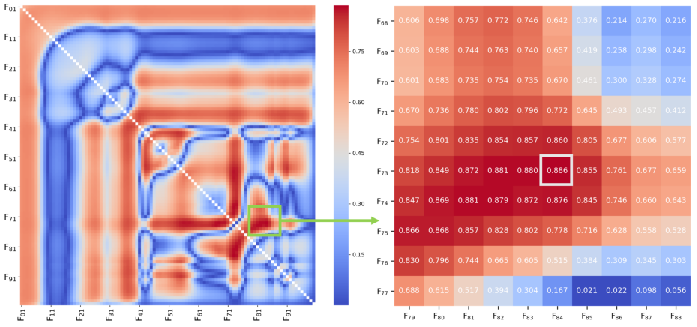}}
	\subfigure[]{\label{fig6:subfig2}\includegraphics[width=0.8\textwidth]{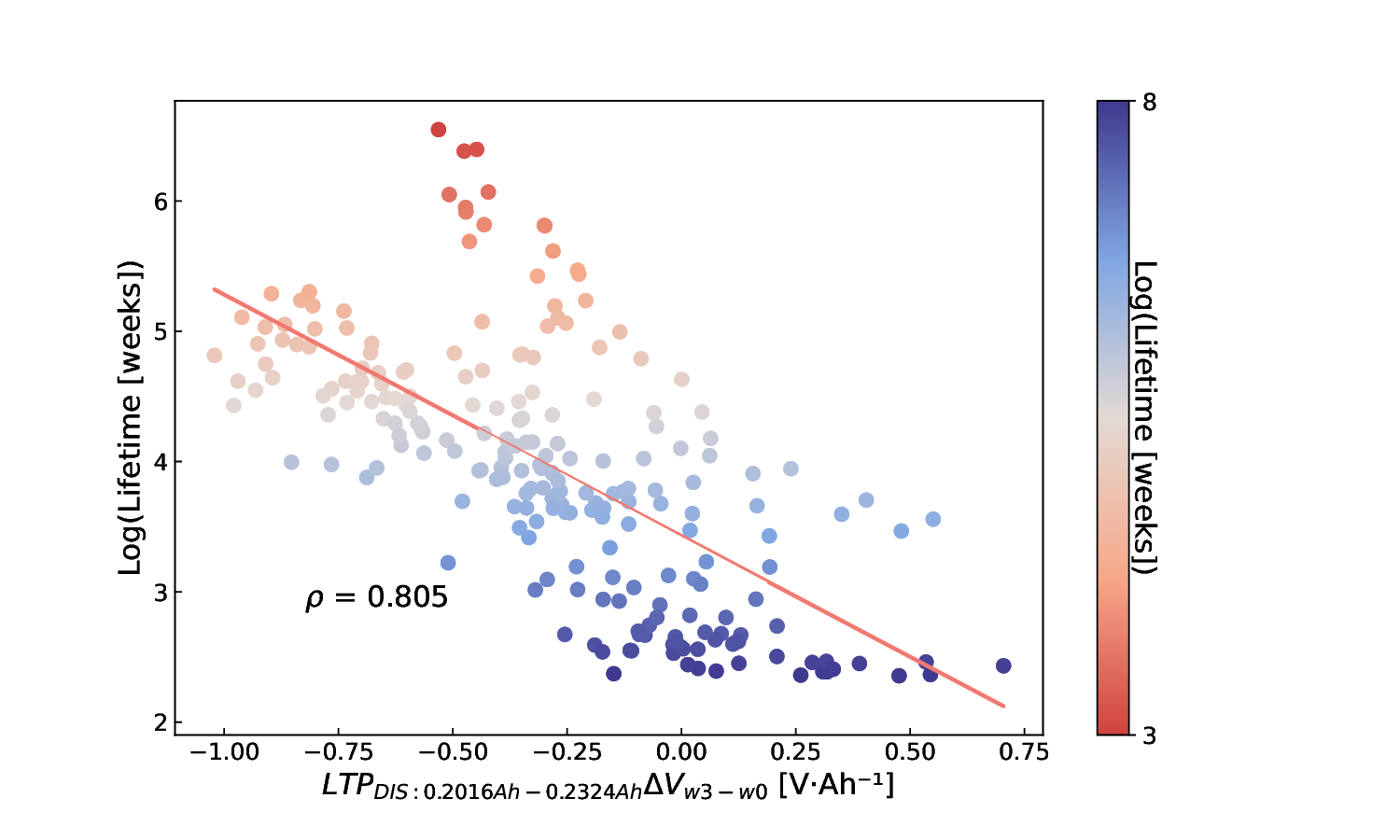}}
	\caption{(a) Selection of the BTPF using the Pearson correlation coefficient between each candidate two-point feature and the week life of 116 cells in the training set. F01 to F100 represent 100 values of discharge capacity from 0 Ah to 0.28 Ah at intervals of 2.8 mAh. The colors are determined based on the Pearson correlation coefficient values. (b) Week life of all 225 cells in Dataset 2 plotted as a function of $PTP_{DIS:0.2016Ah-0.2324Ah}\bigtriangleup dV_{w3-w0}/dQ(Q)$, with a Pearson correlation coefficient of 0.805. The colors are determined based on the week life of cells.}
	\label{fig6}
\end{figure}


\textbf{Week life prediction:} Combining the BTPF $PTP_{DIS:0.2016Ah-0.2324Ah}\bigtriangleup dV_{w3-w0}/dQ(Q)$ and the XGBoost regression model, the cell week life prediction results are shown in Figure 7 and Table 2. More detailed results are shown in \textbf{Supplementary Tables 12 to 14}.

Furthermore, $PTP_{DIS:0.2016Ah-0.2324Ah}\bigtriangleup dV_{w3-w0}/dQ(Q)$ is combined with the other two features proposed in \cite{Li2024Predicting}, $log(|\bigtriangleup CV Time_{w0}|)$ and $DoD$, to predict the week life of cells in Dataset 2. The prediction results are shown in \textbf{Supplementary Figure 13} and \textbf{Supplementary Tables 12 to 14}. It can be found that $PTP_{DIS:0.2016Ah-0.2324Ah}\bigtriangleup dV_{w3-w0}/dQ(Q)$ achieves a comparable accuracy to that of $log(mean(\bigtriangleup dQ/dV_{w3-w0}^{3.6V-3.9V}(V)))$ \cite{Li2024Predicting}, while $log(mean(\bigtriangleup dQ/dV_{w3-w0}^{3.6V-3.9V}(V)))$ uses about 250 discharge $Q/V$ data points, which is about 125 times that of $PTP_{DIS:0.2016Ah-0.2324Ah}\bigtriangleup dV_{w3-w0}/dQ(Q)$. Fewer data points mean less data collection, storage, and computational costs.

\begin{figure}
	\centering
	\includegraphics[width=10cm]{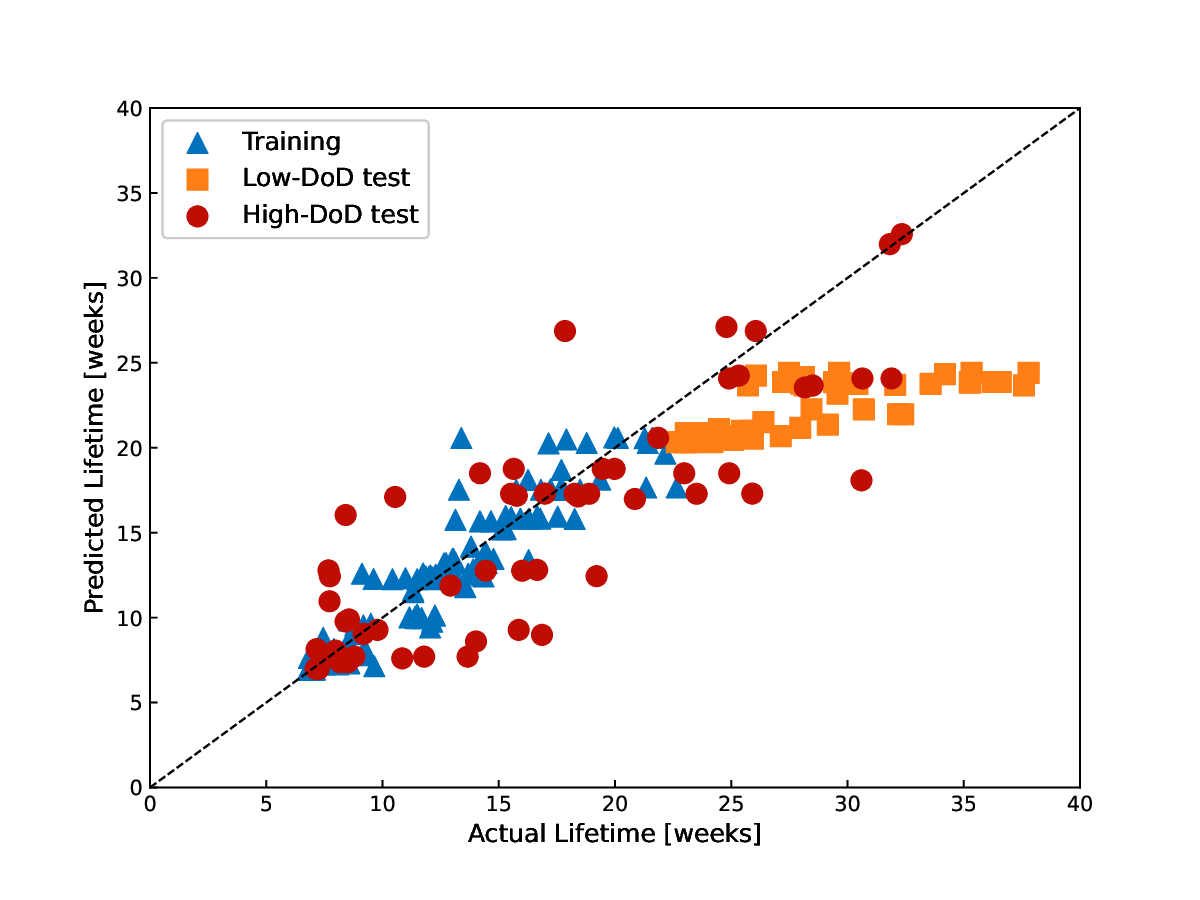}
	\caption{Week life prediction results of of $PTP_{DIS:0.2016Ah-0.2324Ah}\bigtriangleup dV_{w3-w0}/dQ(Q)$ on 225 cells in Dataset 2. The input of XGBoost regression model is $PTP_{DIS:0.2016Ah-0.2324Ah}\bigtriangleup dV_{w3-w0}/dQ(Q)$, and the output is the cell week life.}
	\label{fig7}
\end{figure}


\subsubsection{\textbf{Rationalization of prognostic performance}}

The aging of Li-ion batteries mainly include changes in the amount of active materials (called $LAM$), which may occur at both the anode and cathode) and changes in the amount of reacted lithium (called $LLI$). Results in \cite{severson2019data} and \cite{Guo2023Battery} show that for the incremental capacity curve and differential voltage curve calculated from the charge/discharge $Q/V$ curves, the change in the main peak of the incremental capacity curve can quantitatively characterize $LAM$, and the movement trend of the differential voltage curve can quantitatively characterize $LLI$. Similar to the features $var(\bigtriangleup Q_{100-10}(V))$ and $mean(\bigtriangleup dQ/dV_{w3-w0}^(3.6V-3.9V)(V))$ proposed in \cite{severson2019data, Li2024Predicting}, the BTPFs proposed in this paper is also extracted from two discharge $Q/V$ curves in different cycles. We attribute the success of BTPFs in the prognosis task to capturing the changing trends of $LAM$ and $LLI$.

\subsection{\textbf{Performance on the diagnosis task}}

\subsubsection{\textbf{Dataset 3}}

(1) \quad BTPF extraction from EIS data

For Dataset 3, the BTPF extraction method for diagnosis task is shown in \textbf{Supplementary Figure 14}. To make a fair comparison with SOAT features [raw EIS impedances + future protocols] proposed by Jones et al. \cite{Jones2022Impedance}, this paper uses the same training and test sets as \cite{Jones2022Impedance}. Specifically, Dataset 3 contains 88 commercial Li-ion coin cells with nominal capacities of 35 mAh (40 cells) and 40 mAh (48 cells). The sub-dataset 1 containing 40 commercial Li-ion coin cells with nominal capacity of 35 mAh is divided into a training set and a primary test set. The training set contains all 24 cells with variable discharge, and the primary test set contains all 16 cells with fixed discharge. Then, the sub-dataset 2 containing 48 commercial Li-ion coin cells with nominal capacity of 40 mAh is divided into a secondary test set and a third test set. The secondary test set contains all 32 cells cycled at the temperature of $23 \pm 2^{\circ}C$, and the third test set contains all 16 cells cycled at the temperature of $35 \pm 1^{\circ}C$. The training set is utilized to train the ML regression model, and the test sets are utilized to verify and test the estimation performance of the trained ML regression model. 
To be consistent with \cite{Jones2022Impedance}, we first trained the ML regression model with the training set (24 cells) and tested it with the primary test set (16 cells). Then, we test the BTPF extraction method by using the secondary test set (32 cells). Here we assess how the ML regression model performs under 16-split validations, where two cells are randomly held out in each split. The final results are the average of the results obtained from 16-split validations. After that, we test the BTPF extraction method by using the third test set (16 cells). The ML regression model is trained on the secondary test set and tested on the third test set. For more details about the training and test sets, please refer to \cite{Jones2022Impedance} and will not be repeated here.

\textbf{Data collection:} We collect the EIS data after full discharge in each cycle (a total of 114 real and imaginary impedances corresponding to 57 frequencies between 0.02Hz and 20kHz) for feature extraction. 

\textbf{Difference calculation:} We subtract the real impedance vs. frequency ($Re/f$) data and imaginary impedance vs. frequency ($Im/f$) data collected in the first cycle from the corresponding data in higher cycles to obtain the $\bigtriangleup Re_{n-1}/f$ and $\bigtriangleup Im_{n-1}/f$ curves respectively. It should be noted here that since the frequency of EIS data collection in each cycle is the same, there is no need to perform spline function fitting and unified linear interpolation like Dataset 1 and Dataset 2. Instead, the $\bigtriangleup Re_{n-1}/f$ and $\bigtriangleup Im_{n-1}/f$ curves can be obtained by directly subtracting the EIS data in the first cycle from the EIS data in higher cycles. As shown in \textbf{Supplementary Figures 15(a) and 15(b)}, the obtained $\bigtriangleup Re_{10-1}/f$ and $\bigtriangleup Im_{10-1}/f$ curves are the shaded parts between the two $Re/f$ curves and the two $Im/f$ curves in the 1st and 10th cycles, respectively. The $\bigtriangleup Re_{n-1}/f$ and $\bigtriangleup Im_{n-1}/f$ curves of the representative cell in Dataset 3 are provided in \textbf{Supplementary Figures 15(c) and 15(d)}.

\textbf{Feature extraction:} The real impedance difference ($\bigtriangleup Re$) values corresponding to any two frequency values on the $\bigtriangleup Re_{n-1}/f$ curve are subtracted and the absolute value is taken as a candidate two-point feature, as shown in \textbf{Supplementary Figure 16(a)}. Since the frequency is divided into 57 values between 0.02Hz and 20kHz during EIS measurement, a total of $(57^2-57)/2=1596$ candidate two-point features can be obtained by traversing all combinations of two frequency values on the $\bigtriangleup Re_{n-1}/f$ curve, as shown in Figure 8(a). Further, all candidate two-point features of all cells in the training set on the $\bigtriangleup Re_{n-1}/f$ curve are traversed. We can use the same method in \textbf{Supplementary Figure 16(b)} to obtain all candidate two-point features on the $\bigtriangleup Im_{n-1}/f$ curve of all cells in the training set, as shown in Figure 8(b).

\textbf{Feature selection:} The Pearson correlation coefficient between each candidate two-point feature and capacities of 24 cells in the training set is calculated, and the candidate two-point feature corresponding to the correlation coefficient with the largest absolute value of 0.822 is selected as the BTPF $DTP_{ADIS: 1922.08Hz-4.36Hz}\bigtriangleup Re_{n-1}/f(f)$ on the $\bigtriangleup Re_{n-1}/f$ curve, as shown in Figure 8(a). The two data points on the $\bigtriangleup Re_{n-1}/f$ curve that utilized to calculate $DTP_{ADIS: 1922.08Hz-4.36Hz}\bigtriangleup Re_{n-1}/f(f)$ are shown in \textbf{Supplementary Figure 15(c)}. Similarly, the BTPF $DTP_{ADIS: 17.79Hz-1.07Hz}\bigtriangleup Im_{n-1}/f(f)$ on the $\bigtriangleup Im_{n-1}/f$ curve is selected, and the corresponding Pearson correlation coefficient is 0.797, as shown in Figure 8(b). The two data points on the $\bigtriangleup Im_{n-1}/f$ curve that utilized to calculate $DTP_{ADIS: 17.79Hz-1.07Hz}\bigtriangleup Im_{n-1}/f(f)$ are shown in \textbf{Supplementary Figure 15(d)}.
The distribution relationship between capacities of all 40 cells in the sub-dataset 1 of Dataset 3 and the BTPFs ($DTP_{ADIS: 1922.08Hz-4.36Hz}\bigtriangleup Re_{n-1}/f(f)$ and $DTP_{ADIS: 17.79Hz-1.07Hz}\bigtriangleup Im_{n-1}/f(f)$) is shown in Figures 8(c) and 8(d). The Pearson correlation coefficients are 0.815 and 0.782, respectively.

\begin{figure}[htbp]
	\centering
	\subfigure[]{\label{fig8:subfig1}\includegraphics[width=0.6\textwidth]{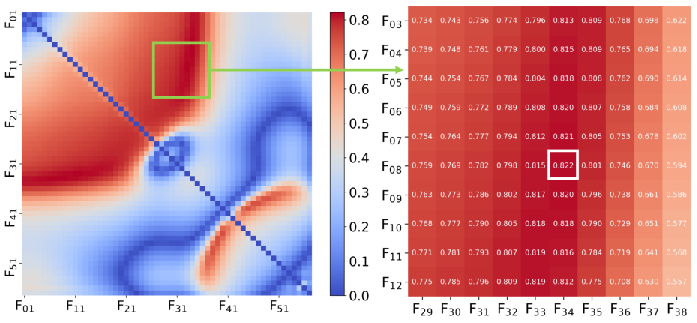}}
	\subfigure[]{\label{fig8:subfig2}\includegraphics[width=0.6\textwidth]{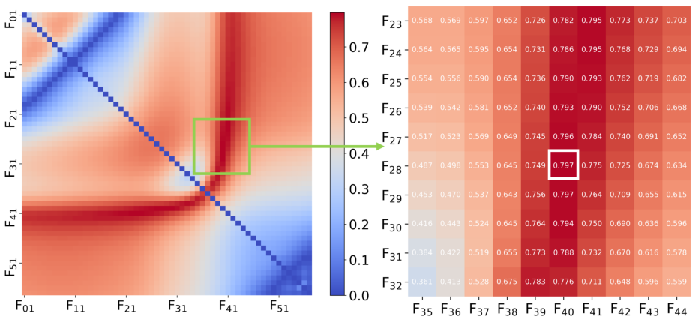}}
	\subfigure[]{\label{fig8:subfig3}\includegraphics[width=0.5\textwidth]{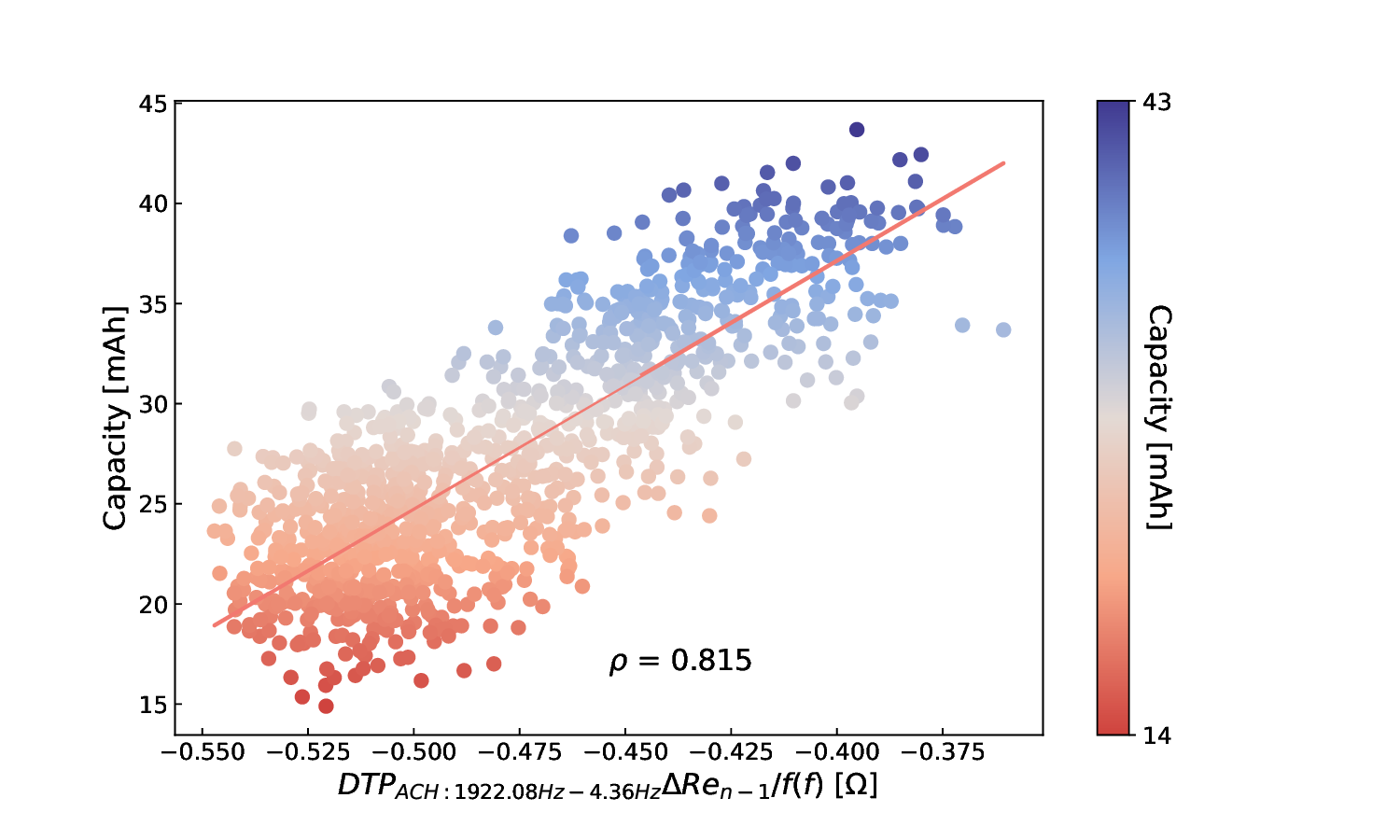}}
	\subfigure[]{\label{fig8:subfig4}\includegraphics[width=0.5\textwidth]{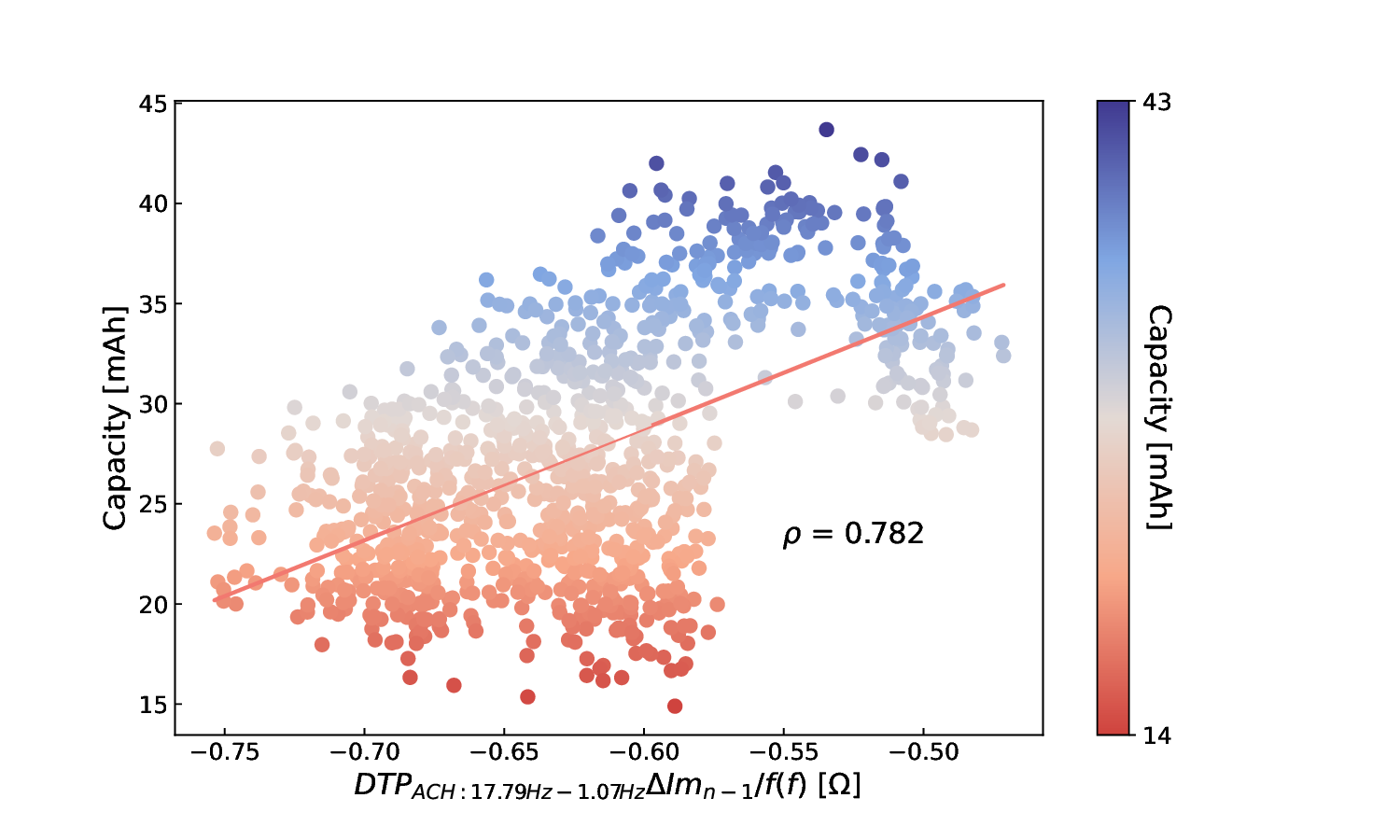}}
	\caption{ (a) Selection of the BTPF on the $\bigtriangleup Re_{n-1}/f$ curve using the Pearson correlation coefficient between each candidate two-point feature and capacities of 24 cells in the training set. F01 to F57 represent 57 frequencies from 0.02Hz to 20kHz. The colors are determined based on the Pearson correlation coefficient values. (b) Selection of the BTPF on the $\bigtriangleup Im_{n-1}/f$ curve using the Pearson correlation coefficient between each candidate two-point feature and capacities of 24 cells in the training set. F01 to F57 represent 57 frequencies from 0.02Hz to 20kHz. The colors are determined based on the Pearson correlation coefficient values. (c) Capacities plotted as a function of $DTP_{ADIS: 1922.08Hz-4.36Hz}\bigtriangleup Re_{n-1}/f(f)$, with a Pearson correlation coefficient of 0.815. The colors are determined based on the capacities of cells. (d) Capacities plotted as a function of $DTP_{ADIS: 17.79Hz-1.07Hz}\bigtriangleup Im_{n-1}/f(f)$, with a Pearson correlation coefficient of 0.782. The colors are determined based on the capacities of cells.}
	\label{fig8}
\end{figure}


\textbf{SoH estimation:} Since the Pearson correlation coefficient corresponding to $DTP_{ADIS: 1922.08Hz-4.36Hz}\bigtriangleup Re_{n-1}/f(f)$ is greater than that of $DTP_{ADIS: 17.79Hz-1.07Hz}\bigtriangleup Im_{n-1}/f(f)$, this paper selects $DTP_{ADIS: 1922.08Hz-4.36Hz}\bigtriangleup Re_{n-1}/f(f)$ for ML regression model training and testing. Combining the BTPF $DTP_{ADIS: 1922.08Hz-4.36Hz}\bigtriangleup Re_{n-1}/f(f)$ and the XGBoost regression model, the cell SoH estimation results are shown in Figure 9(a). The input of the XGBoost regression model is $DTP_{ADIS: 1922.08Hz-4.36Hz}\bigtriangleup Re_{n-1}/f(f)$, and the output is the cell capacity. Combining the raw EIS impedance features \cite{Jones2022Impedance} and the XGBoost regression model, the cell SoH estimation results are shown in Figure 9(b). The input of the XGBoost regression model is 114 raw EIS impedances, and the output is the cell capacity.

\begin{figure}[htbp]
	\centering
	\subfigure[]{\label{fig9:subfig1}\includegraphics[width=0.8\textwidth]{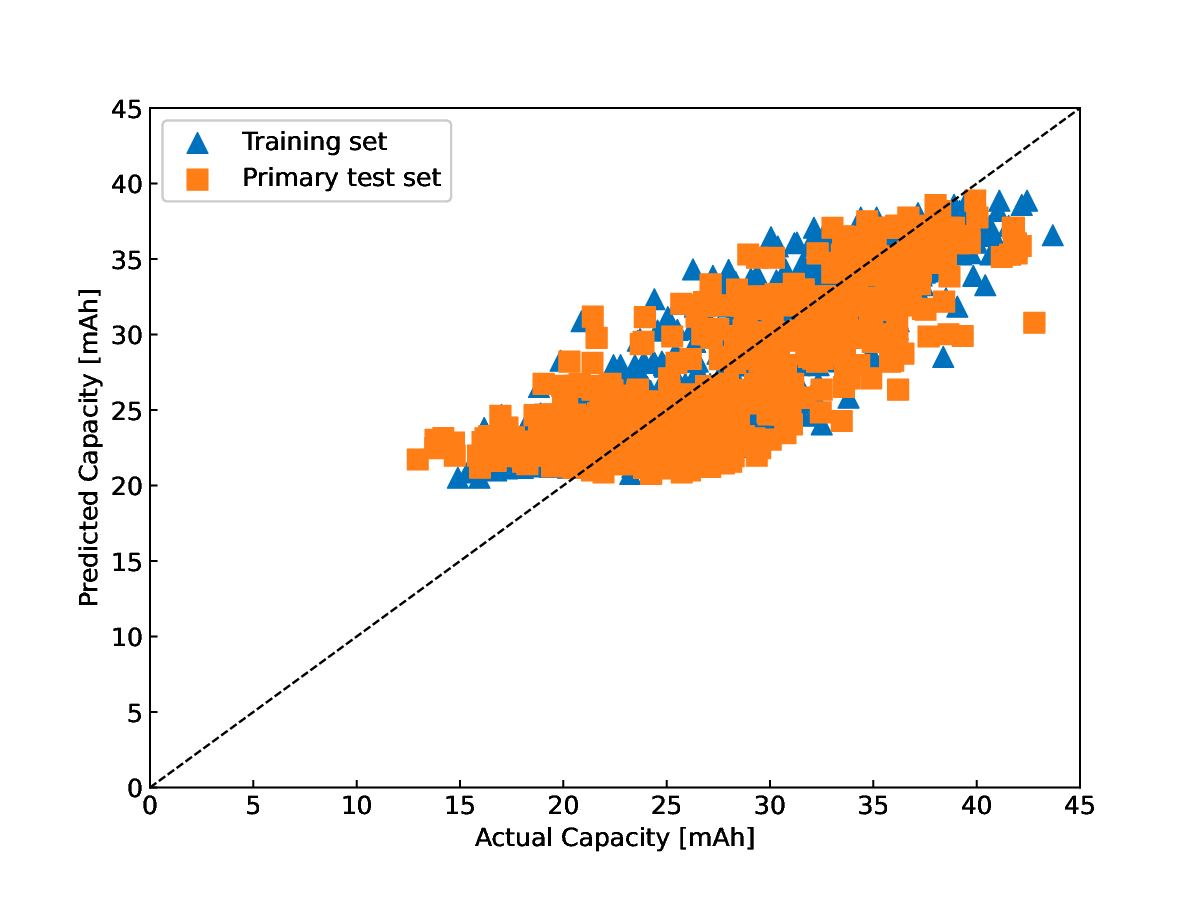}}
	\subfigure[]{\label{fig9:subfig2}\includegraphics[width=0.8\textwidth]{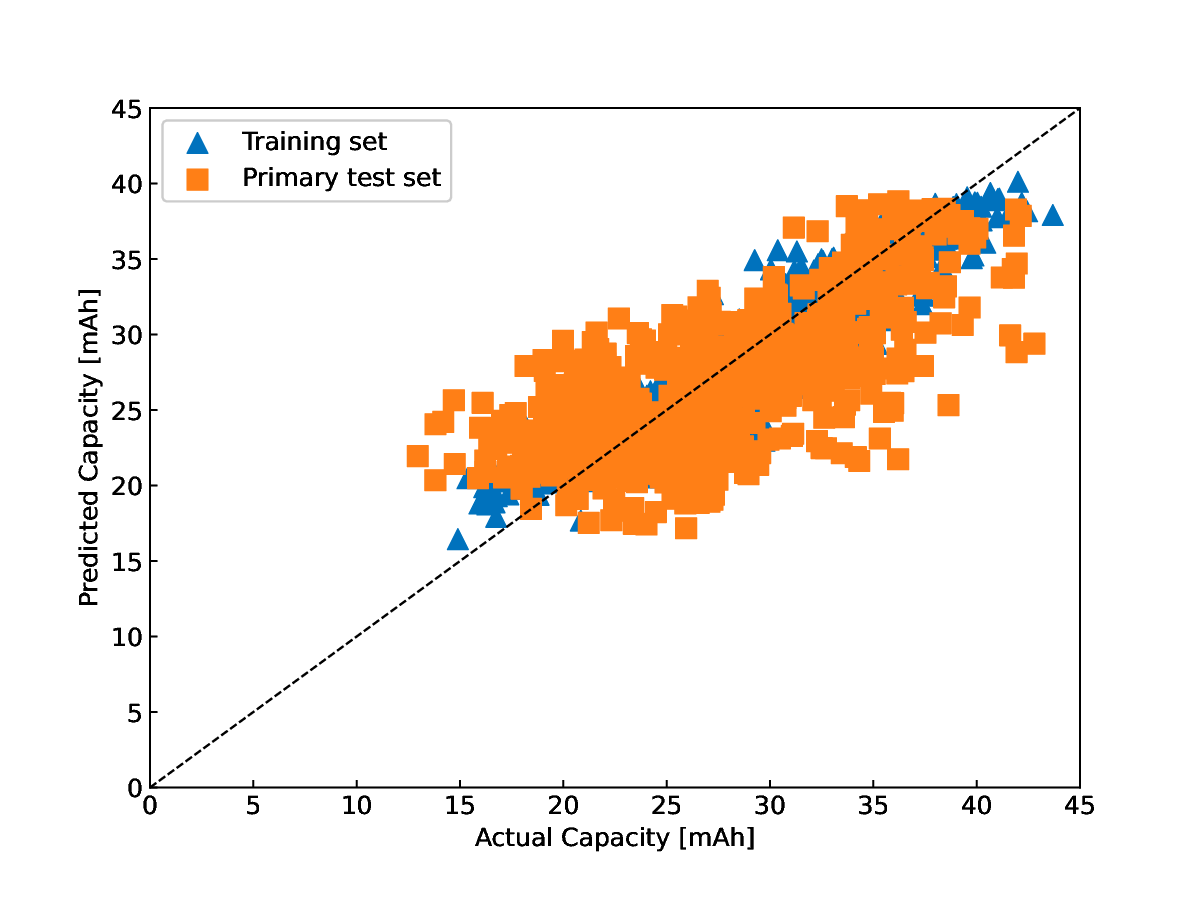}}
	\caption{SoH estimation results of different features on 40 cells in the sub-dataset 1 of Dataset 3. (a) SoH estimation results of $DTP_{ADIS: 1922.08Hz-4.36Hz}\bigtriangleup Re_{n-1}/f(f)$. The input of the XGBoost regression model is $DTP_{ADIS: 1922.08Hz-4.36Hz}\bigtriangleup Re_{n-1}/f(f)$, and the output is the cell capacity. (b) SoH estimation results of raw EIS impedance features \cite{Jones2022Impedance}. The input of the XGBoost regression model is 114 raw EIS impedances, and the output is the cell capacity.}
	\label{fig9}
\end{figure}


SoH estimation results of different features on the training set and primary test set in Dataset 3 are shown in Table 3. More detailed results are shown in \textbf{Supplementary Tables 15 and 16}. It can be found that $DTP_{ADIS: 1922.08Hz-4.36Hz}\bigtriangleup Re_{n-1}/f(f)$ achieves a comparable accuracy to that of raw EIS impedance features \cite{Jones2022Impedance}, while the  raw EIS impedance features use 114 impedance data points, which is 57 times that of $DTP_{ADIS: 1922.08Hz-4.36Hz}\bigtriangleup Re_{n-1}/f(f)$. Fewer data points mean less data collection, storage, and computational costs. 

Furthermore, combining the BTPF $DTP_{ADIS: 1922.08Hz-4.36Hz}\bigtriangleup Re_{n-1}/f(f)$ with the future cycling protocol feature proposed in \cite{Jones2022Impedance} (including two current values of two-stage CC charging protocol and one current value of single stage CC discharge protocol), the SoH estimation results on the sub-dataset 1 of Dataset 3 are shown in \textbf{Supplementary Figure 17(a)}. 
Combining the raw EIS impedance features with the future cycling protocol feature, the SoH estimation results are shown in \textbf{Supplementary Figure 17(b)}. 
After combining different features with the future cycling protocol feature, the SoH estimation results on sub-dataset 1 of Dataset 3 are shown in \textbf{Supplementary Tables 17 and 18}. It can be found that after combining with the future cycling protocol feature, $DTP_{ADIS: 1922.08Hz-4.36Hz}\bigtriangleup Re_{n-1}/f(f)$ achieves a comparable accuracy to that of raw EIS impedance features \cite{Jones2022Impedance}. 

Moreover, based on 48 cells in the sub-dataset 2 of Dataset 3, the generalizability of the proposed BTPF extraction method is tested. For the secondary test set containing 32 cells cycled at the temperature of $23 \pm 2^{\circ}C$, the test results are shown in \textbf{Supplementary Figures 18 to 19} and \textbf{Supplementary Tables 19 to 20}. For the third test set containing 16 cells cycled at the temperature of $35 \pm 1^{\circ}C$, the test results are shown in \textbf{Supplementary Figures 20 to 21} and \textbf{Supplementary Tables 21 to 22}.

(2) \quad BTPF extraction from DRT data

Similar to the BTPF extracted from the EIS data (\textbf{Supplementary Figure 14}), we also can extract the BTPF from the DRT data, as shown in \textbf{Supplementary Figure 22}.

\textbf{Data collection:} By using the open-source software pyDRTtools \cite{Wan2015Influence}, we convert the EIS data after full discharge in each cycle (114 real and imaginary impedances corresponding to 57 frequencies between 0.02Hz and 20kHz) to DRT ($\gamma(\tau)/\tau$) data for feature extraction, where $\gamma(\tau)$ is the actual DRT and $\tau$ is the time. 

\textbf{Difference calculation:} We subtract the DRT data collected in the first cycle from the corresponding data in higher cycles to obtain the $\bigtriangleup \gamma(\tau)_{n-1}/\tau$ curve. It should be noted here that since the time of DRT data calculated in each cycle is the same, there is no need to perform spline function fitting and unified linear interpolation like Dataset 1 and Dataset 2. Instead, the $\bigtriangleup \gamma (\tau)_{n-1}/\tau$ curve can be obtained by directly subtracting the DRT data in the first cycle from the DRT data in higher cycles. As shown in \textbf{Supplementary Figure 23(a)}, the obtained $\bigtriangleup \gamma(\tau)_{10-1}/\tau$ curve is the shaded parts between the two $\gamma (\tau)/\tau$ curves in the 1st and 10th cycles. The $\bigtriangleup \gamma (\tau)_{n-1}/\tau$ curve in all cycles of the representative cell in Dataset 3 are provided in \textbf{Supplementary Figure 23(b)}.

\textbf{Feature extraction:} The $\bigtriangleup \gamma (\tau)$ values corresponding to any two time values on the $\bigtriangleup \gamma (\tau)_{n-1}/\tau$ curve are subtracted and the absolute value is taken as a candidate two-point feature, as shown in \textbf{Supplementary Figure 24}. Since the time is divided into 570 values between $3.19e^{-5}s$ and $1.18e^2s$ during DRT data calculation, a total of $(570^2-570)/2=162165$ candidate two-point features can be obtained by traversing all combinations of two timescale values on the $\bigtriangleup \gamma (\tau)_{n-1}/\tau$ curve. To reduce the computational burden, we only sample 1 time data point from every 10 time data points to extract candidate two-point features. Then, a total of $(57^2-57)/2=1596$ candidate two-point features can be obtained by traversing all combinations of two time values. Increasing the time interval does not affect the fairness of the comparison. Further, all candidate two-point features on the $\bigtriangleup \gamma (\tau)_{n-1}/\tau$ curve of all cells in the training set are traversed.

\textbf{Feature selection:} The Pearson correlation coefficient between each candidate two-point feature on the $\bigtriangleup \gamma (\tau)_{n-1}/\tau$ curve and capacities of all 24 cells in the training set is calculated, and the candidate two-point feature corresponding to the correlation coefficient with the largest absolute value of 0.785 is selected as the BTPF $DTP_{ADIS:1.23e^{-4}-5.41e^{-3}}\bigtriangleup \gamma (\tau)_{n-1}/\tau$, as shown in Figure 10(a). The two data points on the $\bigtriangleup \gamma (\tau)_{n-1}/\tau$ curve that utilized to calculate $DTP_{ADIS:1.23e^{-4}-5.41e^{-3}}\bigtriangleup \gamma (\tau)_{n-1}/\tau$ are shown in \textbf{Supplementary Figure 23(b)}.
The distribution relationship between capacities of 40 cells in Subdataset 1 and the BTPF $DTP_{ADIS:1.23e^{-4}-5.41e^{-3}}\bigtriangleup \gamma (\tau)_{n-1}/\tau$ is shown in Figure 10(b). The Pearson correlation coefficient is 0.773.

\begin{figure}[htbp]
	\centering
	\subfigure[]{\label{fig10:subfig1}\includegraphics[width=1\textwidth]{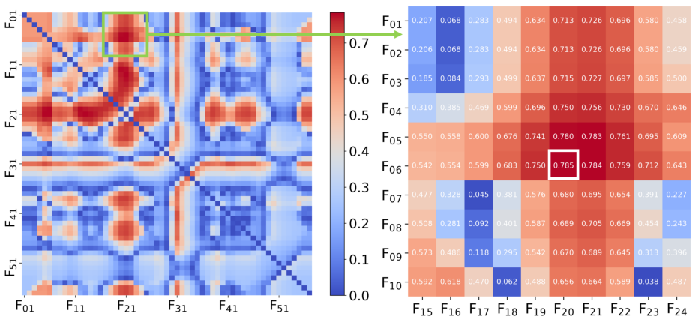}}
	\subfigure[]{\label{fig10:subfig2}\includegraphics[width=0.8\textwidth]{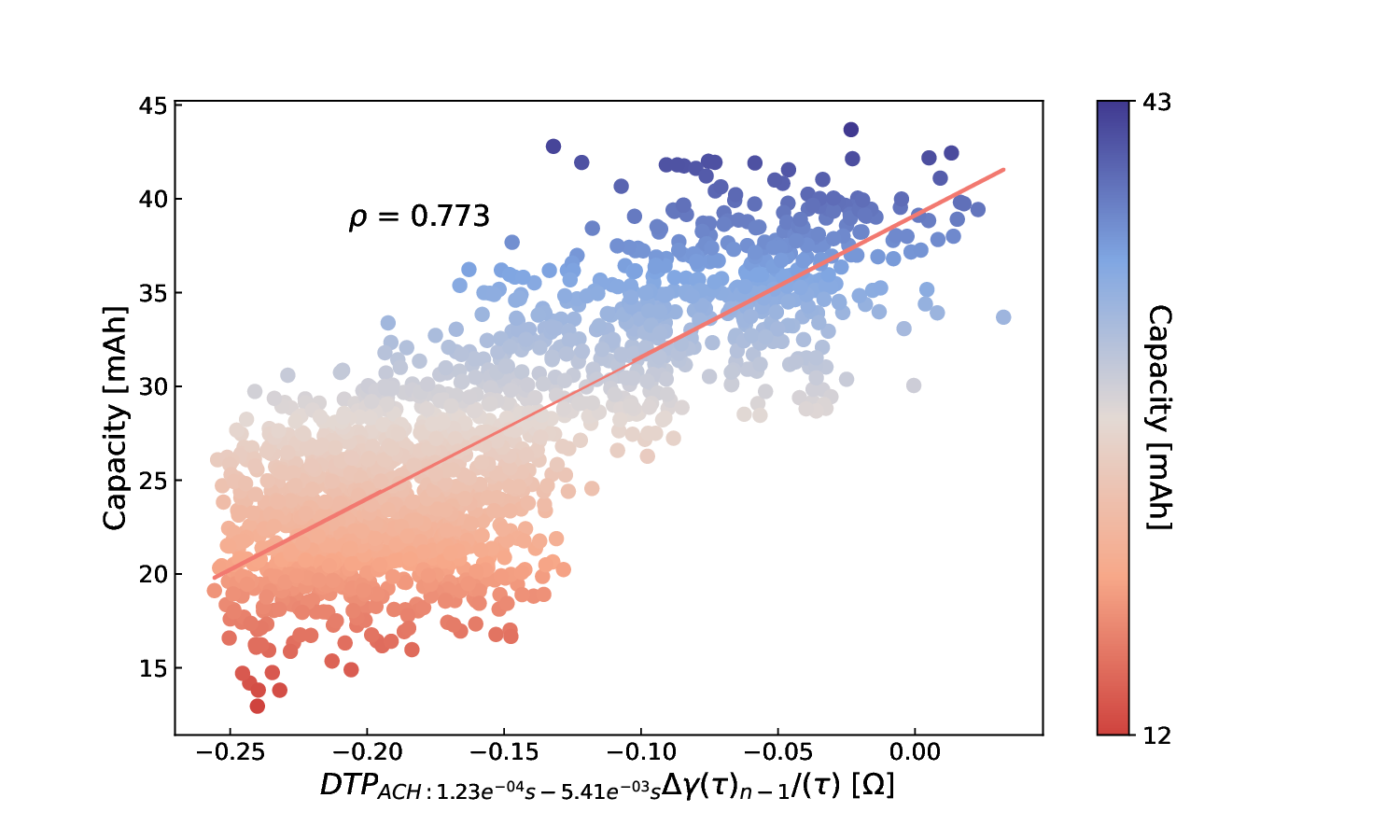}}
	\caption{(a) Selection of the BTPF on the $\bigtriangleup \gamma (\tau)_{n-1}/\tau$ curve using the Pearson correlation coefficient between each candidate two-point feature and capacities of 24 cells in the training set. F01 to F57 represent 57 time values between $3.19e^{-5}s$ and $1.18e^2s$. The colors are determined based on the Pearson correlation coefficient values. (b) Capacities of 40 cells in Subdataset 1 plotted as functions of $DTP_{ADIS:1.23e^{-4}-5.41e^{-3}}\bigtriangleup \gamma (\tau)_{n-1}/\tau$. The colors are determined based on the capacities of cells. }
	\label{fig10}
\end{figure}


\textbf{SoH estimation:} Combining the BTPF $DTP_{ADIS:1.23e^{-4}-5.41e^{-3}}\bigtriangleup \gamma (\tau)_{n-1}/\tau$ and the XGBoost regression model, the battery SoH estimation results are shown in Figure 11(a). Furthermore, combining the BTPF $DTP_{ADIS:1.23e^{-4}-5.41e^{-3}}\bigtriangleup \gamma (\tau)_{n-1}/\tau$ with the future cycling protocol features \cite{Jones2022Impedance}, the battery SoH estimation results of the XGBoost regression model are shown in Figure 11(b). The battery SoH estimation results of different features on Subdataset 1 of Dataset 3 are shown in Table 3. More detailed results are shown in \textbf{Supplementary Tables 23 and 24}.It can be found that $DTP_{ADIS:1.23e^{-4}-5.41e^{-3}}\bigtriangleup \gamma (\tau)_{n-1}/\tau$ achieves a comparable accuracy to that of raw EIS impedance features \cite{Jones2022Impedance}.

\begin{figure}[htbp]
	\centering
	\subfigure[]{\label{fig11:subfig1}\includegraphics[width=0.8\textwidth]{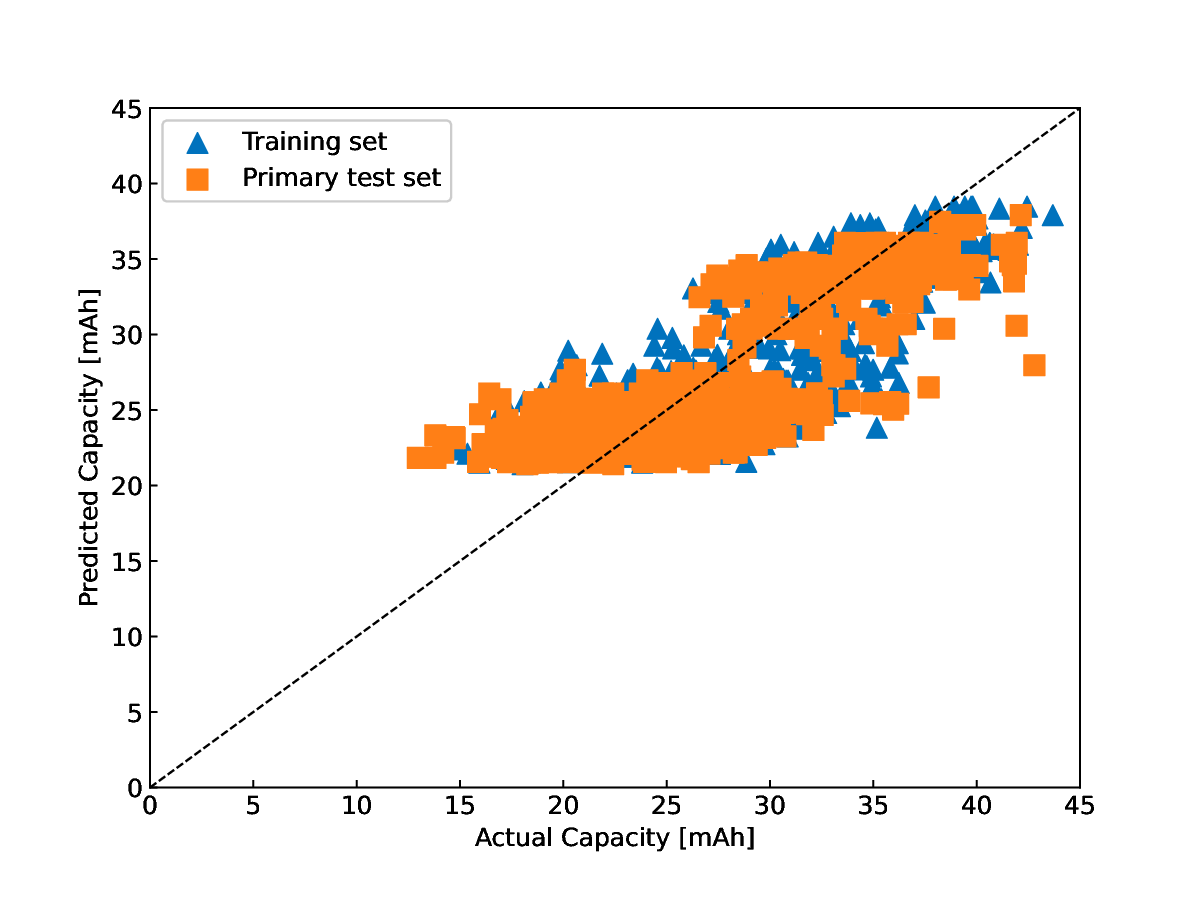}}
	\subfigure[]{\label{fig11:subfig2}\includegraphics[width=0.8\textwidth]{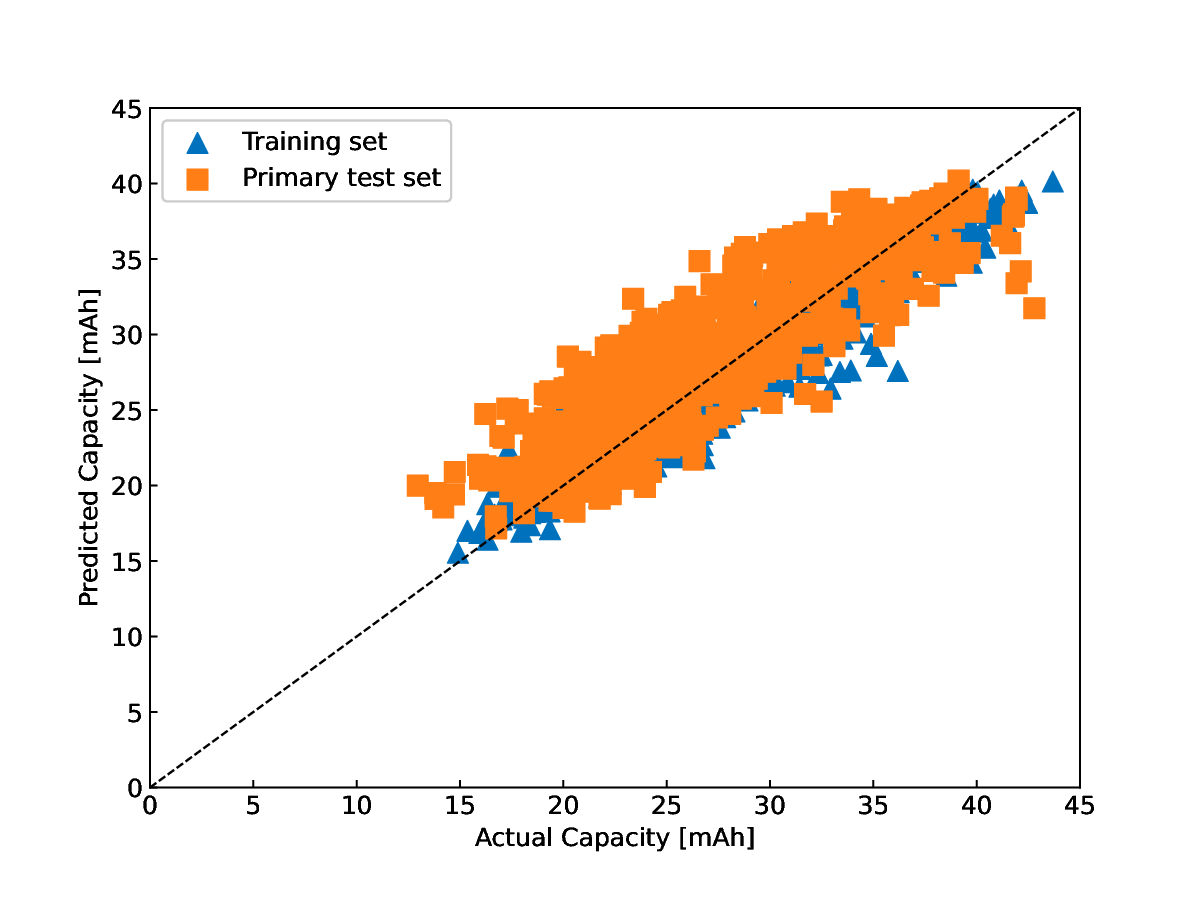}}
	\caption{SoH estimation results of different features on the sub-dataset 1 of Dataset 3 (40 cells). (a) SoH estimation results of $DTP_{ADIS:1.23e^{-4}-5.41e^{-3}}\bigtriangleup \gamma (\tau)_{n-1}/\tau$. The input of the XGBoost regression model is $DTP_{ADIS:1.23e^{-4}-5.41e^{-3}}\bigtriangleup \gamma (\tau)_{n-1}/\tau$, and the output is the cell capacity. (b) SoH estimation results of features [$DTP_{ADIS:1.23e^{-4}-5.41e^{-3}}\bigtriangleup \gamma (\tau)_{n-1}/\tau$, future protocols]. The input of the XGBoost regression model is $DTP_{ADIS:1.23e^{-4}-5.41e^{-3}}\bigtriangleup \gamma (\tau)_{n-1}/\tau$ and the future cycling protocol features, and the output is the cell capacity.}
	\label{fig11}
\end{figure}



\subsubsection{\textbf{Dataset 4}}

(1) \quad BTPF extraction from EIS data

For Dataset 4, the BTPF extraction method for diagnosis task is exactly the same as Dataset 3, as shown in \textbf{Supplementary Figure 14}. To make a fair comparison with SOAT features (raw EIS impedances) proposed by Zhang et al. \cite{Zhang2020Identifying}, this paper uses the same training and test sets as \cite{Zhang2020Identifying}. Specifically, Dataset 4 containing 12 commercial Li-ion coin cells with a nominal capacity of 45 mAh is divided into a training set and a test set. The training set (25C01–25C04, 35C01, and 45C01) and the test set (25C05–25C08, 35C02, and 45C02) both contain 6 cells. The training set is utilized to train the ML regression model, and the test set is utilized to verify and test the estimation performance of the trained ML regression model. To be consistent with \cite{Zhang2020Identifying}, we first trained the ML regression model with four cells in the training set (25C01–25C04) and tested it with four cells in the test set (25C05–25C08). Then, we trained the ML regression model with all cells in the training set and tested it with all cells in the test set. For more details about the training and test sets, please refer to \cite{Zhang2020Identifying} and will not be repeated here.

\textbf{Data collection:} We collect the EIS data after full charge in each cycle (a total of 120 real and imaginary impedances corresponding to 60 frequencies between 0.02Hz and 20kHz) for feature extraction. 

\textbf{Difference calculation:} We subtract the $Re/f$ data and $Im/f$ data collected in the first cycle from the corresponding data in higher cycles to obtain the $\bigtriangleup Re_{n-1}/f$ and $\bigtriangleup Im_{n-1}/f$ curves respectively, as shown in \textbf{Supplementary Figures 25(a) and 25(b)}. The $\bigtriangleup Re_{n-1}/f$ and $\bigtriangleup Im_{n-1}/f$ curves of the representative cell in Dataset 4 are provided in \textbf{Supplementary Figures 25(c) and 25(d)}.

\textbf{Feature extraction:} The real impedance difference ($\bigtriangleup Re$) values corresponding to any two frequency values on the $\bigtriangleup Re_{n-1}/f$ curve are subtracted and the absolute value is taken as a candidate two-point feature, as shown in \textbf{Supplementary Figure 16(a)}. Since the frequency is divided into 60 values between 0.02Hz and 20kHz during EIS measurement, a total of $(60^2-60)/2=1770$ candidate two-point features can be obtained by traversing all combinations of two frequency values on the $\bigtriangleup Re_{n-1}/f$ curve, as shown in Figure 12(a). Further, all candidate two-point features of all cells in the training set on the $\bigtriangleup Re_{n-1}/f$ curve are traversed. We can use the same method in \textbf{Supplementary Figure 16(b)} to obtain all candidate two-point features on the $\bigtriangleup Im_{n-1}/f$ curve of all cells in the training set, as shown in Figure 12(b).

\textbf{Feature selection:} The Pearson correlation coefficient between each candidate two-point feature and capacities of 4 cells in the training set (25C01–25C04) is calculated, and the candidate two-point feature corresponding to the correlation coefficient with the largest absolute value of 0.926 is selected as the BTPF $DTP_{ACH:0.165Hz-0.019Hz}\bigtriangleup Re_{n-1}/f(f)$ of the $\bigtriangleup Re_{n-1}/f$ curve, as shown in Figure 12(a). The two data points on the $\bigtriangleup Re_{n-1}/f$ curve that utilized to calculate $DTP_{ACH:0.165Hz-0.019Hz}\bigtriangleup Re_{n-1}/f(f)$ are shown in \textbf{Supplementary Figure 25(c)}. Similarly, the BTPF $DTP_{ACH:115.778Hz-11.145Hz}\bigtriangleup Im_{n-1}/f(f)$ of the $\bigtriangleup Im_{n-1}/f$ curve is selected, and the corresponding Pearson correlation coefficient is 0.950, as shown in Figure 12(b). The two data points on the $\bigtriangleup Im_{n-1}/f$ curve that utilized to calculate $DTP_{ACH:115.778Hz-11.145Hz}\bigtriangleup Im_{n-1}/f(f)$ are shown in \textbf{Supplementary Figure 25(d)}.
The distribution relationship between capacities of 8 cells in Dataset 4 (25C01–25C08) and the BTPFs ($DTP_{ACH:0.165Hz-0.019Hz}\bigtriangleup Re_{n-1}/f(f)$ and $DTP_{ACH:115.778Hz-11.145Hz}\bigtriangleup Im_{n-1}/f(f)$) is shown in Figures 12(c) and 12(d). The Pearson correlation coefficients are 0.892 and 0.924, respectively.

\begin{figure}[htbp]
	\centering
	\subfigure[]{\label{fig12:subfig1}\includegraphics[width=0.6\textwidth]{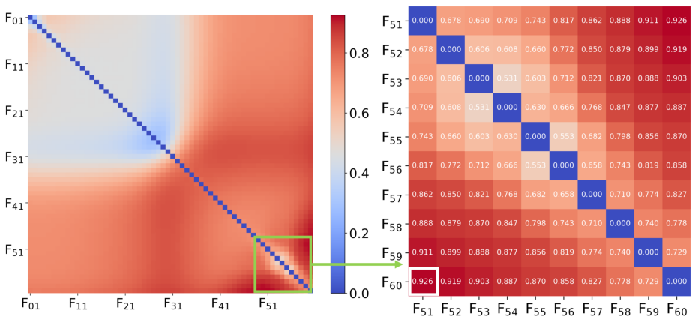}}
	\subfigure[]{\label{fig12:subfig2}\includegraphics[width=0.6\textwidth]{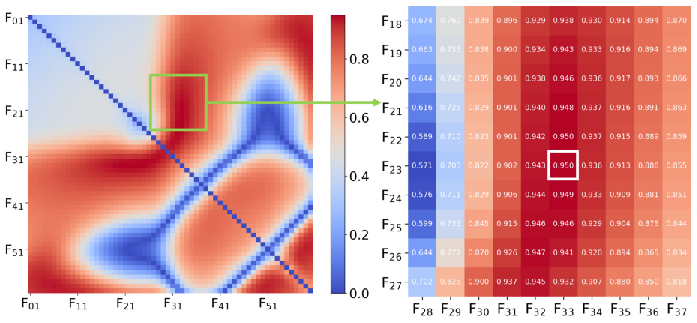}}
	\subfigure[]{\label{fig12:subfig3}\includegraphics[width=0.5\textwidth]{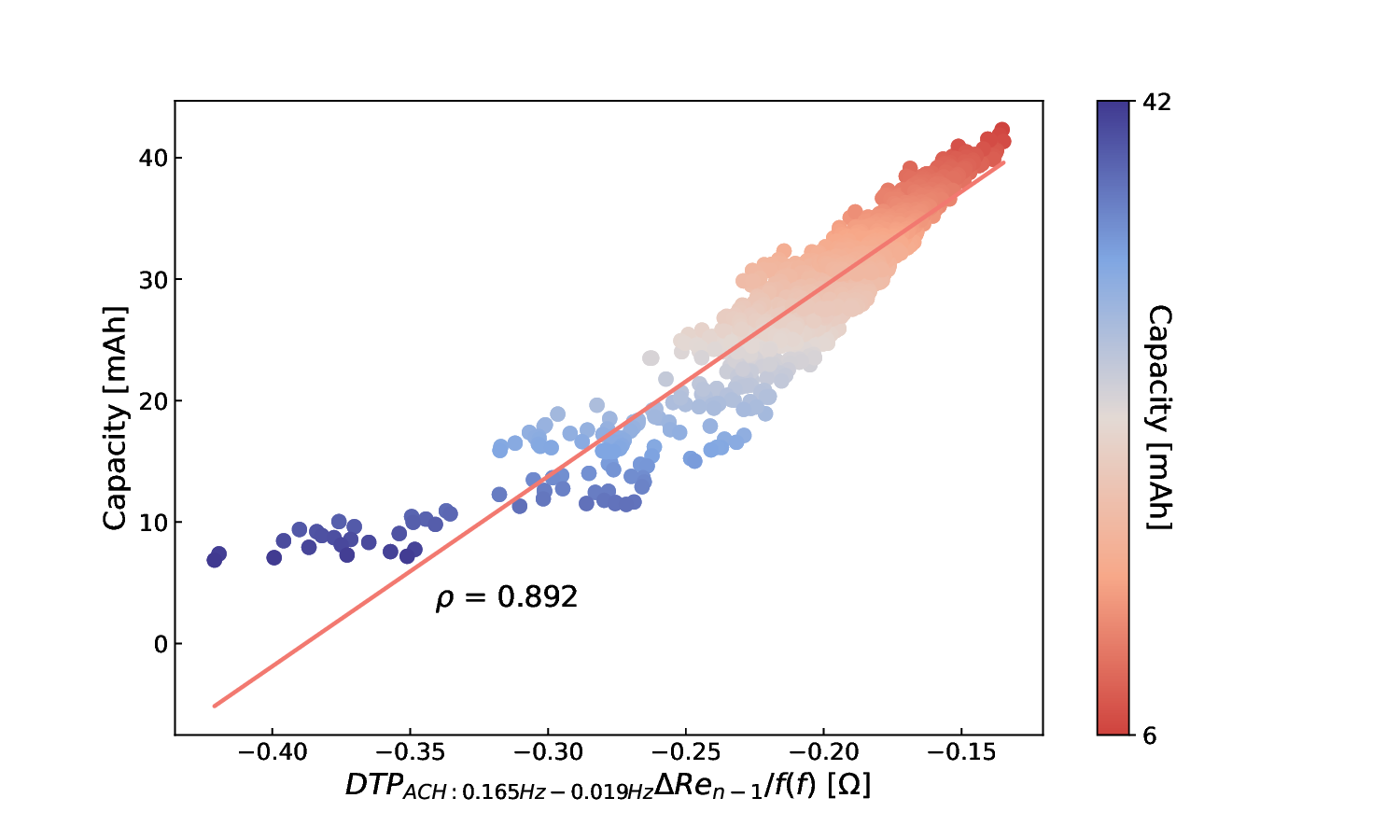}}
	\subfigure[]{\label{fig12:subfig4}\includegraphics[width=0.5\textwidth]{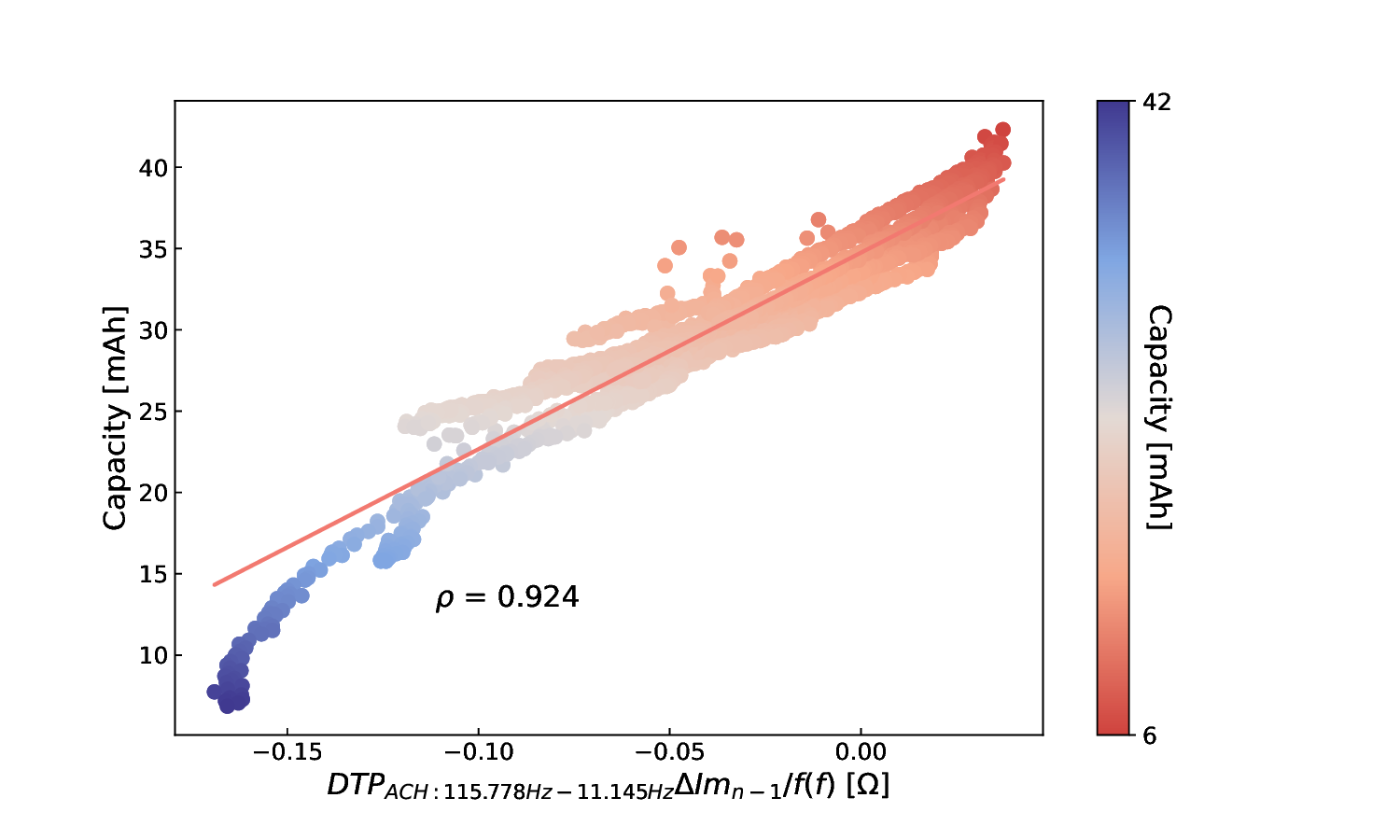}}
	\caption{ Selection of the BTPF using the Pearson correlation coefficient between each candidate two-point feature and capacities of 4 cells in the training set (25C01–25C04). F01 to F60 represent 60 frequencies from 0.02Hz to 20kHz. The colors are determined based on the Pearson correlation coefficient values. (a) Selection of the BTPF on the $\bigtriangleup Re_{n-1}/f$ curve using the Pearson correlation coefficient between each candidate two-point feature and capacities of 4 cells in the training set (25C01–25C04). F01 to F60 represent 60 frequencies from 0.02Hz to 20kHz. The colors are determined based on the Pearson correlation coefficient values. (b) Selection of the BTPF on the $\bigtriangleup Im_{n-1}/f$ curve using the Pearson correlation coefficient between each candidate two-point feature and capacities of 4 cells in the training set (25C01–25C04). F01 to F60 represent 60 frequencies from 0.02Hz to 20kHz. The colors are determined based on the Pearson correlation coefficient values. (c) Capacities 8 cells in Dataset 4 (25C01–25C08) plotted as a function of $DTP_{ACH:f1-f2}\bigtriangleup Re_{n-1}/f(f)$, with a Pearson correlation coefficient of 0.892. The colors are determined based on the capacities of cells. (d) Capacities plotted as a function of $DTP_{ACH:f1-f2}\bigtriangleup Im_{n-1}/f(f)$, with a Pearson correlation coefficient of 0.924. The colors are determined based on the capacities of cells.}
	\label{fig12}
\end{figure}


\textbf{SoH estimation:} Since the Pearson correlation coefficient corresponding to $DTP_{ACH:0.165Hz-0.019Hz}\bigtriangleup Re_{n-1}/f(f)$ is less than that of $DTP_{ACH:115.778Hz-11.145Hz}\bigtriangleup Im_{n-1}/f(f)$, this paper selects $DTP_{ACH:115.778Hz-11.145Hz}\bigtriangleup Im_{n-1}/f(f)$ for ML regression model training and testing. Combining the BTPF $DTP_{ACH:115.778Hz-11.145Hz}\bigtriangleup Im_{n-1}/f(f)$ and the XGBoost regression model, the cell SoH estimation results are shown in Figure 13(a). Combining the raw EIS impedance features \cite{Zhang2020Identifying} and the XGBoost regression model, the cell SoH estimation results are shown in Figure 13(b).

\begin{figure}[htbp]
	\centering
	\subfigure[]{\label{fig13:subfig1}\includegraphics[width=0.6\textwidth]{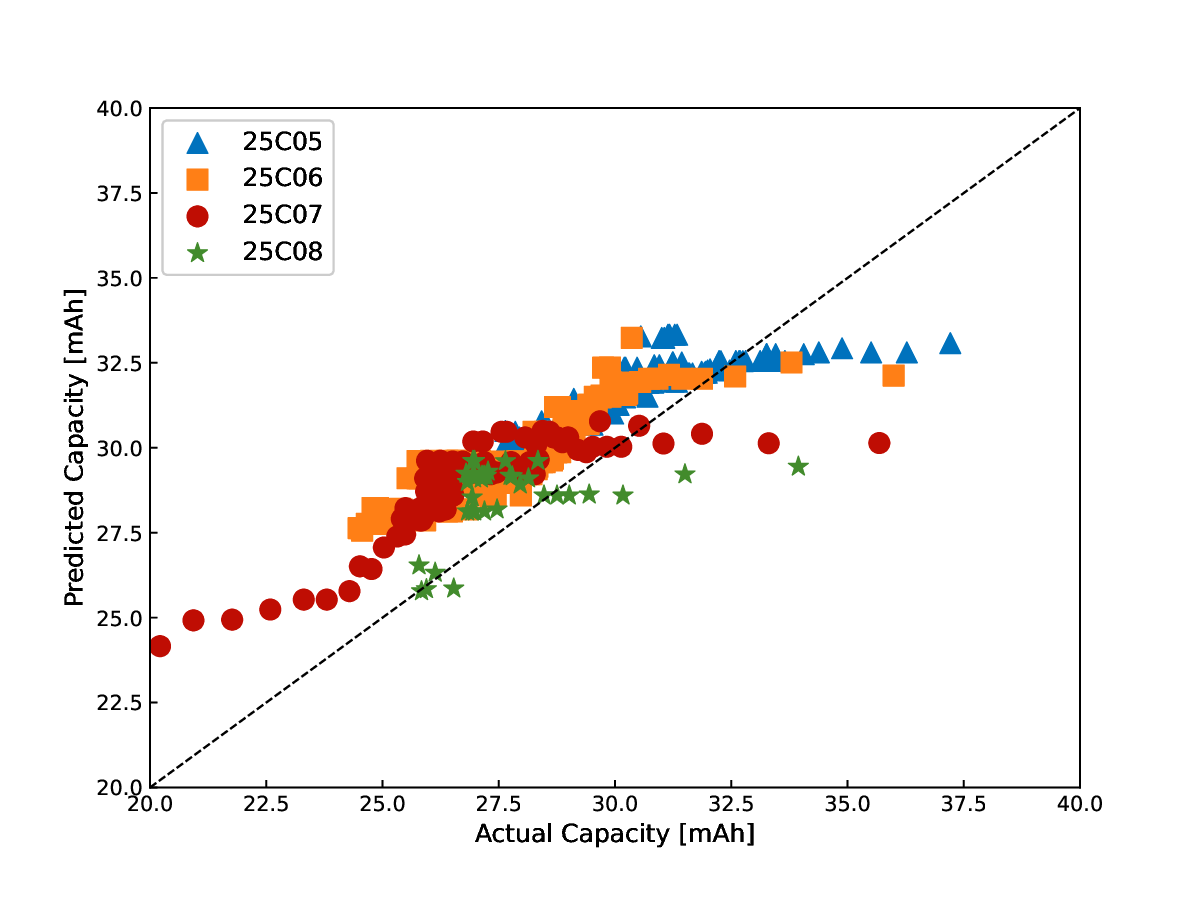}}
	\subfigure[]{\label{fig13:subfig2}\includegraphics[width=0.6\textwidth]{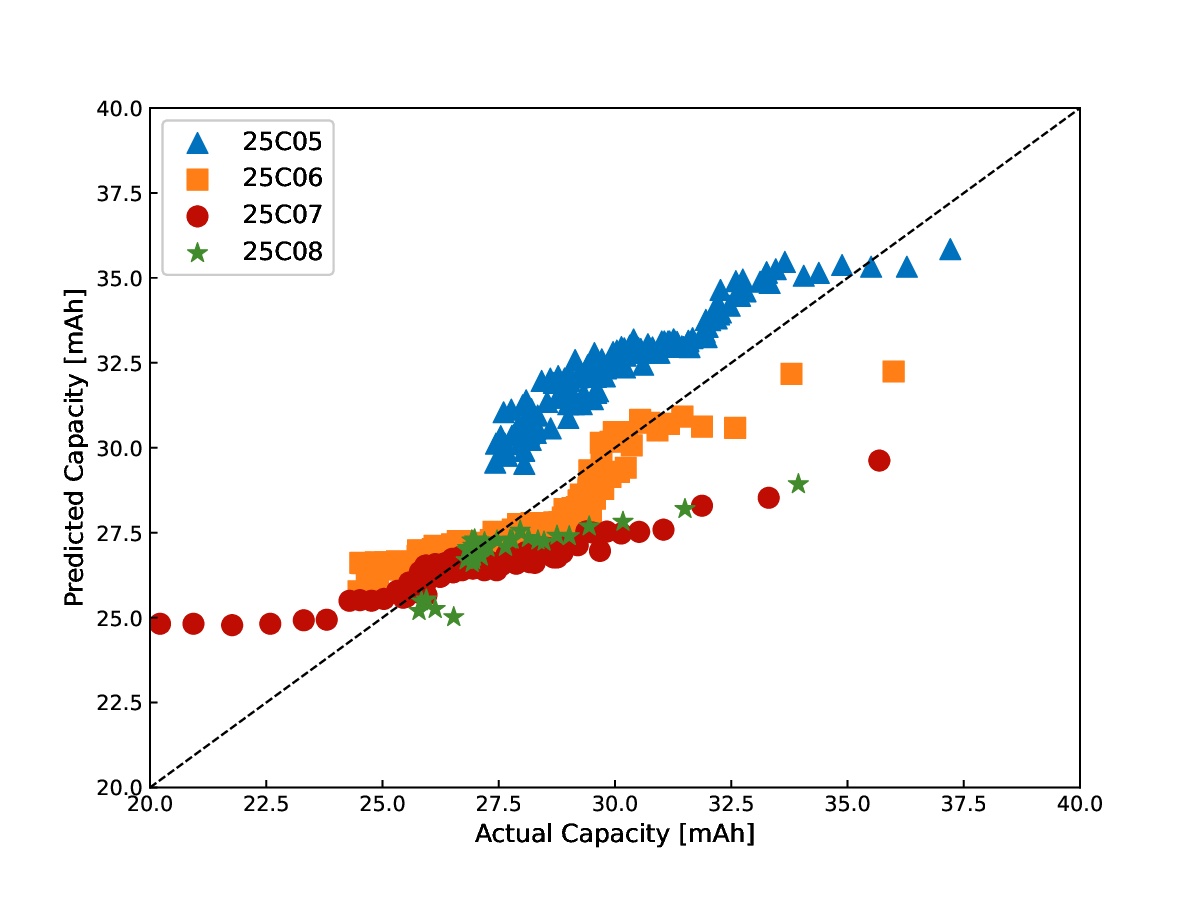}}
	\subfigure[]{\label{fig13:subfig3}\includegraphics[width=0.6\textwidth]{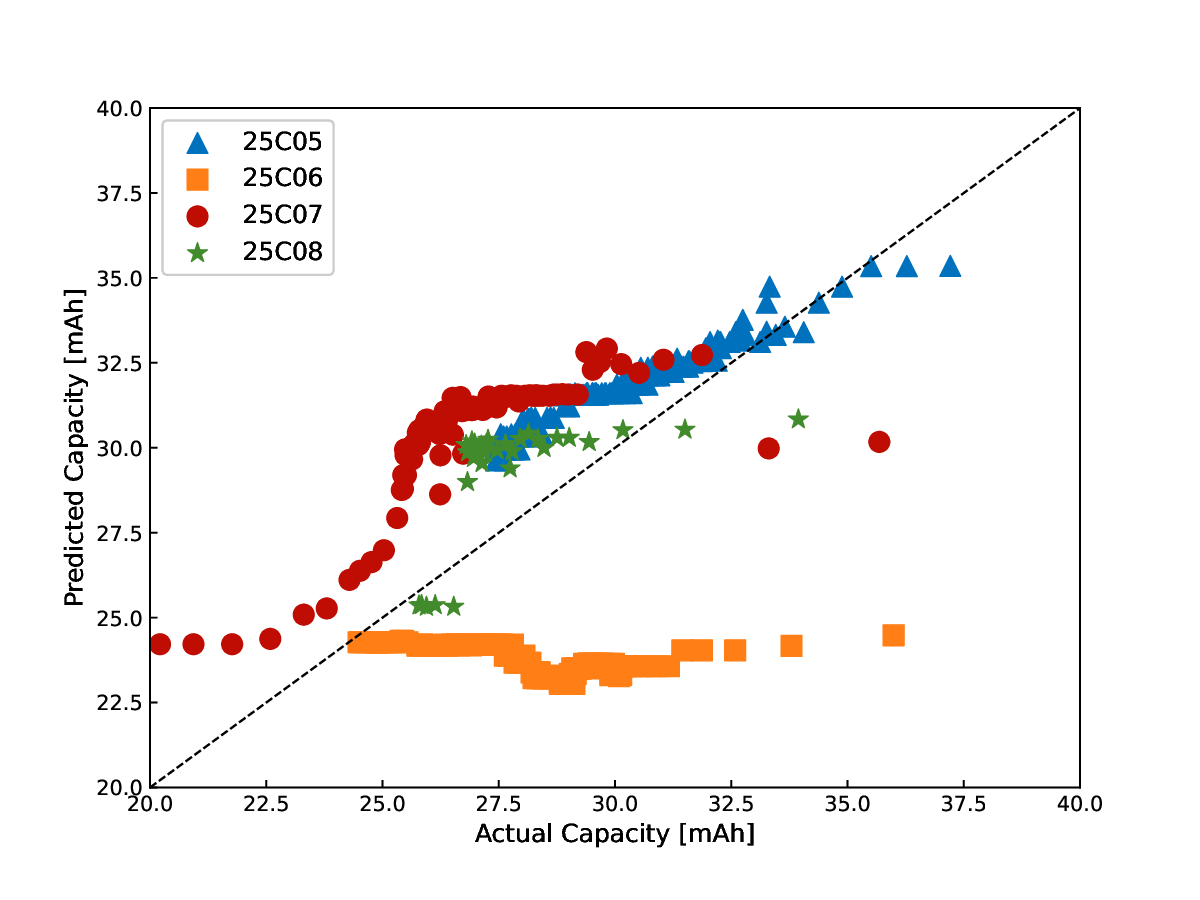}}
	\caption{SoH estimation results of different features on 8 cells in Dataset 4 (25C01–25C08). (a) SoH estimation results of $DTP_{ACH:115.778Hz-11.145Hz}\bigtriangleup Im_{n-1}/f(f)$. The input of the XGBoost regression model is $DTP_{ACH:115.778Hz-11.145Hz}\bigtriangleup Im_{n-1}/f(f)$, and the output is the cell capacity. (b) SoH estimation results of raw EIS impedance features \cite{Zhang2020Identifying}. The inputs of the XGBoost regression model are 120 raw EIS impedances, and the output is the cell capacity. (c) SoH estimation results of two imaginary impedance features corresponding to 17.80 Hz and 2.16 Hz. The inputs of the XGBoost regression model are two imaginary impedances corresponding to 17.80 Hz and 2.16 Hz, and the output is the cell capacity.}
	\label{fig13}
\end{figure}


SoH estimation results of different features on 4 cells in Dataset 4 (25C05–25C08) are shown in Table 4. More detailed results are shown in \textbf{Supplementary Tables 25 to 28}. It can be found that $DTP_{ACH:115.778Hz-11.145Hz}\bigtriangleup Im_{n-1}/f(f)$ achieves a comparable accuracy to that of raw EIS impedance features \cite{Zhang2020Identifying}, while the raw EIS impedance features use 120 impedance data points, which is 60 times that of $DTP_{ACH:115.778Hz-11.145Hz}\bigtriangleup Im_{n-1}/f(f)$. Fewer data points mean less data collection, storage, and computational costs. 

Zhang et al. \cite{Zhang2020Identifying} determined the importance weights of raw EIS impedances at different frequencies through the ARD of the GPR model, and found that the two imaginary impedances, corresponding to 17.80 Hz and 2.16 Hz, have more significant importance weights than other impedances, and the use of these two imaginary impedances can accurately estimate the SoH of the cell. Furthermore, we compare the BTPF $DTP_{ACH:115.778Hz-11.145Hz}\bigtriangleup Im_{n-1}/f(f)$ proposed in this paper with the two imaginary impedance features proposed in \cite{Zhang2020Identifying}, and the SoH estimation results are shown in Figure 12(c) and \textbf{Supplementary Tables 25 to 28}. It can be found that $DTP_{ACH:115.778Hz-11.145Hz}\bigtriangleup Im_{n-1}/f(f)$ achieves more accurate SoH estimation results than two imaginary impedance features \cite{Zhang2020Identifying}.

Furthermore, we trained the ML regression model with all 6 cells in the training set and tested it with all 6 cells in the test set, SoH estimation results of different features on cells 35C02 and 45C02 are shown in \textbf{Supplementary Figure 26} and \textbf{Supplementary Tables 29 to 30}.

(2) \quad BTPF extraction from DRT data

We can also extract the BTPF from the DRT data, which is exactly the same as Dataset 3, as shown in \textbf{Supplementary Figure 22}.

\textbf{Data collection:} By using the open-source software pyDRTtools \cite{Wan2015Influence}, We convert the EIS data after full charge in each cycle (120 real and imaginary impedances corresponding to 60 frequencies between 0.02Hz and 20kHz) to DRT ($\gamma (\tau)/\tau$) data for feature extraction.

\textbf{Difference calculation:} We subtract the DRT data collected in the first cycle from the corresponding data in higher cycles to obtain the $\bigtriangleup \gamma (\tau)_{n-1}/\tau$ curve, as shown in \textbf{Supplementary Figure 27(a)}. The $\bigtriangleup \gamma (\tau)_{n-1}/\tau$ curve in all cycles of the representative cell in Dataset 4 are provided in \textbf{Supplementary Figure 27(b)}.

\textbf{Feature extraction:} The $\bigtriangleup \gamma (\tau)$ values corresponding to any two time values on the $\bigtriangleup \gamma (\tau)_{n-1}/\tau$ curve are subtracted and the absolute value is taken as a candidate two-point feature, as shown in \textbf{Supplementary Figure 24}. Since the time is divided into 600 values between $1.58e^{-5}s$ and $1.24e^2s$ during DRT data calculation, a total of $(600^2-600)/2=179700$ candidate two-point features can be obtained by traversing all combinations of two timescale values on the $\bigtriangleup \gamma (\tau)_{n-1}/\tau$ curve. To reduce the computational burden, we only sample 1 time data point from every 10 time data points to extract candidate two-point features. Then, a total of  $(60^2-60)/2=1770$ candidate two-point features can be obtained by traversing all combinations of two time values. Increasing the time interval does not affect the fairness of the comparison. Further, all candidate two-point features on the $\bigtriangleup \gamma (\tau)_{n-1}/\tau$ curve of 4 cells in the training set (25C01–25C04) are traversed, as shown in Figure 14(a). 

\textbf{Feature selection:} The Pearson correlation coefficient between each candidate two-point feature on the $\bigtriangleup \gamma (\tau)_{n-1}/\tau$ curve and capacities of 4 cells in the training set (25C01–25C04) is calculated, and the candidate two-point feature corresponding to the correlation coefficient with the largest absolute value of 0.961 is selected as the BTPF $DTP_{ACH:0.003s-14.42s}\bigtriangleup \gamma (\tau)_{n-1}/\tau$ of the $\bigtriangleup \gamma (\tau)_{n-1}/\tau$ curve, as shown in Figure 14(a). The two data points on the $\bigtriangleup \gamma (\tau)_{n-1}/\tau$ curve that utilized to calculate $DTP_{ACH:0.003s-14.42s}\bigtriangleup \gamma (\tau)_{n-1}/\tau$ are shown in \textbf{Supplementary Figure 27(b)}.
The distribution relationship between capacities of 8 cells in Dataset 4 (25C01–25C08) and the BTPF $DTP_{ACH:0.003s-14.42s}\bigtriangleup \gamma (\tau)_{n-1}/\tau$ is shown in Figure 14(b). The Pearson correlation coefficient is 0.932.

\begin{figure}[htbp]
	\centering
	\subfigure[]{\label{fig14:subfig1}\includegraphics[width=1\textwidth]{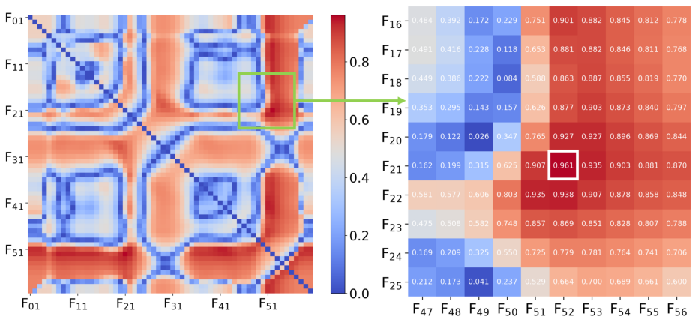}}
	\subfigure[]{\label{fig14:subfig2}\includegraphics[width=0.8\textwidth]{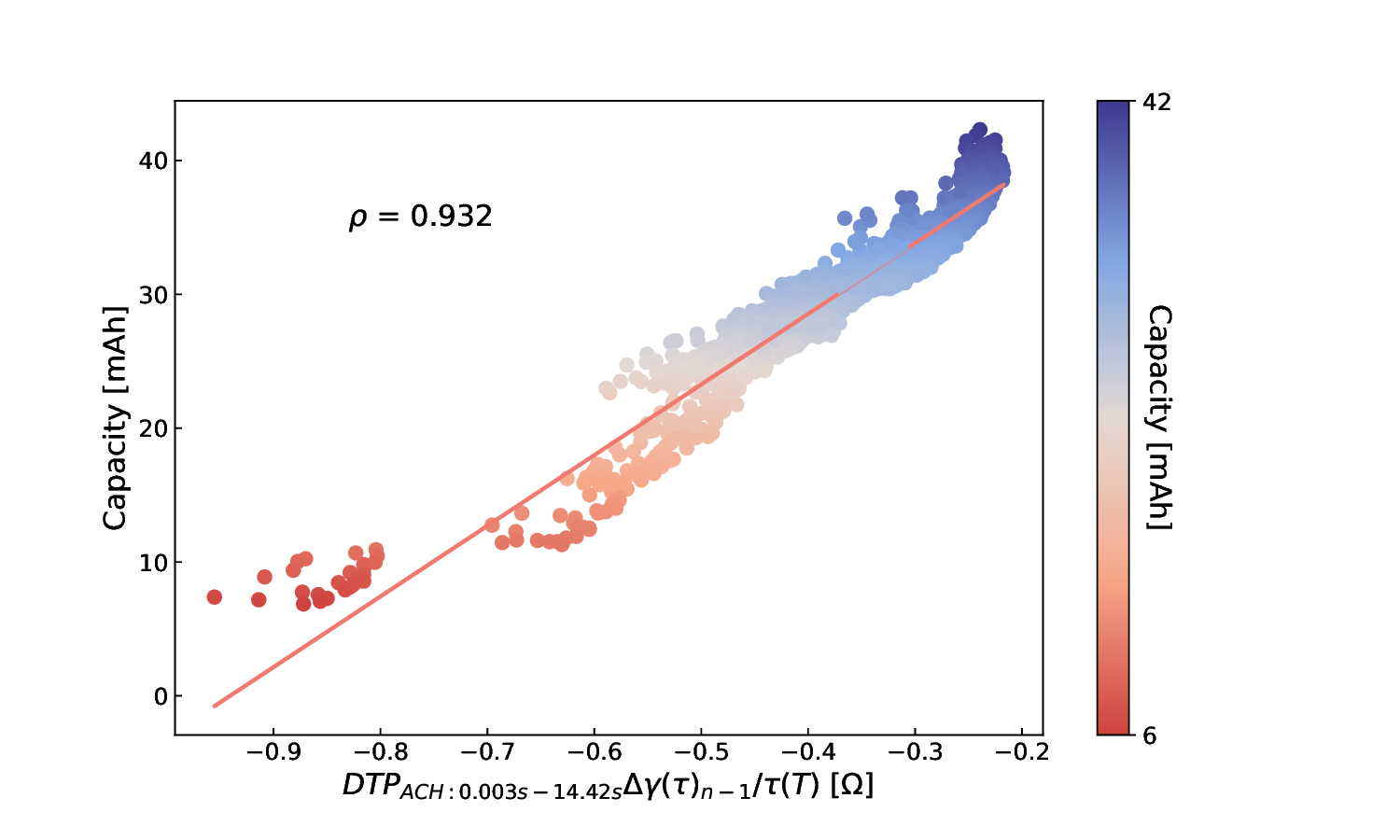}}
	\caption{(a) Selection of the BTPF on the $\bigtriangleup \gamma (\tau)_{n-1}/\tau$ curve using the Pearson correlation coefficient between each candidate two-point feature and capacities of 4 cells in the training set (25C01–25C04). F01 to F60 represent 60 time values between $1.58e^{-5}s$ and $1.24e^2s$. The colors are determined based on the Pearson correlation coefficient values. (b) Capacities of 8 cells in Dataset 3  (25C01–25C08) plotted as functions of $DTP_{ACH:0.003s-14.42s}\bigtriangleup \gamma (\tau)_{n-1}/\tau$. The colors are determined based on the capacities of cells.}
	\label{fig14}
\end{figure}


\textbf{SoH estimation:} Combining the BTPF $DTP_{ACH:0.003s-14.42s}\bigtriangleup \gamma (\tau)_{n-1}/\tau$ and the XGBoost regression model, the battery SoH estimation results of 8 cells in Dataset 3  (25C01–25C08) are shown in Figure 15.

\begin{figure}
	\centering
	\includegraphics[width=10cm]{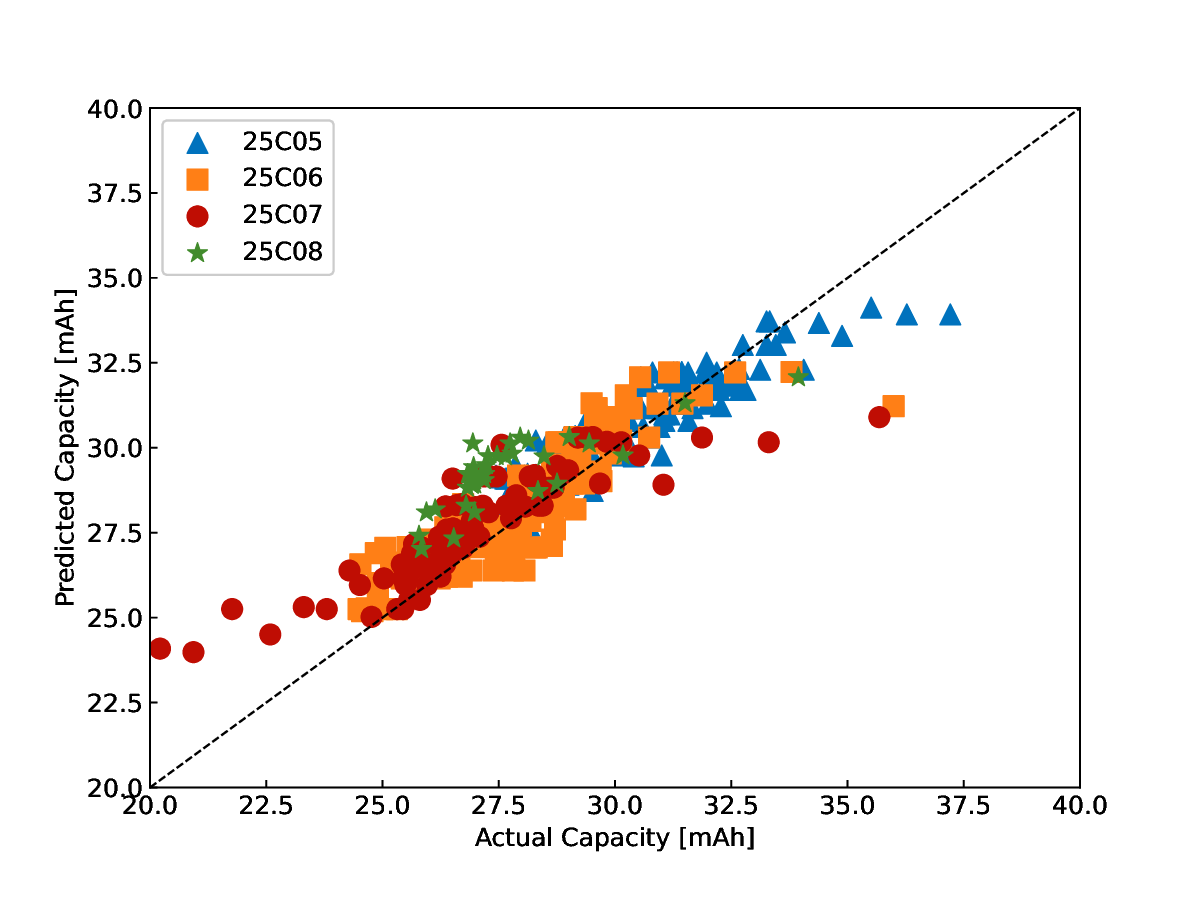}
	\caption{SoH estimation results of $DTP_{ACH:0.003s-14.42s}\bigtriangleup \gamma (\tau)_{n-1}/\tau$ on 8 cells in Dataset 4 (25C01–25C08). The input of the XGBoost regression model is $DTP_{ACH:0.003s-14.42s}\bigtriangleup \gamma (\tau)_{n-1}/\tau$, and the output is the cell capacity.}
	\label{fig15}
\end{figure}


SoH estimation results of $DTP_{ACH:0.003s-14.42s}\bigtriangleup \gamma (\tau)_{n-1}/\tau$ on 4 cells in the test set of Dataset 4 (25C05–25C08) are shown in Table 4. More detailed results are shown in \textbf{Supplementary Tables 31 to 34}. It can be found that $DTP_{ACH:0.003s-14.42s}\bigtriangleup \gamma (\tau)_{n-1}/\tau$ achieves a comparable accuracy to that of raw EIS impedance features and two imaginary impedance features \cite{Zhang2020Identifying}.


\subsubsection{\textbf{Dataset 5}}

For Dataset 5, the BTPF extraction method for diagnosis task is shown in \textbf{Supplementary Figure 28}. To make a fair comparison with SOAT features [Variance, Skewness, Maximum] (abbreviated as [Var, Ske, Max]) proposed by Zhu et al. \cite{Zhu2022Data}, this paper uses the same training set and test set as \cite{Zhu2022Data}. Specifically, the Dataset 5 containing 130 commercial Li-ion cells with different nominal capacities and chemistry types (66 NCA cells with nominal capacity of 3.5 Ah, 55 NCM cells with nominal capacity of 3.5Ah, and 9 NCM+NCA cells with nominal capacity of 2.5 Ah) is divided into three sub-datasets, namely the sub-dataset 1 (66 NCA cells), the sub-dataset 2 (55 NCM cells), and the sub-dataset 3 (9 NCM+NCA cells). Among them, the sub-dataset 1 is divided into a training set (52 NCA cells) and a primary test set (14 NCA cells) by using a stratified sampling method, meaning that the data from the same cell is either in the training set or in the test set. The sub-dataset 2 is used as the secondary test set and the sub-dataset 3 is used as the third test set. To be consistent with \cite{Zhu2022Data}, we first trained the ML regression model with the training set and tested it with the primary test set. Then, the ML regression model is trained on the sub-dataset 1 and tested on the secondary and third test sets. For more details about the training and test sets, please refer to \cite{Zhu2022Data} and will not be repeated here.

\textbf{Data collection:} We collect the relaxation $V/t$ data within 1800s after full charge in each cycle. It should be noted here that in Dataset 5 \cite{Zhu2022Data}, the 1800 s relaxation $V/t$ data of NCA and NCM cells after full charge is collected at a sampling frequency of 120 s, and the 3600 s relaxation $V/t$ data of NCM+NCA cells after full charge is collected at a sampling frequency of 30 s. Therefore, we selected the 1800 s relaxation $V/t$ data after full charge for feature extraction. 

\textbf{Difference calculation:} The relaxation $V/t$ data collected in each cycle is standardized to facilitate subsequent calculations. Specifically, the spline function is utilized to fit the relaxation $V/t$ data, and a unified linear interpolation is performed. The relaxation voltage is fitted as a function of the relaxation time, and the relaxation time linearly is divided into 50 values between 0s and 1800 s at intervals of 36 s, as shown in \textbf{Supplementary Figure 29(a)}. The fitted relaxation $V/t$ data in the 1st cycle is subtracted from the fitted relaxation $V/t$ data in the nth cycles to obtain the relaxation $\bigtriangleup V_{n-1}/t$ curve. As shown in \textbf{Supplementary Figure 29(b)}, the obtained relaxation $\bigtriangleup V_{n-1}/t$ curve is the shaded part between the two relaxation $V/t$ curves in 1st and 10th cycles. The $\bigtriangleup V_{n-1}/t$ curves of the representative cell in Dataset 5 are provided in \textbf{Supplementary Figure 29(c)}.

\textbf{Feature extraction:} The relaxation $\bigtriangleup V$ values corresponding to any two relaxation time values on the $\bigtriangleup V_{n-1}/t$ curve are subtracted and the absolute value is taken as a candidate two-point feature, as shown in \textbf{Supplementary Figure 30}. Since the the relaxation time linearly is divided into 50 values between 0 s and 1800 s at intervals of 36 s, a total of $(50^2-50)/2=1225$ candidate two-point features can be obtained by traversing all combinations of two relaxation time values on the $\bigtriangleup V_{n-1}/t$ curve, as shown in Figure 16(a).

\textbf{Feature selection:} The Pearson correlation coefficient between each candidate two-point feature and capacities of 52 NCA cells in the training set is calculated, and the candidate two-point feature corresponding to the correlation coefficient with the largest absolute value of 0.757 is selected as the BTPF $DTP_{ACH:0s-936s}\bigtriangleup V_{n-1}/t(t)$ on the $\bigtriangleup V_{n-1}/t$ curve, as shown in Figure 16(a). The two data points on the $\bigtriangleup V_{n-1}/t$ curve that utilized to calculate $DTP_{ACH:0s-936s}\bigtriangleup V_{n-1}/t(t)$ are shown in \textbf{Supplementary Figure 29(c)}.
The distribution relationship between capacities of 66 NCA cells in the sub-dataset 1 of Dataset 5 and the BTPF $DTP_{ACH:0s-936s}\bigtriangleup V_{n-1}/t(t)$ on the $\bigtriangleup V_{n-1}/t$ is shown in Figure 16(b). The Pearson correlation coefficient is 0.758.

\begin{figure}[htbp]
	\centering
	\subfigure[]{\label{fig16:subfig1}\includegraphics[width=1\textwidth]{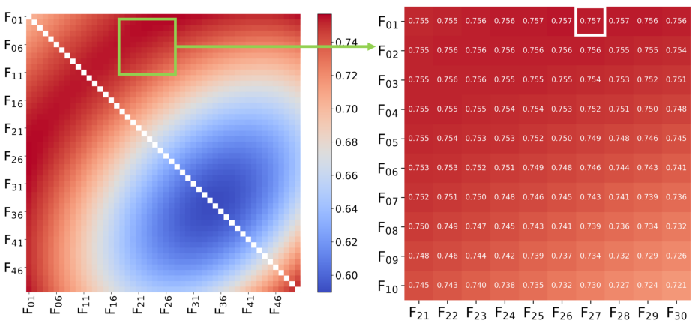}}
	\subfigure[]{\label{fig16:subfig2}\includegraphics[width=0.8\textwidth]{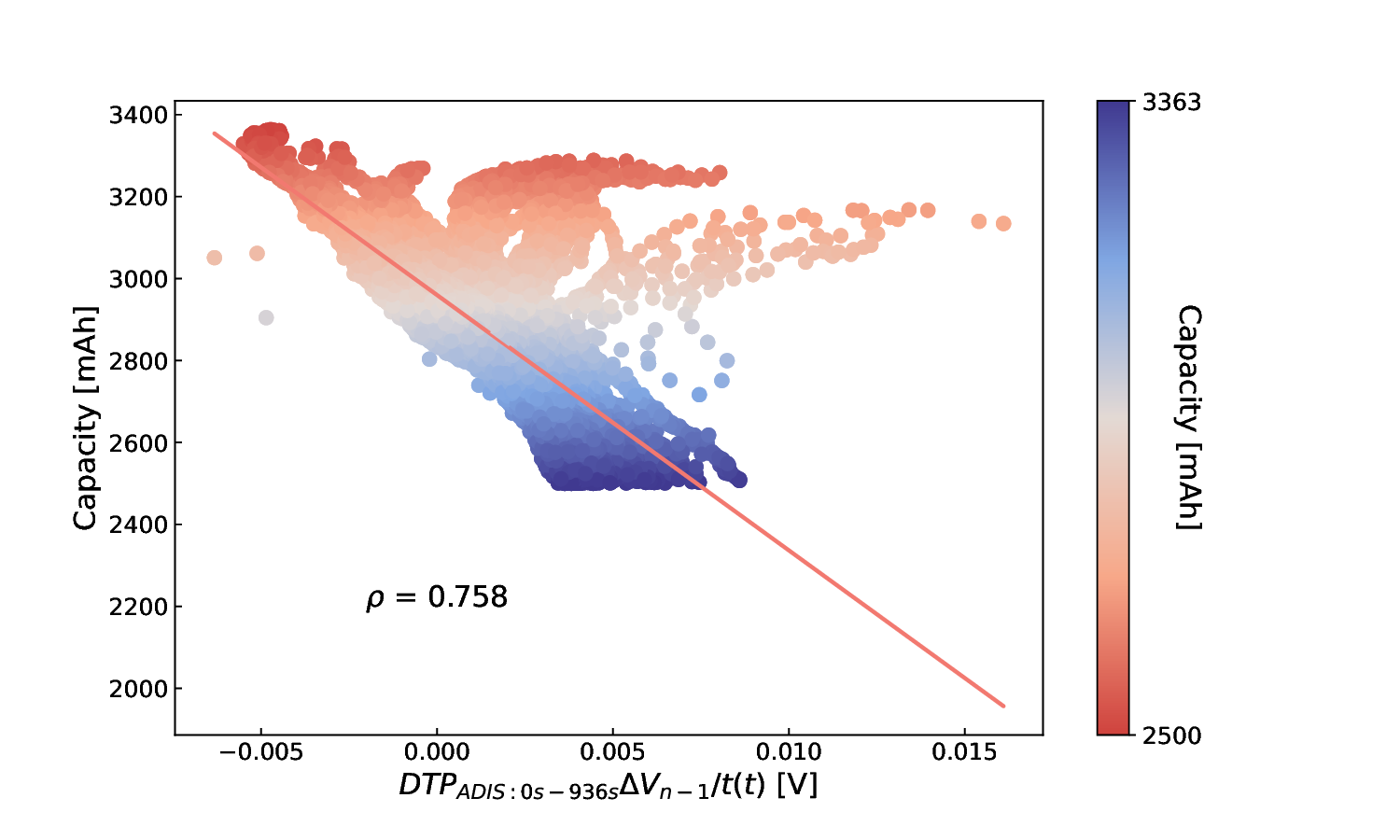}}
	\caption{(a) Selection of the BTPF using the Pearson correlation coefficient between each candidate two-point feature and capacities of 52 NCA cells in the training set. F01 to F50 represent 50 relaxation time values between 0 s and 1800 s at intervals of 36 s. The colors are determined based on the Pearson correlation coefficient values. (b) Capacities of 66 NCA cells in the sub-dataset 1 of Dataset 5 plotted as a function of the BTPF $DTP_{ACH:0s-936s}\bigtriangleup V_{n-1}/t(t)$, with a Pearson correlation coefficient of 0.758.}
	\label{fig16}
\end{figure}


\textbf{SoH estimation:} Combining the BTPF $DTP_{ACH:0s-936s}\bigtriangleup V_{n-1}/t(t)$ and the XGBoost regression model, the battery SoH estimation results of the sub-dataset 1 in Dataset 5 (66 NCA cells) are shown in Figure 17(a). Combining the statistical features [Var, Ske, Max] proposed in \cite{Zhu2022Data} and the XGBoost regression model, the battery SoH estimation results of the sub-dataset 1 in Dataset 5 (66 NCA cells) are shown in Figure 17(b).

\begin{figure}[htbp]
	\centering
	\subfigure[]{\label{fig17:subfig1}\includegraphics[width=0.6\textwidth]{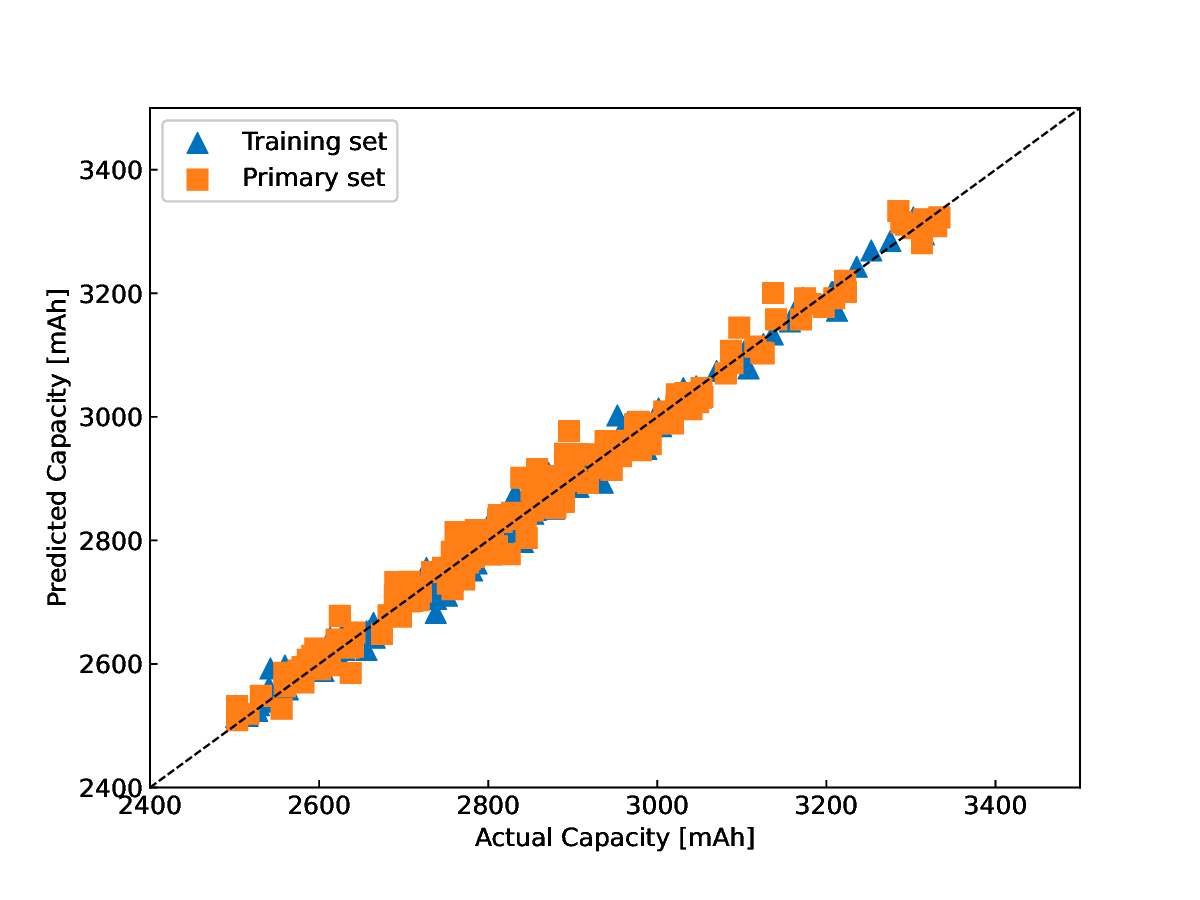}}
	\subfigure[]{\label{fig17:subfig2}\includegraphics[width=0.6\textwidth]{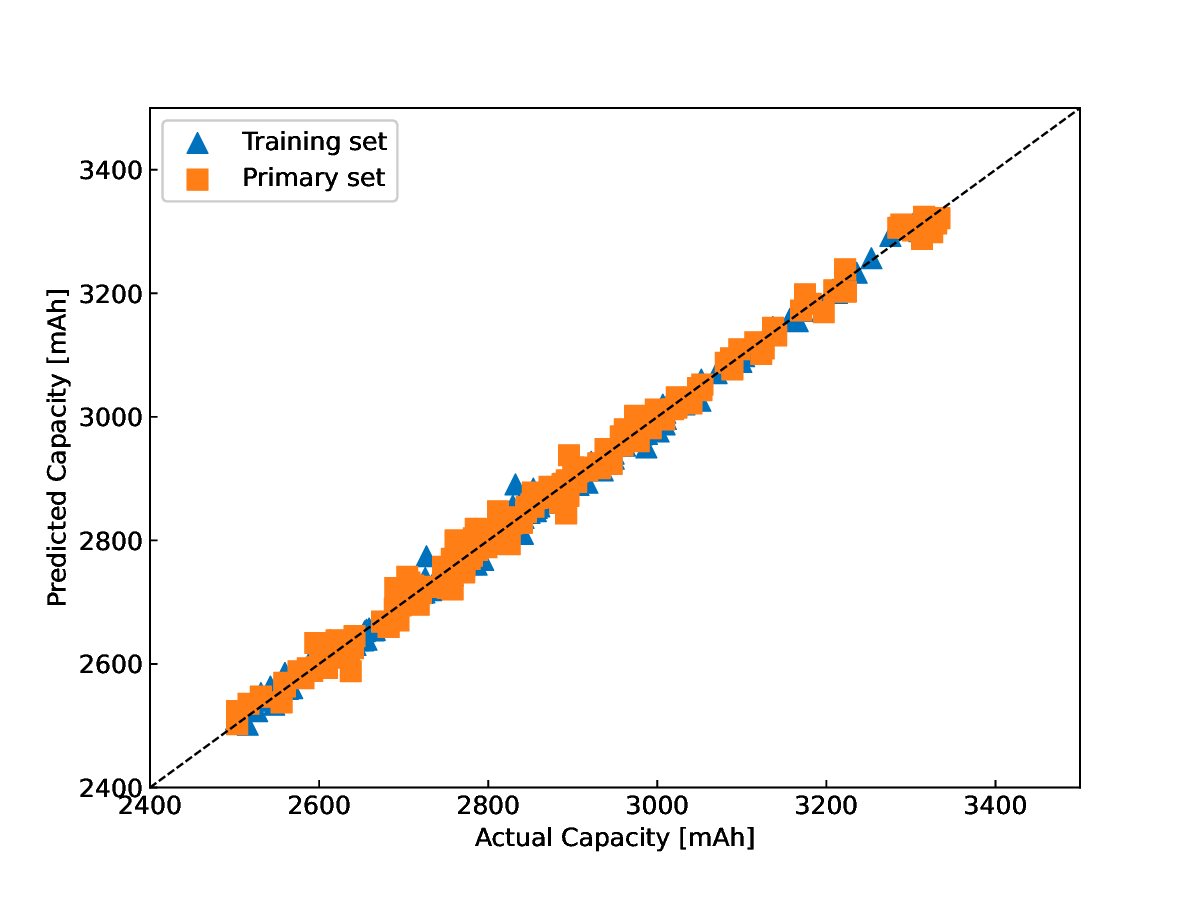}}
	\caption{SoH estimation results of different features on the sub-dataset 1 of Dataset 5 (66 NCA cells). (a) SoH estimation results of $DTP_{ACH:0s-936s}\bigtriangleup V_{n-1}/t(t)$. The input of the XGBoost regression model is $DTP_{ACH:0s-936s}\bigtriangleup V_{n-1}/t(t)$, and the output is the cell capacity. (b) SoH estimation results of statistical features [Var, Ske, Max] \cite{Zhu2022Data}. The inputs of the XGBoost regression model training are statistical features [Var, Ske, Max], and the output is the cell capacity.}
	\label{fig17}
\end{figure}


SoH estimation results of different features on the sub-dataset 1 of Dataset 5 (66 NCA cells) are shown in Table 5. More detailed results are shown in \textbf{Supplementary Tables 35 to 36}. It can be found that $DTP_{ACH:0s-936s}\bigtriangleup V_{n-1}/t(t)$ achieves a comparable accuracy to that of the statistical features [Var, Ske, Max] \cite{Zhu2022Data}. Compared with the statistical features [Var, Ske, Max], $DTP_{ACH:0s-936s}\bigtriangleup V_{n-1}/t(t)$ only needs to collect 2 data points from the relaxation $V/t$ data in each cycle to calculate, while the statistical features [Var, Ske, Max] use 15 data points, which is 7.5 times that of $DTP_{ACH:0s-936s}\bigtriangleup V_{n-1}/t(t)$. Fewer data points mean less data collection, storage, and computing costs.

Moreover, based on 55 NCM cells in the secondary test set of Dataset 3, the generalizability of the proposed BTPF extraction method is tested, and the test results are shown in \textbf{Supplementary Figure 31}. More detailed test results are provided in \textbf{Supplementary Table 37}. Based on 9 NCM+NCA cells in the third test set of Dataset 3, the generalizability of the proposed BTPF extraction method is tested, and the test results are shown in \textbf{Supplementary Figure 32}. More detailed test results are provided in \textbf{Supplementary Table 38}.

\subsubsection{\textbf{Dataset 6}}

For Dataset 6, the BTPF extraction method for diagnosis task is exactly the same as Dataset 5, as shown in \textbf{Supplementary Figure 28}. Since Wildfeuer et al. \cite{Wildfeuer2023Experimental} did not propose features for the battery diagnosis task when publishing the Dataset 6, we still use the SOAT features [Var, Ske, Max] proposed by Zhu et al. \cite{Zhu2022Data} to compare with the BTPF proposed in this paper. Specifically, Dataset 6 containing 196 commercial NCA cells with nominal capacities of 2.5 Ah is randomly divided into a training set and a test ses, where 80\% of the cells are randomly divided into the training set (157 cells) and 20\% of the cells are divided into the test set (39 cells). The data of each cell is either in the training set or in the test set. We trained the ML regression model with the training set and tested it with the test set. For more details about Dataset 6, please refer to \cite{Wildfeuer2023Experimental} and will not be repeated here.

\textbf{Data collection:} We collect the relaxation $V/t$ data within 7200 s after full discharge in each periodic RPT. 

\textbf{Difference calculation:} The relaxation $V/t$ data collected in each RPT is standardized to facilitate subsequent calculations, as shown in \textbf{Supplementary Figure 33(a)}. The fitted relaxation $V/t$ data in the 1st cycle is subtracted from the fitted relaxation $V/t$ data in the nth cycles to obtain the relaxation $\bigtriangleup V_{n-1}/t$ curve, as shown in \textbf{Supplementary Figure 33(b)}. The $\bigtriangleup V_{n-1}/t$ curves of the representative cell in Dataset 6 are provided in \textbf{Supplementary Figure 33(c)}.

\textbf{Feature extraction:} The relaxation $\bigtriangleup V$ values corresponding to any two relaxation time values on the $\bigtriangleup V_{n-1}/t$ curve are subtracted and the absolute value is taken as a candidate two-point feature, as shown in \textbf{Supplementary Figure 30}. Since the the relaxation time linearly is divided into 100 values between 0 s and 7200 s at intervals of 72 s, a total of $(100^2-100)/2=4950$ candidate two-point features can be obtained by traversing all combinations of two relaxation time values on the $\bigtriangleup V_{n-1}/t$ curve, as shown in Figure 18(a).

\textbf{Feature selection:} The Pearson correlation coefficient between each candidate two-point feature and capacities of 157 cells in the training set is calculated, and the candidate two-point feature corresponding to the correlation coefficient with the largest absolute value of 0.896 is selected as the BTPF $DTP_{ADIS:6120s-6984s}\bigtriangleup V_{n-1}/t(t)$ on the $\bigtriangleup V_{n-1}/t$ curve, as shown in Figure 18(a). The two data points on the $\bigtriangleup V_{n-1}/t$ curve that utilized to calculate $DTP_{ADIS:6120s-6984s}\bigtriangleup V_{n-1}/t(t)$ are shown in \textbf{Supplementary Figure 33(c)}. The distribution relationship between capacities of 196 cells in Dataset 6 and the BTPF $DTP_{ADIS:6120s-6984s}\bigtriangleup V_{n-1}/t(t)$ on the $\bigtriangleup V_{n-1}/t$ is shown in Figure 18(b). The Pearson correlation coefficient is 0.804.

\begin{figure}[htbp]
	\centering
	\subfigure[]{\label{fig18:subfig1}\includegraphics[width=1\textwidth]{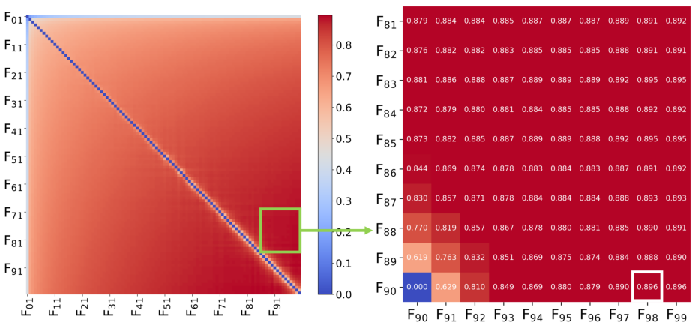}}
	\subfigure[]{\label{fig18:subfig2}\includegraphics[width=0.8\textwidth]{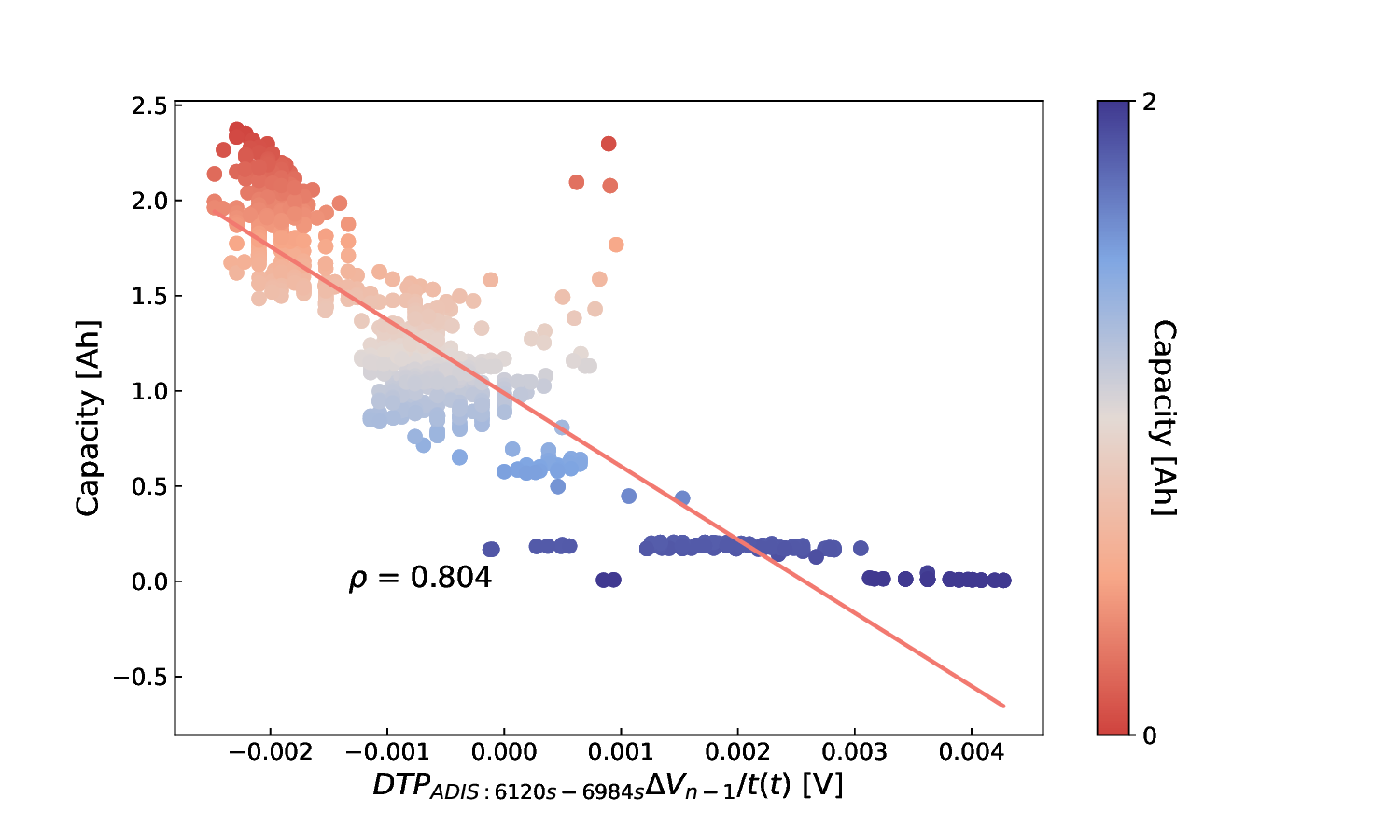}}
	\caption{(a) Selection of the BTPF using the Pearson correlation coefficient between each candidate two-point feature and capacities of 157 cells in the training set. F01 to F100 represent 100 relaxation time values between 0 s and 7200 s at intervals of 72 s. The colors are determined based on the Pearson correlation coefficient values. (b) Capacities of 196 cells in Dataset 6 plotted as a function of $DTP_{ADIS:6120s-6984s}\bigtriangleup V_{n-1}/t(t)$, with a Pearson correlation coefficient of 0.804.}
	\label{fig18}
\end{figure}


\textbf{SoH estimation:} Combining the BTPF $DTP_{ADIS:6120s-6984s}\bigtriangleup V_{n-1}/t(t)$ and the XGBoost regression model, the battery SoH estimation results of 196 cells in Dataset 6 are shown in Figure 19(a). Combining the statistical features [Var, Ske, Max] proposed in \cite{Zhu2022Data} and the XGBoost regression model, the battery SoH estimation results of 196 cells in Dataset 6 are shown in Figure 19(b).

\begin{figure}[htbp]
	\centering
	\subfigure[]{\label{fig19:subfig1}\includegraphics[width=0.8\textwidth]{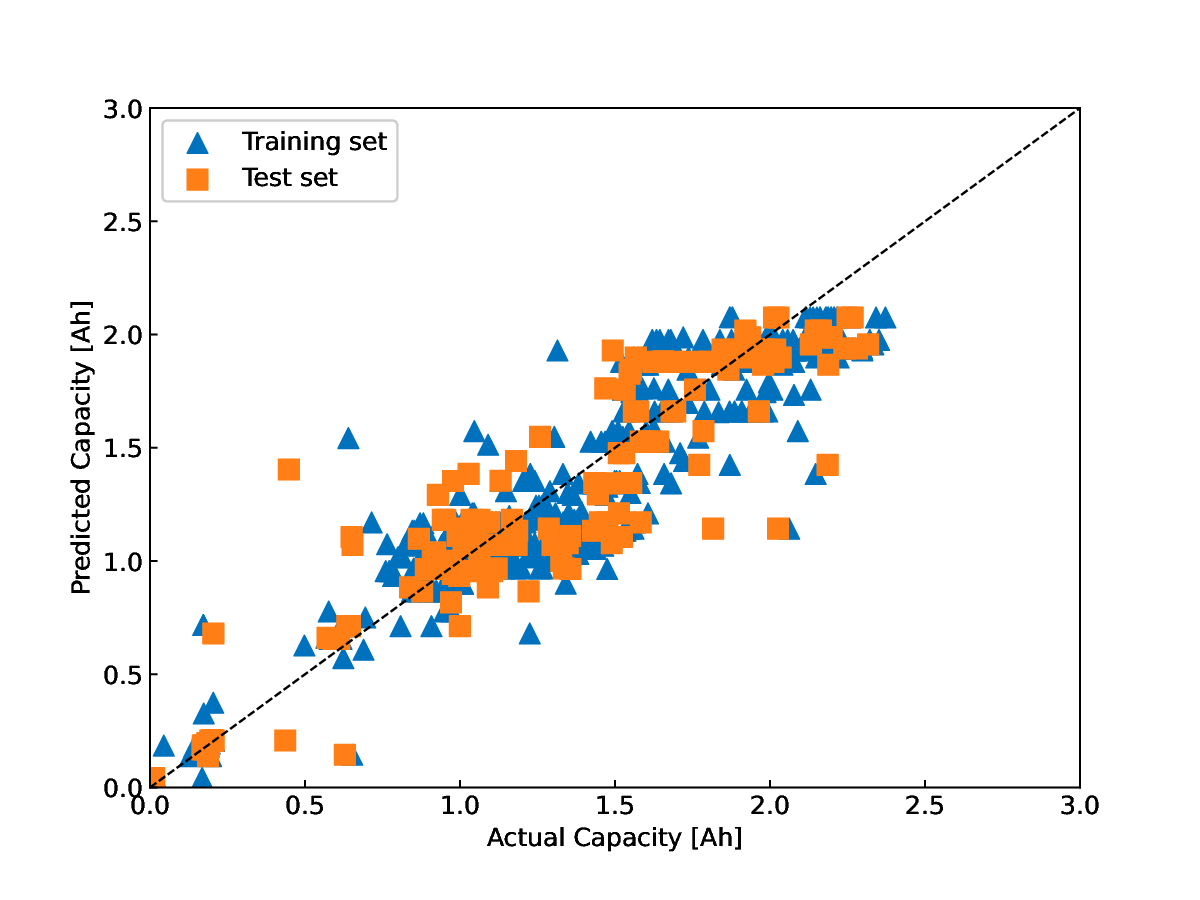}}
	\subfigure[]{\label{fig19:subfig2}\includegraphics[width=0.8\textwidth]{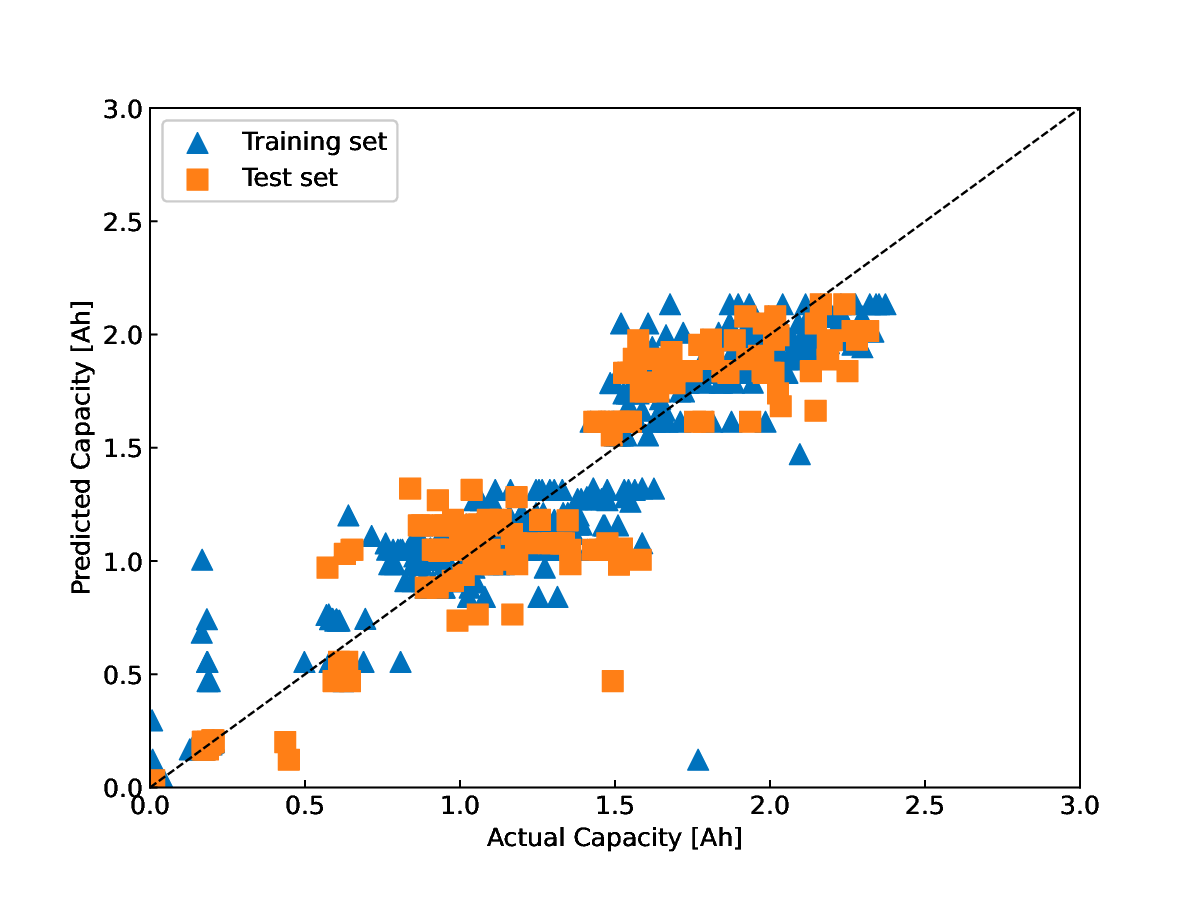}}
	\caption{SoH estimation results of different features on 196 cells in Dataset 6. (a) SoH estimation results of $DTP_{ADIS:6120s-6984s}\bigtriangleup V_{n-1}/t(t)$. The input of the XGBoost regression model is $DTP_{ADIS:6120s-6984s}\bigtriangleup V_{n-1}/t(t)$, and the output is the cell capacity. (b) SoH estimation results of statistical features [Var, Ske, Max]. The input of the XGBoost regression model training is statistical features [Var, Ske, Max], and the output is the cell capacity.}
	\label{fig19}
\end{figure}

\textbf{Figure 19} SoH estimation results of different features on 196 cells in Dataset 6. (a) SoH estimation results of $DTP_{ADIS:6120s-6984s}\bigtriangleup V_{n-1}/t(t)$. The input of the XGBoost regression model is $DTP_{ADIS:6120s-6984s}\bigtriangleup V_{n-1}/t(t)$, and the output is the cell capacity. (b) SoH estimation results of statistical features [Var, Ske, Max]. The input of the XGBoost regression model training is statistical features [Var, Ske, Max], and the output is the cell capacity.

SoH estimation results of different features on 196 cells of Dataset 6 are shown in Table 6. More detailed results are shown in \textbf{Supplementary Tables 39 to 40}. It can be found that $DTP_{ADIS:6120s-6984s}\bigtriangleup V_{n-1}/t(t)$ achieves a comparable accuracy to that of the statistical features [Var, Ske, Max], while the statistical features [Var, Ske, Max] use about 720 data points, which is 360 times that of $DTP_{ADIS:6120s-6984s}\bigtriangleup V_{n-1}/t(t)$. Fewer data points mean less data collection, storage, and computing costs.

\subsubsection{\textbf{Rationalization of diagnostic performance}}

Results in \cite{Iurilli2021On} and \cite{Guo2023Battery} show that EIS curve and its changing trend contain rich information about Li-ion battery aging. In EIS, the high and medium frequency impedance spectrum that represents the solid electrolyte interphase resistance $R_{sei}$ and the charge transfer resistance $R_{ct}$  can be utilized to quantify the LLI, and the low frequency impedance spectrum indicating the diffusion of Li-ions can be utilized to quantify the $LAM$ \cite{Guo2023Battery}. Results in \cite{Chen2021Detection} and \cite{Lu2022The} show that the identified multi-timescale information and the changing trend of DRT curves can be utilized to quantify the LLI and $LAM$ in Li-ion battries. Similar to the raw impedance features and statistical features extracted from EIS \cite{Jones2022Impedance, Zhang2020Identifying} and DRT curves \cite{Zhu2023Adaptive, Su2024Modeling}, we attribute the success of BTPFs extracted from EIS and DRT curves in the diagnosis task to capturing the changing trends of $LAM$ and LLI. 

Results in \cite{Chen2022Battery, Chen2021A} show that relaxation voltage curve and its changing trend can be utilized to characterize the aging modes of Li-ion batteries, such as SEI-driven $LLI$, Li plating related $LLI$, and $LAM_{PE}$. Similar to the raw voltage features and statistical features extracted from relaxation voltage curves \cite{Zhu2022Data, Fan2023Battery}, we attribute the success of BTPFs extracted from relaxation voltage curves in the diagnosis task to capturing the changing trends of $LAM$ and $LLI$.

\section{Conclusions}

Data-driven prognostic and diagnostic methods are promising in accelerating the optimization (including the optimization of design, production, and management) and safer operation of Li-ion batteries. We proposed a BTPF extraction method based on the first principle and conducted comprehensive tests on 820 cells from 6 open source datasets (covering 5 cell chemistries, 7 cell manufacturers, and 3 cell data types). The proposed BTPFs can achieve prognostic and diagnostic accuracy comparable to the SOAT features using only two data points in each cycle. This work well answers an important question that has been unresolved for data-driven battery prognostic and diagnostic methods: What is the minimum amount of data required to extract features for accurate battery prognosis and diagnosis? The answer given by this work is that two points are enough! The success of BTPFs is rationalized by effectively capturing the changing trends of $LLI$ and $LAM$ during cell degradation. In general, this work challenges the cognition of existing studies on the difficulty of battery prognosis and diagnosis tasks, subverts the fixed pattern of establishing prognosis and diagnosis methods for complex dynamic systems through deliberate feature engineering, highlights the promise of data-driven methods for field battery prognosis and diagnosis applications, and provides a new benchmark for future studies.

\section{Data availability}

The Dataset 1 used in this study is an open source dataset published by Severson et al. \cite{severson2019data} (124 LFP cells) and Attia et al. \cite{attia2020closed} (45 LFP cells), and available at https://data.matr.io/1. 
The Dataset 2 used in this study is an open source dataset published by Li et al. \cite{Li2024Predicting} (225 cells) and available at https://doi.org/10.25380/iastate.22582234.
The Dataset 3 used in this study is an open source dataset published by Jones et al. \cite{Jones2022Impedance} (88 cells) and available at  https://doi.org/10.5281/zenodo.6645536.
The Dataset 4 used in this study is an open source dataset published by 
Zhang et al. \cite{Zhang2020Identifying} (12 cells) and available at https://doi.org/10.5281/zenodo.3633835.
The Dataset 5 used in this study is an open source dataset published by 
Zhu et al. \cite{Zhu2022Data} (130 cells) and available at https://doi.org/10.5281/zenodo.6379165.
The Dataset 6 used in this study is an open source dataset published by
Wildfeuer et al. \cite{Wildfeuer2023Experimental} (196 cells) and available at https://mediatum.ub.tum.de/1713382.

\section{Code availability}

The code used in this study is available at https://github.com/Zhao-YB/DatasetX.

\section{Acknowledgements}

This work is supported by the National Natural Science Foundation of Zhejiang Province (Grant No. LQ23E050013).

\section{Author contributions}

H.L. conceived the method. 
H.L., Y.Z., and J.C. conducted the tests. 
H.L. and Y.Z. developed the best two-point feature extraction method. 
H.L. and Y.Z. performed the diagnosis and prognosis modeling. 
H.L., Y.Z., H.Z., Z.D., M.C., X.W., and Z.L performed the data collection, data management, and raw data preprocessing. 
H.L., Y.Z., X. F., J. L., and J.C. performed the visualization and interpreted the results.
H.L., Y.Z., H.Z., and J.C. wrote the original draft.
H.L., Y.Z., H.Z., X. F., J. L., and J.C. edited the manuscript. 
All authors reviewed the manuscript. 
H.L. and J.C. supervised the work.

\section{Competing interests}

The authors declare no competing interests.


\bibliography{mybibfile}

\end{document}